\documentclass[AMA,STIX1COL]{WileyNJD-v2}
\pdfoutput=1
\usepackage{moreverb}
\usepackage{gensymb}
\usepackage{tikz}
\usepackage{alltt}
\usepackage{upquote}
\usepackage{float}
\usepackage{array}
\usepackage{pgfplots}
\usetikzlibrary{positioning}
\usepackage{graphicx}
\usepackage{amsmath,mathtools}
\usetikzlibrary{calc,decorations.markings,fit,shapes,chains,arrows,patterns,shadings}
\usepackage{amsthm}
\usepackage{amsfonts}
\usepackage{enumitem}
\usepackage{subcaption}
\usepackage{caption}
\usepackage{xcolor}
\usepackage{scalerel}
\usepackage{algorithm}
\usepackage{multicol}
\usepackage{mathtools}

\def\vec#1{\mbox{\boldmath $#1$}}

\colorlet{lightBlack}{black!30}

\newcolumntype{M}[1]{>{\centering\arraybackslash}m{#1}}
\newcolumntype{N}{@{}m{0pt}@{}}

\tikzset{
  block3/.style={rectangle, draw, rounded corners, text centered,text width = 9em, minimum height = 10em},
  line/.style={draw, -latex'}
  }

\graphicspath{{Pictures/}}

\newcommand\BibTeX{{\rmfamily B\kern-.05em \textsc{i\kern-.025em b}\kern-.08em
T\kern-.1667em\lower.7ex\hbox{E}\kern-.125emX}}

\articletype{Article Type}%

\received{<day> <Month>, <year>}
\revised{<day> <Month>, <year>}
\accepted{<day> <Month>, <year>}

\begin{document}

\title{A hybrid variational Allen-Cahn/ALE scheme for the coupled analysis of two-phase fluid-structure interaction}

\author[1]{Vaibhav Joshi$^1$}

\author[1]{Rajeev K. Jaiman$^{1,*}$}

\authormark{JOSHI \textsc{et al}}

\address[1]{\orgdiv{Department of Mechanical Engineering}, \orgname{National University of Singapore}, \country{Singapore}}



\corres{*Corresponding author name, Corresponding address. \email{mperkj@nus.edu.sg}}

\presentaddress{Present address}

\abstract[Abstract]{We present a novel partitioned iterative formulation for modeling of fluid-structure interaction in two-phase flows. 
The variational formulation consists of a stable and robust integration of three blocks of differential equations, viz., incompressible viscous fluid, a rigid or flexible structure and two-phase indicator field.
The fluid-fluid interface between the two phases, which may have high density and viscosity ratios, is evolved by solving 
the conservative phase-field Allen-Cahn equation in the arbitrary Lagrangian-Eulerian coordinates. 
%
While the Navier-Stokes equations are solved by a stabilized Petrov-Galerkin method, the conservative Allen-Chan phase-field equation is discretized by the positivity preserving variational scheme. 
Fully decoupled implicit solvers for the two-phase fluid and the structure are integrated by the nonlinear iterative force correction 
in a staggered partitioned manner and the generalized-$\alpha$ method is employed for the time marching. 
We assess the accuracy and stability of the new phase-field/ALE 
variational formulation for two- and three-dimensional problems involving the dynamical 
interaction of rigid bodies with free-surface. 
We consider the decay test problems of increasing complexity, namely 
free translational heave decay of a circular cylinder  and 
free rotation  of a rectangular barge. 
Through numerical experiments, we show that the proposed formulation is stable 
and robust for high density ratios across 
fluid-fluid interface and for low structure-to-fluid mass ratio with strong added-mass effects. 
Overall, the proposed variational formulation produces results with high accuracy 
and compare well with available measurements and reference numerical data.
Using three-dimensional unstructured meshes, 
we demonstrate the second-order temporal accuracy of  the coupled phase-field/ALE method via decay test of a circular 
cylinder interacting with the free-surface.
Finally, we demonstrate the three-dimensional phase-field FSI formulation for a practical problem of internal two-phase flow in a flexible circular pipe subjected to vortex-induced vibrations due to external fluid flow.   }


\jnlcitation{\cname{%
\author{V. Joshi},  and 
\author{R. K. Jaiman}} (\cyear{2018}), 
\ctitle{A hybrid variational Allen-Cahn/ALE scheme for the coupled analysis of two-phase fluid-structure interaction}, \cjournal{International Journal for Numerical Methods in Engineering}, \cvol{xxxx;xx:xx--xx}.}

\maketitle

\section{Introduction}
Fluid-structure interaction (FSI) involving two-phase flows finds its applications in offshore pipelines conveying oil or gas \cite{Sorensen, Chakrabarti}, marine vessels exposed to free-surface waves, blood flow through veins and arteries, 
and multiphase flow inside heat exchangers \cite{Vist}. Of particular interest to the present study is the internal flow in offshore risers which are subjected to turbulent external flow associated with the ocean currents as well as the motion of the offshore vessel induced by wave-structure interaction. These offshore structures may undergo self-excited vibrations and fluid-elastic instabilities \cite{blevins, FSI_1}, which may lead to structural failure 
and operational delay due to nonlinear dynamical effects of fluid-structure interaction. High-fidelity numerical 
simulations can play an important role to understand the nonlinear coupled physics as well as to provide guidelines for engineering design and optimization. 
The development of robust, efficient and general integration procedure of two-phase flow interacting with freely moving rigid and 
flexible bodies poses serious challenges from a computational viewpoint. 
In particular, the accurate modeling of free-surface motion with topological changes 
and strong FSI effects at high Reynolds number remain a daunting task in computational science and engineering.

\begin{figure}
\begin{subfigure}[b]{0.5\textwidth}
\centering
\begin{tikzpicture}[decoration={markings,mark=at position 1.0 with {\arrow{>}}},scale=0.22,every node/.style={scale=0.833}]
	\draw[fill=white,draw=none] plot [smooth] coordinates {  (-8,7) (-3,9)  (-0.5,9)  (2.5,8)} -- (8,8) --plot [smooth] coordinates { (15,5.7) (16.5,6)  (19,6)  (24,8)} --(24,16) --(-8,16) -- (-8,7)  ;
	\draw[fill={lightBlack},draw=none] plot [smooth] coordinates {  (-8,6.6) (-3,8.6)  (-0.5,8.6)  (2.5,7.6)} -- (8,7.6) -- plot [smooth] coordinates { (15,5.3) (16.5,5.6)  (19,5.6)  (24,7.6)} -- (24,0)--(-8,0) -- (-8,7)  ;
	
	\foreach \i[evaluate={\col=(\i+1)/30*100}] in {0,...,30}
      \fill[color=white!\col!lightBlack]
      plot [smooth] coordinates {  (-8,6.6+\i*0.4/30) (-3,8.6+\i*0.4/30)  (-0.5,8.6+\i*0.4/30)  (2.5,7.6+\i*0.4/30)} -- (8,7.6+\i*0.4/30) -- plot [smooth] coordinates { (15,5.3+\i*0.4/30) (16.5,5.6+\i*0.4/30)  (19,5.6+\i*0.4/30)  (24,7.6+\i*0.4/30)} -- (24,7.6+\i*0.4/30+0.4/30) --plot [smooth] coordinates { (24,7.6+\i*0.4/30+0.4/30) (19,5.6+\i*0.4/30+0.4/30)  (16.5,5.6+\i*0.4/30+0.4/30)  (15,5.3+\i*0.4/30+0.4/30)} -- (8,7.6+\i*0.4/30+0.4/30) -- plot [smooth] coordinates {  (2.5,7.6+\i*0.4/30+0.4/30) (-0.5,8.6+\i*0.4/30+0.4/30)  (-3,8.6+\i*0.4/30+0.4/30)  (-8,6.6+\i*0.4/30+0.4/30)} -- cycle  ;
	\draw (24,0) -- (-8,0)--(-8,16)--(24,16)--(24,0);
	\draw[fill={black! 2}] plot [smooth,scale=0.7] coordinates {(7.7,3.3) (5.8,4.0) (3.1,10.9) (6.7,13.8) (10.4,16.5) (17.0,15.5) (20.7,12.3) (21.5,7) (21.2,5.5) (19.6,4.0) (15.0,4.2) (12.3,2.7) (10.1,2.9) (7.7,3.3) };
	\draw (9,6.5) node(A){$\Omega^\mathrm{s}$};
	\draw (19,13) node(A){$\Omega_1^\mathrm{f}(0)$};
	\draw (19,3) node(A){$\Omega_2^\mathrm{f}(0)$};
	\draw (21,7.5) node(A){$\Gamma^\mathrm{ff}$};
	\draw (9,12.5) node(A){$\Gamma^\mathrm{fs}$};
\end{tikzpicture}
\caption{}
\end{subfigure}%
\begin{subfigure}[b]{0.5\textwidth}
\centering
\begin{tikzpicture}[decoration={markings,mark=at position 1.0 with {\arrow{>}}},scale=0.22,every node/.style={scale=0.833}]
	\draw[fill=white,draw=none] plot [smooth] coordinates {  (-8,6) (-3,8)  (-0.5,8)  (2.5,7) (5,7.5)} -- (8,8) -- plot [smooth] coordinates { (15,6.7) (16.5,7)  (19,7)  (24,9)} --(24,16) --(-8,16) -- (-8,6)  ;
	\draw[fill=lightBlack,draw=none] plot [smooth] coordinates {  (-8,5.6) (-3,7.6)  (-0.5,7.6)  (2.5,6.6) (5,7.1)} -- (8,7.6) -- plot [smooth] coordinates { (15,6.3) (16.5,6.6)  (19,6.6)  (24,8.6)} --(24,0) --(-8,0) -- (-8,5.6)  ;

	\foreach \i[evaluate={\col=(\i+1)/30*100}] in {0,...,30}
      \fill[color=white!\col!lightBlack]
        plot [smooth] coordinates {  (-8,5.6+\i*0.4/30) (-3,7.6+\i*0.4/30)  (-0.5,7.6+\i*0.4/30)  (2.5,6.6+\i*0.4/30) (5,7.1+\i*0.4/30)} -- (8,7.6+\i*0.4/30) -- plot [smooth] coordinates { (15,6.3+\i*0.4/30) (16.5,6.6+\i*0.4/30)  (19,6.6+\i*0.4/30)  (24,8.6+\i*0.4/30)} -- (24,8.6+\i*0.4/30+0.4/30) -- plot [smooth] coordinates { (24,8.6+\i*0.4/30+0.4/30) (19,6.6+\i*0.4/30+0.4/30)  (16.5,6.6+\i*0.4/30+0.4/30)  (15,6.3+\i*0.4/30+0.4/30)} -- (8,7.6+\i*0.4/30+0.4/30) -- plot [smooth] coordinates {  (5,7.1+\i*0.4/30+0.4/30) (2.5,6.6+\i*0.4/30+0.4/30)  (-0.5,7.6+\i*0.4/30+0.4/30)  (-3,7.6+\i*0.4/30+0.4/30) (-8,5.6+\i*0.4/30+0.4/30)} -- cycle;	
	\draw (24,0) -- (-8,0)--(-8,16)--(24,16)--(24,0);
	\draw[fill={black! 2}] plot [smooth,scale=0.7] coordinates {(9.5,3.3) (8,4.3) (6,9) (8.5,13.8) (12.2,16.7) (18.8,14) (22.5,12.3) (23.3,6.5) (23,5.5) (21.4,4.0) (16.8,3.0) (14.1,2.7) (11.9,2.9) (9.5,3.3) };
		\draw[dashed] plot [smooth,scale=0.7] coordinates {(7.7,3.3) (5.8,4.0) (3.1,10.9) (6.7,13.8) (10.4,16.5) (17.0,15.5) (20.7,12.3) (21.5,7) (21.2,5.5) (19.6,4.0) (15.0,4.2) (12.3,2.7) (10.1,2.9) (7.7,3.3) };
	\draw[dashed] plot [smooth] coordinates {  (-8,7) (-3,9)  (-0.5,9)  (2.35,8) };
	\draw[dashed] plot [smooth] coordinates {  (15,5.7) (16.5,6)  (19,6)  (24,8) };
	\draw (10,6.5) node(A){$\Omega^\mathrm{s}(t)$};
	\draw (19,13) node(A){$\Omega_1^\mathrm{f}(t)$};
	\draw (19,3) node(A){$\Omega_2^\mathrm{f}(t)$};
	\draw (20,9) node(A){$\Gamma^\mathrm{ff}(t)$};
	\draw (9,12.5) node(A){$\Gamma^\mathrm{fs}(t)$};
\end{tikzpicture}
\caption{}
\end{subfigure}
\begin{subfigure}[b]{0.5\textwidth}
\centering
\begin{tikzpicture}[decoration={markings,mark=at position 1.0 with {\arrow{>}}},scale=0.25,every node/.style={scale=0.833}]
	\draw [fill=white,draw=none] (-2,0) -- (-2,13)--(-7,13)--(-7,0)--(-2,0);
	\draw [fill=black!2,draw=none] (-2,0) -- (-2,13)--(3,13)--(3,0)--(-2,0);
    \draw (-7,0) -- (3,0) -- (3,13) -- (-7,13) -- (-7,0);
    \draw (-2,0) -- (-2,13);
	\draw (-5,7) node{$\Omega^\mathrm{f}$};
	\draw (1,7) node{$\Omega^\mathrm{s}$};
	\draw (-3,12) node {$\Gamma^\mathrm{fs}$};
	\draw [->,thick] (-7,0) -- (-5,0);
	\draw (-5,-1) node {$x$};
	\draw [->,thick] (-7,0) -- (-7,2);
	\draw (-8,2) node {$y$};
	
	\draw [->] (7,0) -- (20,0);
	\draw [->] (7,0) -- (7,13);
	\draw[red,thick] (7, 1) -- (13,1) -- (13,12) -- (19,12) ;
	\draw (6,1) node {$\rho^\mathrm{f}$};
	\draw (6,12) node {$\rho^\mathrm{s}$};
	\draw [dashed] (13,12) -- (7,12);
	\draw (20,-1) node {$x$};
\end{tikzpicture}
\caption{}
\end{subfigure}
\begin{subfigure}[b]{0.5\textwidth}
\centering
\begin{tikzpicture}[decoration={markings,mark=at position 1.0 with {\arrow{>}}},scale=0.25,every node/.style={scale=0.833}]
	\draw [fill=white,draw=none] (-2.5,0) -- (-2.5,13)--(-7,13)--(-7,0)--(-2.5,0);
	\draw [fill=lightBlack,draw=none] (-1.5,0) -- (-1.5,13)--(3,13)--(3,0)--(-1.5,0);
	\foreach \i[evaluate={\col=(\i+1)/30*100}] in {0,...,30}
      \fill[color=white!\col!lightBlack]
      (-1.5-\i*1/30,0) -- (-1.5-\i*1/30,13) -- (-1.5-\i*1/30-1/30,13) -- (-1.5-\i*1/30-1/30,0) -- cycle  ;
    \draw (-7,0) -- (3,0) -- (3,13) -- (-7,13) -- (-7,0);
	\draw (-5,7) node{$\Omega^\mathrm{f}_{1}$};
	\draw (1,7) node{$\Omega^\mathrm{f}_{2}$};
	\draw[->] (-4,5) -- (-2.5,5);
	\draw[<-] (-1.5,5)--(0,5);
	\draw (0,5)--(1,4);
	\draw (-2.5,6)--(-2.5,4);
	\draw (-1.5,6)--(-1.5,4);
	\draw (1,3.5) node{$\sim 4\varepsilon$};
	\draw (-2,12) node {$\Gamma^\mathrm{ff}$};
	\draw [->,thick] (-7,0) -- (-5,0);
	\draw (-5,-1) node {$x$};
	\draw [->,thick] (-7,0) -- (-7,2);
	\draw (-8,2) node {$y$};
	
	\draw [->] (7,0) -- (20,0);
	\draw [->] (7,0) -- (7,13);
	\draw[red,thick] (7, 1) -- (11,1) ;
	\draw[red,thick] (14,12) -- (19,12) ;
	\draw (6,1) node {$\rho^\mathrm{f}_{1}$};
	\draw (6,12) node {$\rho^\mathrm{f}_{2}$};
	\draw [dashed] (13,12) -- (7,12);
	\draw (20,-1) node {$x$};
	\draw[red,thick,scale=5.5,shift={(2.35cm, 1.182cm)}] plot [smooth] coordinates {(-0.500000,-1.000000)(-0.495000,-1.000000)(-0.490000,-1.000000)(-0.485000,-1.000000)(-0.480000,-1.000000)(-0.475000,-1.000000)(-0.470000,-1.000000)(-0.465000,-1.000000)(-0.460000,-1.000000)(-0.455000,-1.000000)(-0.450000,-1.000000)(-0.445000,-1.000000)(-0.440000,-1.000000)(-0.435000,-1.000000)(-0.430000,-1.000000)(-0.425000,-1.000000)(-0.420000,-1.000000)(-0.415000,-1.000000)(-0.410000,-1.000000)(-0.405000,-1.000000)(-0.400000,-1.000000)(-0.395000,-1.000000)(-0.390000,-1.000000)(-0.385000,-1.000000)(-0.380000,-1.000000)(-0.375000,-1.000000)(-0.370000,-1.000000)(-0.365000,-1.000000)(-0.360000,-1.000000)(-0.355000,-1.000000)(-0.350000,-1.000000)(-0.345000,-1.000000)(-0.340000,-1.000000)(-0.335000,-1.000000)(-0.330000,-1.000000)(-0.325000,-1.000000)(-0.320000,-1.000000)(-0.315000,-1.000000)(-0.310000,-1.000000)(-0.305000,-1.000000)(-0.300000,-1.000000)(-0.295000,-1.000000)(-0.290000,-1.000000)(-0.285000,-1.000000)(-0.280000,-1.000000)(-0.275000,-1.000000)(-0.270000,-1.000000)(-0.265000,-1.000000)(-0.260000,-1.000000)(-0.255000,-1.000000)(-0.250000,-1.000000)(-0.245000,-1.000000)(-0.240000,-1.000000)(-0.235000,-1.000000)(-0.230000,-1.000000)(-0.225000,-1.000000)(-0.220000,-1.000000)(-0.215000,-1.000000)(-0.210000,-1.000000)(-0.205000,-1.000000)(-0.200000,-1.000000)(-0.195000,-1.000000)(-0.190000,-1.000000)(-0.185000,-1.000000)(-0.180000,-1.000000)(-0.175000,-1.000000)(-0.170000,-1.000000)(-0.165000,-1.000000)(-0.160000,-1.000000)(-0.155000,-1.000000)(-0.150000,-1.000000)(-0.145000,-1.000000)(-0.140000,-1.000000)(-0.135000,-1.000000)(-0.130000,-1.000000)(-0.125000,-1.000000)(-0.120000,-1.000000)(-0.115000,-1.000000)(-0.110000,-1.000000)(-0.105000,-0.999999)(-0.100000,-0.999999)(-0.095000,-0.999997)(-0.090000,-0.999994)(-0.085000,-0.999988)(-0.080000,-0.999976)(-0.075000,-0.999950)(-0.070000,-0.999900)(-0.065000,-0.999796)(-0.060000,-0.999587)(-0.055000,-0.999163)(-0.050000,-0.998303)(-0.045000,-0.996561)(-0.040000,-0.993037)(-0.035000,-0.985929)(-0.030000,-0.971668)(-0.025000,-0.943364)(-0.020000,-0.888386)(-0.015000,-0.785916)(-0.010000,-0.608859)(-0.005000,-0.339523)(0.000000,0.000000)(0.005000,0.339523)(0.010000,0.608859)(0.015000,0.785916)(0.020000,0.888386)(0.025000,0.943364)(0.030000,0.971668)(0.035000,0.985929)(0.040000,0.993037)(0.045000,0.996561)(0.050000,0.998303)(0.055000,0.999163)(0.060000,0.999587)(0.065000,0.999796)(0.070000,0.999900)(0.075000,0.999950)(0.080000,0.999976)(0.085000,0.999988)(0.090000,0.999994)(0.095000,0.999997)(0.100000,0.999999)(0.105000,0.999999)(0.110000,1.000000)(0.115000,1.000000)(0.120000,1.000000)(0.125000,1.000000)(0.130000,1.000000)(0.135000,1.000000)(0.140000,1.000000)(0.145000,1.000000)(0.150000,1.000000)(0.155000,1.000000)(0.160000,1.000000)(0.165000,1.000000)(0.170000,1.000000)(0.175000,1.000000)(0.180000,1.000000)(0.185000,1.000000)(0.190000,1.000000)(0.195000,1.000000)(0.200000,1.000000)(0.205000,1.000000)(0.210000,1.000000)(0.215000,1.000000)(0.220000,1.000000)(0.225000,1.000000)(0.230000,1.000000)(0.235000,1.000000)(0.240000,1.000000)(0.245000,1.000000)(0.250000,1.000000)(0.255000,1.000000)(0.260000,1.000000)(0.265000,1.000000)(0.270000,1.000000)(0.275000,1.000000)(0.280000,1.000000)(0.285000,1.000000)(0.290000,1.000000)(0.295000,1.000000)(0.300000,1.000000)(0.305000,1.000000)(0.310000,1.000000)(0.315000,1.000000)(0.320000,1.000000)(0.325000,1.000000)(0.330000,1.000000)(0.335000,1.000000)(0.340000,1.000000)(0.345000,1.000000)(0.350000,1.000000)(0.355000,1.000000)(0.360000,1.000000)(0.365000,1.000000)(0.370000,1.000000)(0.375000,1.000000)(0.380000,1.000000)(0.385000,1.000000)(0.390000,1.000000)(0.395000,1.000000)(0.400000,1.000000)(0.405000,1.000000)(0.410000,1.000000)(0.415000,1.000000)(0.420000,1.000000)(0.425000,1.000000)(0.430000,1.000000)(0.435000,1.000000)(0.440000,1.000000)(0.445000,1.000000)(0.450000,1.000000)(0.455000,1.000000)(0.460000,1.000000)(0.465000,1.000000)(0.470000,1.000000)(0.475000,1.000000)(0.480000,1.000000)(0.485000,1.000000)(0.490000,1.000000)(0.495000,1.000000)(0.500000,1.000000)} ;
	\draw[->] (11,5) -- (12.4,5);
	\draw[<-] (13.4,5)--(15,5);
	\draw (12.4,6)--(12.4,4);
	\draw (13.4,6)--(13.4,4);
	\draw (15,5)--(16,4);
	\draw (16,3.5) node{$\sim 4\varepsilon$};
\end{tikzpicture}
\caption{}
\end{subfigure}
\caption{Schematic of two-phase fluid-structure interaction at (a) the initial configuration $t=0$, and (b) some deformed configuration of the structure at time $t>0$. $\Omega^\mathrm{f}(0)$, $\Omega^\mathrm{s}$ and $\Omega^\mathrm{f}(t)$, $\Omega^\mathrm{s}(t)$ are the fluid and the structural domains at $t=0$ and some time $t>0$ respectively with (c) sharp fluid-structure interface, and (d) diffused fluid-fluid interface, smeared using the internal length scale 
parameter $\varepsilon$. $\Gamma^\mathrm{fs}$ and $\Gamma^\mathrm{ff}$ denote the fluid-structure 
and fluid-fluid interfaces, respectively. }
\label{schematic_definition}
\end{figure}
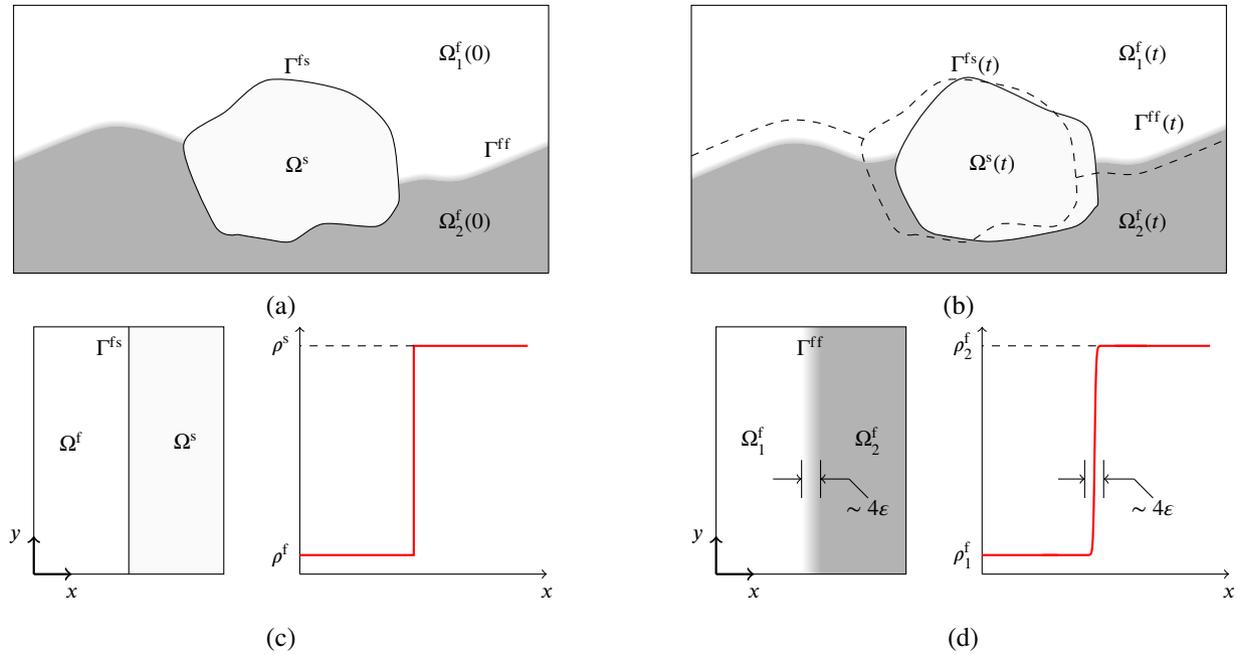

Two-phase FSI involves complex nonlinear interface dynamics and the difficulties associated with the boundary conditions, viz., the evolution of the fluid-fluid interface with complex surface phenomena, the no-slip condition at the structure in the neighborhood of highly-deformable fluid-fluid interface, and the precise movement of the fluid-structure interface. A typical schematic of the problem is shown in Fig.~\ref{schematic_definition}. It can be observed that fluid-fluid interface has to be evolved with the deformation of the structure while satisfying the no-slip condition at the fluid-structure interface. This can be achieved by either considering the structural domain as Lagrangian and then solving the two-phase flow equations in the arbitrary Lagrangian Eulerian (ALE) coordinates, or with the help of an immersed boundary technique where the equations are solved in an Eulerian grid with boundary conditions represented by a fictitious force field \cite{Mittal, SOTIROPOULOS20141}. The modeling of boundary-layer vorticity flux and  near-wall turbulence at the fluid-structure interface is accurate if one considers a boundary-conforming grid 
(e.g., ALE framework) to track the fluid-structure interface. Due to the accuracy 
consideration along the interface for the two-phase FSI analysis, 
we employ the ALE framework for the moving fluid domain i.e., 
the physical properties vary as a sharp discontinuity at the fluid-structure interface as shown in Fig. \ref{schematic_definition}(c) and the Eulerian fluid mesh follows the moving interface at all time with the precise satisfaction of the boundary conditions. 

The robust and efficient modeling of the three-dimensional fluid-fluid interface using a sharp interface description via ALE-type interface tracking technique is a non-trivial task, especially in problems involving any topological changes of the interface. Herein, the fluid-fluid interface is described by a diffuse interface description based phase-field method where the interface is distributed over a finite width $(\mathcal{O}(\varepsilon))$ across which the physical properties vary gradually as a function of the order parameter $\phi$ as shown in Fig. \ref{schematic_definition}(d). Unlike the widely used level-set, volume-of-fluid (VOF), and front-tracking methods, the phase-field method is based on the minimization of a free energy functional which drives the evolution of the interface. The phase-field method is a type of interface-capturing method (i.e., interface is solved over a fixed Eulerian domain), which has an advantage in dealing with any topological change in the fluid-fluid interface without involving any complex re-initialization (level-set) or geometric reconstruction (VOF). A recent review of the phase-field methods for multiphase flows can be found in \cite{Kim}. 

Most of the works carried out in the literature involving vortex-induced vibrations (VIV) have focused on the FSI involving single-phase flows \cite{SARPKAYA2004389}. Extensive experiments and numerical simulations have been performed to study the VIV dynamics of different kinds of structures subjected to external single-phase flows. However, there are very few works on the FSI involving internal flow \cite{Pettigrew, Riverin} which are mainly concerned with the experiments. Some of the recent works based on two-phase FSI simulations are \cite{Calderer} and \cite{ZHENG201619} where an immersed-boundary based level-set approach and a spectral/hp element method based phase-field technique are employed, respectively. The numerical study of \cite{ZHENG201619} solves the Cahn-Hilliard equation \cite{Cahn_Hilliard} for the interface evolution. While the high order of the Cahn-Hilliard equation poses numerical challenges, the second-order Allen-Cahn equation \cite{Allen_Cahn} is relatively simpler to implement for complex phase-field FSI modeling in three dimensions using the standard finite element framework.  The present work presents the two-phase FSI formulation by considering 
the conservative Allen-Cahn phase-field equation for the interface evolution. To the best of our knowledge, 
such variational FSI formulation with two-phase flows does not exist in literature.    

In this paper, we propose a robust and efficient variational formulation for the fluid-structure interaction involving incompressible viscous two-phase flows. The rigid/flexible body equations are solved in the Lagrangian framework with the two-phase flow equations written in the ALE reference coordinate system. 
The coupled two-phase fluid-structure equations are solved in the nonlinear partitioned iterative format by consistent variational finite element formulation in arbitrarily complex three-dimensional domains.
The fluid-fluid interface is evolved by the conservative Allen-Cahn equation written in the ALE framework which takes care of the moving fluid-structure interface due to the structural deformation. While the  two-phase fluid domain with the Navier-Stokes equations is discretized using Petrov-Galerkin finite-element 
and semi-discrete time stepping, the conservative phase-field Allen-Cahn is discretized by the positivity 
preserving variational scheme which provides the boundedness, the mass conservation and the energy stability 
in the underlying discretization \cite{PPV,AC_JCP}.
Identical order of interpolation has been used for all the fluid and phase-field variables, which implies their collocated
arrangement at discrete nodes of unstructured finite element mesh.
A nonlinear iterative force correction (NIFC) scheme \cite{NIFC_1, NIFC_2} is employed for updating the hydrodynamic forces from the fluid flow to the structure in strongly-coupled fluid-structure interaction. To achieve stability at low structure-to-fluid mass ratios, the approximate interface force generated through iterations  is corrected 
via nonlinear force transformation in the predictor-corrector format.
The temporal discretizations of the fluid, the structure and the phase-field equations are performed by 
the generalized-$\alpha$ method together with the partitioned iterative solution strategy.
The moving mesh/ALE characteristic of the fluid-structure interface with the interface capturing technique for the fluid-fluid interface forms our new hybrid ALE/phase-field formulation, which is presented 
for the first time in this paper.  
The salient features of the phase-field FSI formulation are: (i) accurate and stable variational FSI formulation for low structure-to-fluid mass ratios, (ii) consistent two-phase flow formulation for high density and viscosity ratios, 
(iii) robustness to handle topological changes in the fluid-fluid interface with FSI, and (iv) ease of implementation and flexibility in existing variational solvers 
due to the partitioned block-iterative coupling.
These desirable features of the proposed variational framework together with increasing complexity of two-phase FSI 
problems are covered in the present paper.

The organization of the paper is as follows. Section \ref{formulation} reviews the governing equations for the two-phase fluid flow (Navier-Stokes and Allen-Cahn) and the structure with their corresponding variational formulations. The proposed partitioned iterative coupling of the fluid-structure interaction and two-phase flows is presented in Section \ref{proposed_scheme}. The proposed formulation is then assessed numerically in Section \ref{tests} via decay tests involving translation of a circular cylinder and rotation of a rectangular barge. The results are compared with the experiments and simulations from the literature. For a practical application, we demonstrate the vortex-induced vibration of a flexible riser exposed to a uniform current with an internal two-phase flow in Section \ref{demonstration}. Finally, we conclude the paper with some of the key findings in Section \ref{conclusion}.

\section{Coupled fluid-structure formulation for two-phase flows}
\label{formulation}
In this section, we describe the governing equations of the two-phase flow and the structure and their corresponding variational formulations. We first review the Navier-Stokes equations for two-phase flows along with the conservative Allen-Cahn equation which evolves the fluid-fluid interface. Thereafter, we discuss the structural equation and the treatments of the fluid-structure and the fluid-fluid interfaces. 

\subsection{The Navier-Stokes equations for two-phase flow}
We present the differential equation for the two-phase flow at the continuous level and then review its semi-discrete and variational formulation using the stabilized finite element framework.  
\subsubsection{Strong differential form}
Consider a spatial domain $\Omega^\mathrm{f}(t)$ consisting of the spatial points $\boldsymbol{x}^\mathrm{f}$ at temporal coordinate $t$. The boundary to the domain $\Gamma^\mathrm{f}(t)$ consists of three components, the Dirichlet boundary $\Gamma^\mathrm{f}_{D}(t)$, the Neumann boundary $\Gamma^\mathrm{f}_{H}(t)$ and the fluid-structure boundary $\Gamma^\mathrm{fs}(t)$ at time $t$. We write the one-fluid formulation for two-phase incompressible and immiscible fluid flow in the arbitrary Lagrangian-Eulerian framework with the boundary conditions as
\begin{align} 
	\rho^\mathrm{f}\frac{\partial {\boldsymbol{u}}^\mathrm{f}}{\partial t}\bigg|_{\boldsymbol{\chi}} + \rho^\mathrm{f}(\boldsymbol{u}^\mathrm{f}-\boldsymbol{u}^\mathrm{m})\cdot\nabla{\boldsymbol{u}}^\mathrm{f} &= \nabla\cdot {\boldsymbol{\sigma}}^\mathrm{f} + \mathbf{sf} + \boldsymbol{b}^\mathrm{f},&&\mathrm{on}\ \Omega^\mathrm{f}(t),\label{NS_OF_mom}\\
	\nabla\cdot{\boldsymbol{u}}^\mathrm{f} &= 0,&&\mathrm{on}\ \Omega^\mathrm{f}(t), \label{NS_OF_cont}\\
	\boldsymbol{u}^\mathrm{f} &= \boldsymbol{u}^\mathrm{f}_{D},&&\forall \boldsymbol{x}^\mathrm{f}\in \Gamma^\mathrm{f}_{D}(t),\\
	\boldsymbol{\sigma}^\mathrm{f}\cdot\mathbf{n}^\mathrm{f} &= \boldsymbol{h}^\mathrm{f},&&\forall \boldsymbol{x}^\mathrm{f}\in \Gamma^\mathrm{f}_{H}(t),\\
	\boldsymbol{u}^\mathrm{f} &= \boldsymbol{u}^\mathrm{f}_{0},&&\mathrm{on}\ \Omega^\mathrm{f}(0),
\end{align}
where $\boldsymbol{u}^\mathrm{f}$ and $\boldsymbol{u}^\mathrm{m}$ represent the fluid velocity and the mesh velocity defined for each spatial point $\boldsymbol{x}^\mathrm{f}$ in $\Omega^\mathrm{f}(t)$, $\rho^\mathrm{f}$ is the density of the fluid, $\mathbf{sf}$ is the surface tension singular force replaced by the continuum surface force in the diffuse interface description, $\boldsymbol{b}^\mathrm{f}$ is the body force on the fluid such as gravity $(\boldsymbol{b}^\mathrm{f}=\rho^\mathrm{f}\boldsymbol{g})$, $\boldsymbol{g}$ being the acceleration due to gravity, $\boldsymbol{u}^\mathrm{f}_{D}$ and $\boldsymbol{h}^\mathrm{f}$ denote the boundary conditions at the Dirichlet and Neumann boundaries respectively, $\mathbf{n}^\mathrm{f}$ is the unit outward normal to the Neumann boundary and $\boldsymbol{u}^\mathrm{f}_{0}$ represents the initial velocity field at $t=0$. The partial derivative of the velocity field with respect to time is evaluated with the ALE referential coordinate $\boldsymbol{\chi}$ fixed. The Cauchy stress tensor for a Newtonian fluid is given as 
\begin{align}
\boldsymbol{\sigma}^\mathrm{f}=-p\boldsymbol{I} + \boldsymbol{T}^\mathrm{f},\quad \boldsymbol{T}^\mathrm{f}=2\mu^\mathrm{f}\boldsymbol{\epsilon}^\mathrm{f}(\boldsymbol{u}^\mathrm{f}),\quad \boldsymbol{\epsilon}^\mathrm{f}(\boldsymbol{u}^\mathrm{f})= \frac{1}{2}\big[ \nabla\boldsymbol{u}^\mathrm{f} + (\nabla\boldsymbol{u}^\mathrm{f})^T \big],
\end{align}
where $p$ is the pressure field,  $\boldsymbol{T}^\mathrm{f}$ and $\boldsymbol{\epsilon}^\mathrm{f}$ represent the shear stress tensor and the fluid strain rate tensor respectively and $\mu^\mathrm{f}$ denotes the dynamic viscosity of the fluid. The physical parameters of the fluid such as $\rho^\mathrm{f}$ and $\mu^\mathrm{f}$ are dependent on the order parameter $\phi$ (which evolves with the fluid-fluid interface) as
\begin{align}
	\rho^\mathrm{f}(\phi) &= \frac{1+\phi}{2}\rho^\mathrm{f}_1 + \frac{1-\phi}{2}\rho^\mathrm{f}_2, \label{dens}\\
	\mu^\mathrm{f}(\phi) &= \frac{1+\phi}{2}\mu^\mathrm{f}_1 + \frac{1-\phi}{2}\mu^\mathrm{f}_2, \label{visc}
\end{align} 
where $\rho^\mathrm{f}_i$ and $\mu^\mathrm{f}_i$ are the density and dynamic viscosity of the $i$th phase of the fluid respectively. 

\subsubsection{Semi-discrete variational form}
The temporal discretization of the two-phase incompressible Navier-Stokes equations is carried out by the generalized-$\alpha$ method \cite{Gen_alpha}. It enables a user-controlled high frequency damping which is desirable for coarser discretization in space and time. This is achieved by a single parameter called the spectral radius $\rho_{\infty}$. The following expressions are employed for the temporal discretization:
\begin{align}
	\boldsymbol{u}^\mathrm{f,n+1} &= \boldsymbol{u}^\mathrm{f,n} + \Delta t\partial_t\boldsymbol{u}^\mathrm{f,n} + \gamma^\mathrm{f}\Delta t (\partial_t\boldsymbol{u}^\mathrm{f,n+1} - \partial_t\boldsymbol{u}^\mathrm{f,n}),\\
	\partial_t\boldsymbol{u}^\mathrm{f,n+\alpha^f_m} &= \partial_t\boldsymbol{u}^\mathrm{f,n} + \alpha^\mathrm{f}_\mathrm{m}(\partial_t\boldsymbol{u}^\mathrm{f,n+1} - \partial_t\boldsymbol{u}^\mathrm{f,n}),\\
	\boldsymbol{u}^\mathrm{f,n+\alpha^f} &= \boldsymbol{u}^\mathrm{f,n} + \alpha^\mathrm{f}(\boldsymbol{u}^\mathrm{f,n+1} - \boldsymbol{u}^\mathrm{f,n}),
\end{align} 
where $\alpha^\mathrm{f}$, $\alpha^\mathrm{f}_\mathrm{m}$ and $\gamma^\mathrm{f}$ are the generalized-$\alpha$ parameters dependent on the user-defined spectral radius $\rho_{\infty}$. The time step size is denoted by $\Delta t$ and $\partial_t$ denotes the partial differentiation with respect to time.

Suppose $\mathcal{S}^\mathrm{h}_{\boldsymbol{u}^\mathrm{f}}$ and $\mathcal{S}^\mathrm{h}_{p}$ denote the space of trial solution such that
\begin{align}
	\mathcal{S}^\mathrm{h}_{\boldsymbol{u}^\mathrm{f}} &= \big\{ \boldsymbol{u}^\mathrm{f}_\mathrm{h}\ |\ \boldsymbol{u}^\mathrm{f}_\mathrm{h} \in (H^1(\Omega^\mathrm{f}(t)))^{d}, \boldsymbol{u}^\mathrm{f}_\mathrm{h} = \boldsymbol{u}^\mathrm{f}_{D}\ \mathrm{on}\ \Gamma^\mathrm{f}_{D}(t) \big\},\\
	\mathcal{S}^\mathrm{h}_{p} &= \big\{ p_\mathrm{h}\ |\ p_\mathrm{h} \in L^2(\Omega^\mathrm{f}(t)) \big\},
\end{align}
where $(H^1(\Omega^\mathrm{f}(t)))^{d}$ denotes the space of square-integrable $\mathbb{R}^{d}$-valued functions with square-integrable derivatives on $\Omega^\mathrm{f}(t)$ and $L^2(\Omega^\mathrm{f}(t))$ is the space of the scalar-valued functions that are square-integrable on $\Omega^\mathrm{f}(t)$. Similarly, we define $\mathcal{V}^\mathrm{h}_{\boldsymbol{\psi}^\mathrm{f}}$ and $\mathcal{V}^\mathrm{h}_{q}$ as the space of test functions such that
\begin{align}
	\mathcal{V}^\mathrm{h}_{\boldsymbol{\psi}^\mathrm{f}} &= \big\{ \boldsymbol{\psi}^\mathrm{f}_\mathrm{h}\ |\ \boldsymbol{\psi}^\mathrm{f}_\mathrm{h} \in (H^1(\Omega^\mathrm{f}(t)))^{d}, \boldsymbol{\psi}^\mathrm{f}_\mathrm{h} = \boldsymbol{0}\ \mathrm{on}\ \Gamma^\mathrm{f}_{D}(t) \big\},\\
	\mathcal{V}^\mathrm{h}_{q} &= \big\{ q_\mathrm{h}\ |\ q_\mathrm{h} \in L^2(\Omega^\mathrm{f}(t)) \big\}.
\end{align}
The variational statement of the Navier-Stokes equations can thus be written as: find $[\boldsymbol{u}^\mathrm{f}_\mathrm{h}(t^\mathrm{n+\alpha^f}),p_\mathrm{h}(t^\mathrm{n+1})]\in \mathcal{S}^\mathrm{h}_{\boldsymbol{u}^\mathrm{f}} \times \mathcal{S}^\mathrm{h}_{p}$ such that $\forall [\boldsymbol{\psi}^\mathrm{f}_\mathrm{h},q_\mathrm{h}]\in \mathcal{V}^\mathrm{h}_{\boldsymbol{\psi}^\mathrm{f}} \times \mathcal{V}^\mathrm{h}_{q}$,
\begin{align}
&\int_{\Omega^\mathrm{f}(t)} \rho^\mathrm{f}(\phi) ( \partial_t{\boldsymbol{u}}_\mathrm
{h}^\mathrm{f} + ({\boldsymbol{u}}_{\mathrm{h}}^{\mathrm{f}}-{\boldsymbol{u}}_{\mathrm{h}}^{\mathrm{m}}) \cdot\nabla{\boldsymbol{u}}_{\mathrm{h}}^{\mathrm{f}})\cdot\boldsymbol{\psi}^\mathrm{f}_{\mathrm{h}} \mathrm{d\Omega} +
\int_{\Omega^\mathrm{f}(t)} {\boldsymbol{\sigma}}_{\mathrm{h}}^{\mathrm{f}}
: \nabla\boldsymbol{\psi}^\mathrm{f}_{\mathrm{h}} \mathrm{d\Omega} \nonumber \\
- &\int
_{\Omega^\mathrm{f}(t)} \mathbf{sf}_{\mathrm{h}}(\phi)\cdot
\boldsymbol{\psi}^\mathrm{f}_{\mathrm{h}} \mathrm{d\Omega} + \displaystyle\sum_\mathrm{e=1}^\mathrm{n_{el}}\int_{\Omega
^{\mathrm{e}}} \frac{\tau_\mathrm{m}}{\rho^\mathrm{f}(\phi)} (\rho^\mathrm{f}(\phi
)({\boldsymbol{u}}_{\mathrm{h}}^{\mathrm{f}}-{\boldsymbol{u}}_{\mathrm{h}}^{\mathrm{m}})\cdot\nabla
\boldsymbol{\psi}^\mathrm{f}_{\mathrm{h}}+ \nabla q_{\mathrm{h}} )\cdot
\boldsymbol{\mathcal{R}}_\mathrm{m} \mathrm
{d\Omega^e} \nonumber \\
+ &\int_{\Omega^\mathrm{f}(t)}q_{\mathrm{h}}(\nabla\cdot{\boldsymbol{u}}_{\mathrm{h}}^{\mathrm{f}}) \mathrm{d\Omega} + \displaystyle\sum_\mathrm{e=1}^\mathrm{n_{el}}\int
_{\Omega^{\mathrm{e}}} \nabla\cdot\boldsymbol{\psi}^\mathrm{f}_{\mathrm{h}}\tau
_\mathrm{c}\rho^\mathrm{f}(\phi) \boldsymbol{\mathcal{R}}_\mathrm
{c} \mathrm{d\Omega^e}\nonumber\\
-&\displaystyle\sum_\mathrm{e=1}^\mathrm{n_{el}}\int_{\Omega
^{\mathrm{e}}} \tau_\mathrm{m} \boldsymbol{\psi}^\mathrm{f}_{\mathrm{h}}\cdot
(\boldsymbol{\mathcal{R}}_\mathrm{m} \cdot
\nabla{\boldsymbol{u}}_{\mathrm{h}}^{\mathrm{f}}) \mathrm
{d\Omega^e} -\displaystyle\sum_\mathrm{e=1}^\mathrm{n_{el}}\int
_{\Omega^{\mathrm{e}}} \frac{\nabla\boldsymbol{\psi}^\mathrm{f}_\mathrm
{h}}{\rho^\mathrm{f}(\phi)}:(\tau_\mathrm{m}\boldsymbol{\mathcal{R}}_\mathrm
{m} \otimes\tau_\mathrm{m}\boldsymbol
{\mathcal{R}}_\mathrm{m}) \mathrm{d\Omega
^e}\nonumber\\
= &\int_{\Omega^\mathrm{f}(t)} \boldsymbol{b}^\mathrm{f}(t^{\mathrm{n}+\alpha^\mathrm{f}})\cdot
\boldsymbol{\psi}^\mathrm{f}_{\mathrm{h}} \mathrm{d\Omega} + \int_{\Gamma
^\mathrm{f}_{H}} \boldsymbol{h}^\mathrm{f}\cdot\boldsymbol{\psi}^\mathrm{f}_{\mathrm{h}}
\mathrm{d\Gamma},
\end{align}
where the first and the second lines represent the Galerkin terms and the Petrov-Galerkin stabilization terms for the momentum equation, the third line depicts the Galerkin and the stabilization term for the continuity equation, the fourth line consists of the terms which are introduced as the approximation of the fine scale velocity on the element interiors based on the multi-scale argument \cite{Akkerman, Hughes_conserve, Hsu} and the fifth line represents the body forces and the Neumann boundary conditions. The element-wise residuals of the momentum and the continuity equations are represented by $\boldsymbol{\mathcal{R}}_\mathrm{m}$ and $\boldsymbol{\mathcal{R}}_\mathrm{c}$ respectively. The stabilization parameters  $\tau_\mathrm{m}$ and $\tau_\mathrm{c}$ are the least-squares metrics added to the element-level integrals in the stabilized formulation \cite{Hughes_X, Brooks, Tezduyar_1, France_II} and are defined as
\begin{align}
	\tau_\mathrm{m} = \bigg[ \bigg( \frac{2}{\Delta t}\bigg)^2 + (\boldsymbol{u}^\mathrm{f}_\mathrm{h}-\boldsymbol{u}^\mathrm{m}_\mathrm{h})\cdot\boldsymbol{G}(\boldsymbol{u}^\mathrm{f}_\mathrm{h}-\boldsymbol{u}^\mathrm{m}_\mathrm{h}) + C_I \bigg( \frac{\mu^\mathrm{f}(\phi)}{\rho^\mathrm{f}(\phi)}\bigg)^2 \boldsymbol{G}:\boldsymbol{G} \bigg]^{-1/2},\qquad \tau_\mathrm{c} = \frac{1}{\mathrm{tr}(\boldsymbol{G})\tau_\mathrm{m}},
\end{align} 
where $C_I$ is a constant derived from the element-wise inverse estimates \cite{Hughes_inv_est}, $\boldsymbol{G}$ is the element contravariant metric tensor and $\mathrm{tr}(\boldsymbol{G})$ is the trace of the contravariant metric tensor. This stabilization in the variational form circumvents the Babu$\mathrm{\check{s}}$ka-Brezzi condition that is required to be satisfied by any standard mixed Galerkin method \cite{Johnson}. 

\subsection{The Allen-Cahn equation}
The modeling of the order parameter which distinguishes the two phases is achieved by solving the conservative Allen-Cahn equation and evolving the fluid-fluid interface. The governing equation in its strong form and the recently proposed positivity preserving variational formulation in the present context have been described in this section.

\subsubsection{Strong differential form}
Consider the spatial domain $\Omega^\mathrm{f}(t)$ with Dirichlet and Neumann boundaries for the order parameter denoted by $\Gamma^\mathrm{\phi}_{D}(t)$ and $\Gamma^\mathrm{\phi}_{H}(t)$ respectively. The phase-field order parameter $\phi$ which represents the different phases of the fluid is evolved by solving the conservative Allen-Cahn equation in the ALE framework with the boundary conditions given as
\begin{align}
	\frac{\partial\phi}{\partial t}\bigg|_{\boldsymbol{\chi}} + (\boldsymbol{u}^\mathrm{f}-\boldsymbol{u}^\mathrm{m})\cdot\nabla\phi- \gamma
\big(\varepsilon^2\nabla^2\phi- F'(\phi) + \beta(t)\sqrt{F(\phi
)}\big) &= 0,\ \ &&\mathrm{on}\ \Omega^\mathrm{f}(t),\\
	\phi &= \phi_{D},\ \ &&\forall \boldsymbol{x}^\mathrm{f}\in \Gamma^\mathrm{\phi}_{D}(t),\\
	\nabla\phi\cdot\mathbf{n}^\mathrm{\phi} &= 0,\ \ &&\forall \boldsymbol{x}^\mathrm{f}\in \Gamma^\mathrm{\phi}_{H}(t),\\
	\phi &= \phi_{0},\ \ &&\mathrm{on}\ \Omega^\mathrm{f}(0),
\end{align}
where $\gamma$ is a mobility parameter (taken as $1$ for the present study), $\varepsilon$ is the parameter which represents the thickness of the interface between the phases, $F(\phi)$ is the double-well energy potential which represents the free energy of mixing or bulk energy. It has two minima corresponding to the two stable phases of the fluid. The value of the order parameter at the Dirichlet boundary is denoted by $\phi_{D}$, the initial condition is represented by $\phi_{0}$ and $\mathbf{n}^{\phi}$ denotes the unit normal to the Neumann boundary where zero flux condition is satisfied. The mass conservation is enforced in the Allen-Cahn equation by a Lagrange multiplier $\beta(t)\sqrt{F(\phi)}$ where $\beta(t) = \int_{\Omega^\mathrm{f}(t)} F'(\phi)\mathrm{d}\Omega / \int_{\Omega^\mathrm{f}(t)}\sqrt{F(\phi)}\mathrm{d}\Omega$, $F'(\phi)$ is the derivative of the energy potential with respect to the order parameter. 

\subsubsection{Semi-discrete variational form}
Following the generalized-$\alpha$ temporal discretization of the above equation for consistency, we write the expressions used in the variational statement as
\begin{align}
	\phi^\mathrm{n+1} &= \phi^\mathrm{n} + \Delta t\partial_t\phi^\mathrm{n} + \gamma^\mathrm{f}\Delta t(\partial_t\phi^\mathrm{n+1} - \partial_t\phi^\mathrm{n}),\\
	\partial_t\phi^\mathrm{n+\alpha^f_m} &= \partial_t\phi^\mathrm{n} + \alpha^\mathrm{f}_\mathrm{m}(\partial_t\phi^\mathrm{n+1} - \partial_t\phi^\mathrm{n}),\\
	\phi^\mathrm{n+\alpha^f} &= \phi^\mathrm{n} + \alpha^\mathrm{f}(\phi^\mathrm{n+1} - \phi^\mathrm{n}),
\end{align}
where the generalized-$\alpha$ parameters depend on the spectral radius $\rho_{\infty}$.
The temporal discretized Allen-Cahn equation can be transformed into a convection-diffusion-reaction equation as follows:
\begin{align}
	\partial_t\phi^\mathrm{n+\alpha^f_m} + \hat{\boldsymbol{u}}\cdot\nabla\phi^\mathrm{n+\alpha^f}- \hat{k}\nabla^2\phi^\mathrm{n+\alpha^f} + \hat{s}\phi^\mathrm{n+\alpha^f} - \hat{f}(t^\mathrm{n+\alpha^f}) = 0,\ \ &\mathrm{on}\ \Omega^\mathrm{f}(t)
\end{align}
where $\hat{\boldsymbol{u}}$, $\hat{k}$, $\hat{s}$ and $\hat{f}$ are the modified convection velocity, diffusion coefficient, reaction coefficient and the source respectively given as
\begin{align}
\hat{\boldsymbol{u}} &= \boldsymbol{u}^\mathrm{f}-\boldsymbol{u}^\mathrm{m},\\
\hat{k} &= \varepsilon^2,\\
\hat{s} &= \frac{1}{4}\bigg[ \frac{(\phi
^\mathrm{n+\alpha^f})^2}{(\alpha^\mathrm{f})^3} - \bigg(\frac{3}{(\alpha^\mathrm{f})^3} -
\frac{4}{(\alpha^\mathrm{f})^2}\bigg)\phi^\mathrm{n+\alpha^f}\phi^{\mathrm{n}} +
\bigg( \frac{3}{(\alpha^\mathrm{f})^3} - \frac{8}{(\alpha^\mathrm{f})^2} + \frac{6}{\alpha^\mathrm{f}
} \bigg) (\phi^{\mathrm{n}})^2 - \frac{2}{\alpha^\mathrm{f}} \bigg]\nonumber
\\ &- \frac{\beta(t^\mathrm{n+\alpha^f})}{2}\bigg[ \frac{\phi^\mathrm{n+\alpha^f
}}{3(\alpha^\mathrm{f})^2} + \frac{1}{3}\bigg( -\frac{2}{(\alpha^\mathrm{f})^2} + \frac
{3}{\alpha^\mathrm{f}} \bigg)\phi^{\mathrm{n}} \bigg],\\
\hat{f} &= -\frac{1}{4}\bigg[ \bigg(-\frac
{1}{(\alpha^\mathrm{f})^3} +\frac{4}{(\alpha^\mathrm{f})^2} - \frac{6}{\alpha^\mathrm{f}} + 4 \bigg
)(\phi^{\mathrm{n}})^3 + \bigg( \frac{2}{\alpha^\mathrm{f}} - 4\bigg)\phi
^{\mathrm{n}} \bigg] \nonumber\\
&+ \frac{\beta(t^\mathrm{n+\alpha^f})}{2}\bigg[ \frac{1}{3}\bigg( \frac{1}{(\alpha^\mathrm{f})^2}
- \frac{3}{\alpha^\mathrm{f}} + 3 \bigg)(\phi^{\mathrm{n}})^2 - 1\bigg].
\end{align}

Defining the space of trial solution as $\mathcal{S}^\mathrm{h}_{\phi}$ and that of the test function as $\mathcal{V}^\mathrm{h}_{\phi}$ such that
\begin{align}
	\mathcal{S}^\mathrm{h}_{\phi} &= \big\{ \phi_\mathrm{h}\ |\ \phi_\mathrm{h} \in H^1(\Omega^\mathrm{f}(t)), \phi_\mathrm{h} = \phi_{D}\ \mathrm{on}\ \Gamma^\mathrm{\phi}_{D}(t) \big\},\\
	\mathcal{V}^\mathrm{h}_{\phi} &= \big\{ \hat{w}_\mathrm{h}\ |\ \hat{w}_\mathrm{h} \in H^1(\Omega^\mathrm{f}(t)), \hat{w}_\mathrm{h} = 0\ \mathrm{on}\ \Gamma^\mathrm{\phi}_{D}(t) \big\},
\end{align}
the variational statement for the Allen-Cahn equation is given as: find $\phi
_{\mathrm{h}}(\boldsymbol{x}^\mathrm{f},t^{\mathrm{n}+\alpha^\mathrm{f}}) \in\mathcal
{S}^{\mathrm{h}}_{\phi}$ such that $\forall \hat{w}_{\mathrm{h}} \in\mathcal
{V}^{\mathrm{h}}_{\phi}$,
\begin{align} \label{PPV_AC}
&\int_{\Omega^\mathrm{f}(t)}\bigg( \hat{w}_{\mathrm{h}}\partial_t{\phi}_{\mathrm{h}} +
\hat{w}_{\mathrm{h}}(\hat{\boldsymbol{u}}\cdot\nabla\phi_{\mathrm{h}}) + \nabla
\hat{w}_{\mathrm{h}}\cdot(\hat{k}\nabla\phi_{\mathrm{h}} ) + \hat{w}_{\mathrm{h}}\hat{s}\phi
_{\mathrm{h}} - \hat{w}_{\mathrm{h}}\hat{f} \bigg) \mathrm{d}\Omega\nonumber\\
&+ \displaystyle\sum_\mathrm{e=1}^\mathrm{n_{el}}\int_{\Omega
^{\mathrm{e}}}\bigg( \big(\hat{\boldsymbol{u}}\cdot\nabla \hat{w}_{\mathrm{h}}
\big)\tau\big( \partial_t{\phi}_{\mathrm{h}} + \hat{\boldsymbol{u}}\cdot
\nabla\phi_{\mathrm{h}} - \nabla\cdot(\hat{k}\nabla\phi_{\mathrm{h}}) +
\hat{s}\phi_{\mathrm{h}} -\hat{f} \big) \bigg) \mathrm{d}\Omega^{\mathrm{e}}
\nonumber\\
&+ \displaystyle\sum_\mathrm{e=1}^\mathrm{n_{el}}\int_{\Omega
^{\mathrm{e}}} \chi\frac{|\mathcal{R}(\phi_{\mathrm{h}})|}{|\nabla
\phi_{\mathrm{h}}|}k_s^\mathrm{add} \nabla \hat{w}_{\mathrm{h}}\cdot\bigg(
\frac{\hat{\boldsymbol{u}}\otimes\hat{\boldsymbol{u}}}{|\hat{\boldsymbol{u}}|^2}
\bigg) \cdot\nabla\phi_{\mathrm{h}} \mathrm{d}\Omega^{\mathrm{e}} +
\sum_\mathrm{e=1}^\mathrm{n_{el}} \int_{\Omega^{\mathrm{e}}}\chi
\frac{|\mathcal{R}(\phi_{\mathrm{h}})|}{|\nabla\phi_{\mathrm{h}}|}
k^\mathrm{add}_{c} \nabla \hat{w}_{\mathrm{h}} \cdot\bigg( \mathbf{I} -
\frac{\hat{\boldsymbol{u}}\otimes\hat{\boldsymbol{u}}}{|\hat{\boldsymbol{u}}|^2}
\bigg) \cdot\nabla\phi_{\mathrm{h}} \mathrm{d}\Omega^{\mathrm{e}}
\nonumber\\
&= 0,
\end{align}
where the first line represents the Galerkin terms for the Allen-Cahn equation, the second line is the streamline upwind Petrov-Galerkin terms and the third line depicts the positivity preserving terms. Here, the stabilization parameter $\tau$ is given by
\begin{align}
	\tau = \bigg[ \bigg(\frac{2}{\Delta t} \bigg)^2 + \hat{\boldsymbol{u}}\cdot\boldsymbol{G}\hat{\boldsymbol{u}} + 9\hat{k}^2 \boldsymbol{G}:\boldsymbol{G} + \hat{s}^2 \bigg]^{-1/2}.
\end{align}
The positivity preserving stabilization terms are derived for the multi-dimensional convection-diffusion-reaction equation by satisfying the positivity condition at the element level matrix of the variationally discretized equation in \cite{PPV}. For one-dimensional explicit scheme, the positivity preserving property reduces to the conditional inequality of the Harten's coefficients \cite{Harten}. This has been shown for some particular cases in \cite{AC_adaptive_JCP}. For an implicit matrix form of the scheme, the positivity condition can be imparted by transforming the left-hand-side matrix $\boldsymbol{A} = \{a_{ij}\}$ to an M-matrix which ensures positivity and convergence \cite{Kuzmin} satisfying the following properties
\begin{align}
	a_{ii} &> 0, \forall i,\\
	a_{ij} &\leq 0, \forall j\neq i,\\
	\sum_{j} a_{ij} &= 0, \forall i.
\end{align}
This transformation is carried out by the addition of the discrete upwind matrix, which renders the variational scheme positivity preserving and monotone \cite{FCT}. The factor $\chi |\mathcal{R}(\phi_\mathrm{h})|/|\nabla\phi_\mathrm{h}|$ acts as a limiter to the upwinding near the regions of high solution gradients which is adjusted by the nonlinear corrections.  Several test cases have been performed to assess the effectiveness of the PPV technique in \cite{PPV}. The details of the derivation of the added diffusions $k_s^\mathrm{add}$, $k_c^\mathrm{add}$ and $\chi$ can be found in \cite{PPV}, which are given for the present context by \cite{AC_JCP}
\begin{align}
	\chi &= \frac{2}{|\hat{s}|h + 2|\hat{\boldsymbol{u}}|},\\
	k_s^\mathrm{add} &= \mathrm{max} \bigg\{ \frac{||\hat{\boldsymbol{u}}| - \tau|\hat{\boldsymbol{u}}|\hat{s}|h}{2} - (\hat{k} + \tau|\hat{\boldsymbol{u}}|^2) + \frac{\hat{s}h^2}{6}, 0 \bigg\},\\
	k_c^\mathrm{add} &= \mathrm{max} \bigg\{ \frac{|\hat{\boldsymbol{u}}|h}{2} - \hat{k} + \frac{\hat{s}h^2}{6}, 0 \bigg\},
\end{align} 
where $|\hat{\boldsymbol{u}}|$ is the magnitude of the modified convection velocity and $h$ is the characteristic element length defined in \cite{PPV}. 

\subsection{The structural equation}
We describe the structural equation and its weak formulation in this section. The structural equation is solved in the Lagrangian framework where the fluid-structure interface is a sharp boundary and the mesh surrounding the structure deforms in the ALE framework.

\subsubsection{Strong differential form}
Consider a $d$-dimensional structural domain $\Omega^\mathrm{s}\subset \mathbb{R}^{d}$ with a piecewise smooth boundary $\Gamma^\mathrm{s}$ consisting of the material coordinates $\boldsymbol{x}^\mathrm{s}$ at time $t=0$. The boundary $\Gamma^\mathrm{s}$  can be decomposed into three disjoint sections consisting of the Dirichlet boundary $\Gamma^\mathrm{s}_{D}$, the Neumann boundary $\Gamma^\mathrm{s}_{H}$ and the fluid-structure interface $\Gamma^\mathrm{fs}$. We define a one-to-one mapping function $\boldsymbol{\varphi}^\mathrm{s}(\boldsymbol{x}^\mathrm{s},t):\Omega^\mathrm{s} \to \Omega^\mathrm{s}(t)$ which denotes the position vector and maps the reference coordinates of the structure $\boldsymbol{x}^\mathrm{s}$ at $t=0$ to its position in the deformed configuration $\Omega^\mathrm{s}(t)$. Let $\boldsymbol{\eta}^\mathrm{s}(\boldsymbol{x}^\mathrm{s},t)$ be the structural displacement due to the deformation. The position vector mapping is thus given by
\begin{align} \label{varphi_eqn}
	\boldsymbol{\varphi}^\mathrm{s}(\boldsymbol{x}^\mathrm{s},t) = \boldsymbol{\eta}^\mathrm{s}(\boldsymbol{x}^\mathrm{s},t) + \boldsymbol{x}^\mathrm{s}.
\end{align}
The velocity of the body at the deformed configuration is defined by
\begin{align}
	\boldsymbol{u}^\mathrm{s} = \frac{\partial\boldsymbol{\varphi}^\mathrm{s}}{\partial t} = \frac{\partial\boldsymbol{\eta}^\mathrm{s}}{\partial t},\qquad \ \frac{\partial\boldsymbol{u}^\mathrm{s}}{\partial t} = \frac{\partial^2\boldsymbol{\varphi}^\mathrm{s}}{\partial t^2} = \frac{\partial^2\boldsymbol{\eta}^\mathrm{s}}{\partial t^2}.
\end{align}
The structural equations can be written in the most general form as
\begin{align}
	\rho^\mathrm{s}\frac{\partial^2\boldsymbol{\varphi}^\mathrm{s}}{\partial t^2} + \nabla\cdot\boldsymbol{\sigma}^\mathrm{s} &= \boldsymbol{b}^\mathrm{s},	&&\mathrm{on}\ \Omega^\mathrm{s},\\
	\boldsymbol{u}^\mathrm{s} &= \boldsymbol{u}^\mathrm{s}_{D},  &&\forall \boldsymbol{x}^\mathrm{s}\in \Gamma^\mathrm{s}_{D},\\
	\boldsymbol{\sigma}^\mathrm{s}\cdot\mathbf{n}^\mathrm{s} &= \boldsymbol{h}^\mathrm{s}, &&\forall \boldsymbol{x}^\mathrm{s}\in \Gamma^\mathrm{s}_{H},\\
	\boldsymbol{\varphi}^\mathrm{s} &= \boldsymbol{\varphi}^\mathrm{s}_{0}, &&\mathrm{on}\ \Omega^\mathrm{s},\\
	\boldsymbol{u}^\mathrm{s} &= \boldsymbol{u}^\mathrm{s}_{0}, &&\mathrm{on}\ \Omega^\mathrm{s},
\end{align}
where $\rho^\mathrm{s}$, $\boldsymbol{\sigma}^\mathrm{s}$ and $\boldsymbol{b}^\mathrm{s}$ denote the density, stress tensor and the body forces acting on the structure respectively. The quantities $\boldsymbol{u}^\mathrm{s}_{D}$ and $\boldsymbol{h}^\mathrm{s}$ denote the Dirichlet and Neumann conditions on the structural velocity respectively and $\boldsymbol{\varphi}^\mathrm{s}_{0}$ and $\boldsymbol{u}^\mathrm{s}_{0}$ represent the initial position vector and the initial velocity of the structure respectively. The unit normal to the Neumann boundary is denoted by $\mathbf{n}^\mathrm{s}$.

\subsubsection{Semi-discrete variational form}
Following the consistency in the time discretization using the generalized-$\alpha$ framework \cite{Gen_alpha}, we can write the following expressions
\begin{align}
	\boldsymbol{\varphi}^\mathrm{s,n+1} &= \boldsymbol{\varphi}^\mathrm{s,n} + \Delta t \boldsymbol{u}^\mathrm{s,n} + \Delta t^2 \bigg( \big( \frac{1}{2} - \beta^\mathrm{s} \big)\partial_{t}\boldsymbol{u}^\mathrm{s,n} + \beta^\mathrm{s}\partial_{t}\boldsymbol{u}^\mathrm{s,n+1} \bigg),\\
	\boldsymbol{u}^\mathrm{s,n+1} &= \boldsymbol{u}^\mathrm{s,n} + \Delta t\bigg( \big( 1-\gamma^\mathrm{s}\big)\partial_{t}\boldsymbol{u}^\mathrm{s,n} + \gamma^\mathrm{s}\partial_{t}\boldsymbol{u}^\mathrm{s,n+1} \bigg),\\
	\partial_{t}\boldsymbol{u}^\mathrm{s,n+\alpha^s_m} &= \partial_{t}\boldsymbol{u}^\mathrm{s,n} + \alpha^\mathrm{s}_\mathrm{m}( \partial_{t}\boldsymbol{u}^\mathrm{s,n+1} - \partial_{t}\boldsymbol{u}^\mathrm{s,n}), \\
	\boldsymbol{u}^\mathrm{n+\alpha^s} &= \boldsymbol{u}^\mathrm{s,n} + \alpha^\mathrm{s}(\boldsymbol{u}^\mathrm{s,n+1} - \boldsymbol{u}^\mathrm{s,n}),\\
	\boldsymbol{\varphi}^\mathrm{n+\alpha^\mathrm{s}} &= \boldsymbol{\varphi}^\mathrm{s,n} + \alpha^\mathrm{s}(\boldsymbol{\varphi}^\mathrm{s,n+1} - \boldsymbol{\varphi}^\mathrm{s,n}),
\end{align} 
where $\alpha^\mathrm{s}$, $\alpha^\mathrm{s}_\mathrm{m}$, $\beta^\mathrm{s}$ and $\gamma^\mathrm{s}$ are the generalized-$\alpha$ parameters which are selected as $\alpha^\mathrm{s}=\alpha^\mathrm{s}_\mathrm{m}=\gamma^\mathrm{s} = 0.5$ and $\beta^\mathrm{s}=0.25$.

Considering the space of trial solution $\mathcal{S}^\mathrm{h}_{\boldsymbol{u}^\mathrm{s}}$ and that of the test function $\mathcal{V}^\mathrm{h}_{\boldsymbol{\psi}^\mathrm{s}}$ which are defined as
\begin{align}
		\mathcal{S}^\mathrm{h}_{\boldsymbol{u}^\mathrm{s}} &= \big\{ \boldsymbol{u}^\mathrm{s}_\mathrm{h}\ |\ \boldsymbol{u}^\mathrm{s}_\mathrm{h} \in (H^1(\Omega^\mathrm{s}))^{d}, \boldsymbol{u}^\mathrm{s}_\mathrm{h} = \boldsymbol{u}^\mathrm{s}_{D}\ \mathrm{on}\ \Gamma^\mathrm{s}_{D} \big\},\\
		\mathcal{V}^\mathrm{h}_{\boldsymbol{\psi}^\mathrm{s}} &= \big\{ \boldsymbol{\psi}^\mathrm{s}_\mathrm{h}\ |\ \boldsymbol{\psi}^\mathrm{s}_\mathrm{h} \in (H^1(\Omega^\mathrm{s}))^{d}, \boldsymbol{\psi}^\mathrm{s}_\mathrm{h} = \boldsymbol{0}\ \mathrm{on}\ \Gamma^\mathrm{s}_{D} \big\},
\end{align}
the variational statement for the structural equation is given as: find $\boldsymbol{u}^\mathrm{s}_\mathrm{h} \in\mathcal
{S}^{\mathrm{h}}_{\boldsymbol{u}^\mathrm{s}}$ such that $\forall \boldsymbol{\psi}^\mathrm{s}_\mathrm{h} \in\mathcal{V}^{\mathrm{h}}_{\boldsymbol{\psi}^\mathrm{s}}$,
\begin{align}
	\int_{\Omega^\mathrm{s}} (\rho^\mathrm{s}\partial_{t}\boldsymbol{u}^\mathrm{s,n+\alpha^f_m}_\mathrm{h})\cdot\boldsymbol{\psi}^\mathrm{s}_\mathrm{h} \mathrm{d}\Omega + \int_{\Omega^\mathrm{s}} \boldsymbol{\sigma}^\mathrm{s}_\mathrm{h}: \nabla\boldsymbol{\psi}^\mathrm{s}_\mathrm{h} \mathrm{d}\Omega = \int_{\Gamma^\mathrm{s}_\mathrm{H}} \boldsymbol{h}^\mathrm{s}\cdot\boldsymbol{\psi}^\mathrm{s}_\mathrm{h} \mathrm{d}\Gamma  + \int_{\Omega^\mathrm{s}} \boldsymbol{b}^\mathrm{s}\cdot\boldsymbol{\psi}^\mathrm{s}_\mathrm{h} \mathrm{d}\Omega.
\end{align}
In the present study, we consider rigid and linear flexible motions of the structure. The respective equations and their matrix form are reviewed in Appendix \ref{app_structure}.  
We next describe the treatment of the fluid-structure and the fluid-fluid interfaces, i.e., how the boundary conditions are satisfied with the help of the equilibrium conditions at those interfaces.

\subsection{The fluid-structure interface}
The coupling between the fluid and the structural equations is achieved by the velocity continuity and the equilibrium of the tractions along the fluid-structure interface. Suppose $\Gamma^\mathrm{fs} = \Gamma^\mathrm{f}(0) \cap \Gamma^\mathrm{s}$ denotes the fluid-structure interface at $t=0$. The interface at time $t$ will then be denoted by $\Gamma^\mathrm{fs}(t) = \boldsymbol{\varphi}^\mathrm{s}(\Gamma^\mathrm{fs},t)$. The required conditions to be satisfied at the interface can be mathematically formulated as
\begin{align}
	\boldsymbol{u}^\mathrm{f}(\boldsymbol{\varphi}^\mathrm{s}(\boldsymbol{x}^\mathrm{s},t),t) &= \boldsymbol{u}^\mathrm{s}(\boldsymbol{x}^\mathrm{s},t),\ &&\forall \boldsymbol{x}^\mathrm{s}\in \Gamma^\mathrm{fs}, \label{kin_cond}\\
	\int_{\boldsymbol{\varphi}^\mathrm{s}(\gamma,t)} \boldsymbol{\sigma}^\mathrm{f}(\boldsymbol{x}^\mathrm{f},t)\cdot \mathbf{n}^\mathrm{f} \mathrm{d}\Gamma + \int_\gamma \boldsymbol{\sigma}^\mathrm{s}(\boldsymbol{x}^\mathrm{s},t)\cdot \mathbf{n}^\mathrm{s} \mathrm{d}\Gamma &= 0,\ &&\forall \gamma \subset \Gamma^\mathrm{fs}, \label{dyn_cond}
\end{align}
where $\mathbf{n}^\mathrm{f}$ and $\mathbf{n}^\mathrm{s}$ are the unit normals to the deformed fluid element $\boldsymbol{\varphi}^\mathrm{s}(\gamma,t)$ and its corresponding undeformed structural element $\gamma$ respectively. Here, $\gamma$ is any part of the interface $\Gamma^\mathrm{fs}$ in the reference configuration.
 
\subsection{The fluid-fluid interface}
In the sharp fluid-fluid interface description, the velocity continuity and the pressure-jump condition are required to be satisfied at the interface,
\begin{align}
	\boldsymbol{u}^\mathrm{f}_{\Omega^\mathrm{f}_{1}} &= \boldsymbol{u}^\mathrm{f}_{\Omega^\mathrm{f}_{2}}, &&\forall \boldsymbol{x}^\mathrm{f}\in \Gamma^\mathrm{ff}(t), \label{FFIC_1}\\
	(\boldsymbol{\sigma}^\mathrm{f}_{\Omega^\mathrm{f}_{1}} - \boldsymbol{\sigma}^\mathrm{f}_{\Omega^\mathrm{f}_{2}})\cdot\mathbf{n}_{\Gamma^\mathrm{ff}} &= \sigma\kappa\mathbf{n}_{\Gamma^\mathrm{ff}}, &&\forall \boldsymbol{x}^\mathrm{f}\in\Gamma^\mathrm{ff}(t), \label{FFIC_2}
\end{align}
where $(\cdot)_{\Omega^\mathrm{f}_{i}}$ denotes the argument in the fluid phase $i$, $\mathbf{n}_{\Gamma^\mathrm{ff}}$ is the normal to the fluid-fluid interface, $\sigma$ is the surface tension coefficient between the two fluid phases and $\kappa$ is the curvature of the interface denoted by $\kappa=-\nabla\cdot\mathbf{n}_{\Gamma^\mathrm{ff}}$. The surface tension singular force in the Navier-Stokes equations (Eq.~(\ref{NS_OF_mom})) which models the surface tension is thus written as $\mathbf{sf} = \sigma\kappa\delta_{\Gamma^\mathrm{ff}}\mathbf{n}_{\Gamma^\mathrm{ff}}$, where $\delta_{\Gamma^\mathrm{ff}}$ is the one-dimensional Dirac delta function given as
\begin{align}
	\delta_{\Gamma^\mathrm{ff}} = \begin{dcases} 1, \qquad \mathrm{for}\ \boldsymbol{x}^\mathrm{f}\in \Gamma^\mathrm{ff}(t),\\
												 0, \qquad \mathrm{otherwise}. \end{dcases}
\end{align}
It was pointed out in the introduction that the sharp interface description based on the moving mesh framework is not trivial for complex three-dimensional fluid-fluid interfaces. Therefore, in the present formulation, we employ the diffuse fluid-fluid interface description in which the interface is assumed to have a finite thickness, $\mathcal{O}(\varepsilon)$, on which the physical properties of the two phases vary gradually based on an indicator field $\phi$. The diffuse interface description of the fluid-fluid interface recovers to the classical jump discontinuity conditions (Eqs.~(\ref{FFIC_1}-\ref{FFIC_2})) for the sharp interface description asymptotically as $\varepsilon \to 0$ \cite{Anderson_1}. The singular force in the diffuse interface description is replaced by a continuum surface force (CSF) \cite{Brackbill}, which depends on the order parameter $(\phi)$. Several forms of $\mathbf{sf}(\phi)$ have been used in the literature which are reviewed in \cite{Kim, Kim_3}. In this study, we employ the following definition:
\begin{align} \label{surface_tension}
	\mathbf{sf}(\phi) = \sigma\varepsilon\alpha_\mathrm{sf}\nabla\cdot( |\nabla\phi|^2\mathbf{I} - \nabla\phi \otimes \nabla\phi ), 
\end{align}
where $\varepsilon$ is the interface thickness parameter defined in the Allen-Cahn phase-field equation and $\alpha_\mathrm{sf}=3\sqrt{2}/4$ is a constant. 
This completes the fully-coupled variational formulation for FSI in two-phase flow. In what follows, we propose the nonlinear partitioned iterative formulation 
for the linearized matrix forms of the coupled field equations.
%
 
\section{The nonlinear partitioned iterative coupling} 
\label{proposed_scheme}

\subsection{Coupled linearized matrix form}
We present the coupled linearized matrix form of the variationally discretized two-phase fluid-structure equations formulated in the previous section for non-overlapping decomposition of the fluid and structure domains. The linear system of equations for the formulation can be written as $\boldsymbol{A}\boldsymbol{u}=\boldsymbol{R}$, where $\boldsymbol{u}$ and $\boldsymbol{R}$ are the vector of unknowns and the right-hand side matrix respectively. Corresponding to the domain decomposition, the set of degrees of freedom (DOF) is decomposed into the interior DOFs for the two-phase fluid-structure system and the fluid-structure interface DOFs for the Dirichlet-to-Neumann (DtN) mapping. Using the Newton-Raphson type of linearization, the coupled two-phase fluid-structure system with the DtN mapping along the fluid-structure interface can be expressed as
\begin{align} \label{CLMF}
	\begin{bmatrix} \boldsymbol{A}^\mathrm{ss} & \boldsymbol{0} & \boldsymbol{0} & \boldsymbol{A}^\mathrm{Is} \\ \noalign{\vspace{4pt}}
					\boldsymbol{A}^\mathrm{sI} & \mathbf{I} & \boldsymbol{0} & \boldsymbol{0} \\ \noalign{\vspace{4pt}}
					\boldsymbol{0} & \boldsymbol{A}^\mathrm{If} & \boldsymbol{A}^\mathrm{ff} & \boldsymbol{0} \\ \noalign{\vspace{4pt}}
					\boldsymbol{0} & \boldsymbol{0} & \boldsymbol{A}^\mathrm{fI} & \mathbf{I} \end{bmatrix}  
					\begin{Bmatrix} \Delta \boldsymbol{\eta}^\mathrm{s} \\ \noalign{\vspace{4pt}}
									\Delta \boldsymbol{\eta}^\mathrm{I} \\ \noalign{\vspace{4pt}}
									\Delta \boldsymbol{q}^\mathrm{f} \\ \noalign{\vspace{4pt}}
									\Delta \boldsymbol{f}^\mathrm{I} \end{Bmatrix}
				= \begin{Bmatrix} {\boldsymbol{\mathcal{R}}}^\mathrm{s} \\ \noalign{\vspace{4pt}}
								  {\boldsymbol{\mathcal{R}}}^\mathrm{I}_\mathrm{D} \\ \noalign{\vspace{4pt}}
								  {\boldsymbol{\mathcal{R}}}^\mathrm{f} \\ \noalign{\vspace{4pt}}
								  {\boldsymbol{\mathcal{R}}}^\mathrm{I}_\mathrm{N} \end{Bmatrix}
\end{align}
where $\Delta \boldsymbol{\eta}^\mathrm{s}$ denotes the increment in the structural displacement, $\Delta \boldsymbol{\eta}^\mathrm{I}$ and $\Delta \boldsymbol{f}^\mathrm{I}$ represent the increments in the displacement and the forces along the fluid-structure interface. The increment in the unknowns associated with the two-phase fluid domain is denoted by $\Delta \boldsymbol{q}^\mathrm{f} = (\Delta \boldsymbol{u}^\mathrm{f}, \Delta p, \Delta \phi)$. On the right hand side, $\boldsymbol{\mathcal{R}}^\mathrm{s}$ and $\boldsymbol{\mathcal{R}}^\mathrm{f}$ represent the weighted residual of the structural and stabilized two-phase flow equations respectively, whereas $\boldsymbol{\mathcal{R}}^\mathrm{I}_\mathrm{D}$ and $\boldsymbol{\mathcal{R}}^\mathrm{I}_\mathrm{N}$ denote the residuals corresponding to the imbalances during the enforcement of the kinematic (Dirichlet) condition (Eq.~(\ref{kin_cond})) and the dynamic (Neumann) condition (Eq.~(\ref{dyn_cond})) at the fluid-structure interface respectively. 

The block matrices on the left-hand side can be described as follows. $\boldsymbol{A}^\mathrm{ss}$ represents the matrix consisting of the mass, damping and stiffness matrices of the structural equation for the non-interface structural DOFs and $\boldsymbol{A}^\mathrm{Is}$ is the transformation to obtain the structural force vector from the fluid-structure interface. $\boldsymbol{A}^\mathrm{sI}$ maps the structural displacements to the fluid-structure interface which satisfies the Dirichlet kinematic condition with $\mathbf{I}$ being an identity matrix. $\boldsymbol{A}^\mathrm{fI}$ transfers the fluid forces to the fluid-structure interface to satisfy the Neumann dynamic equilibrium condition. $\boldsymbol{A}^\mathrm{If}$ associates the ALE mapping of the fluid spatial points and $\boldsymbol{A}^\mathrm{ff}$ consists of the stabilized terms for the Navier-Stokes and the Allen-Cahn equations. It can be expanded as
\begin{align} \label{TPFS_LM}
	\boldsymbol{A}^\mathrm{ff} = \begin{bmatrix}
\boldsymbol{K}_{\Omega^\mathrm{f}}& & \boldsymbol{G}_{\Omega^\mathrm{f}} & & \boldsymbol{D}_{\Omega^\mathrm{f}}\\ \noalign{\vspace{4pt}}
-\boldsymbol{G}^T_{\Omega^\mathrm{f}}& &\boldsymbol{C}_{\Omega^\mathrm{f}} & & \boldsymbol{0}\\ \noalign{\vspace{4pt}}
\boldsymbol{G}_{AC} & & \boldsymbol{0} & & \boldsymbol{K}_{AC}
\end{bmatrix}
\end{align}
where $\boldsymbol{K}_{\Omega^\mathrm{f}}$ is the stiffness matrix of the momentum equation consisting of inertia, convection, viscous and stabilization terms, $\boldsymbol{G}_{\Omega^\mathrm{f}}$ is the gradient operator, $\boldsymbol{G}^T_{\Omega^\mathrm{f}}$ is the divergence operator for the continuity equation and $\boldsymbol{C}_{\Omega^\mathrm{f}}$ is the pressure-pressure stabilization term. On the other hand, $\boldsymbol{D}_{\Omega^\mathrm{f}}$ contains the terms in the momentum equation having dependency on the order parameter, $\boldsymbol{G}_{AC}$ consists of the velocity coupled term in the Allen-Cahn equation and $\boldsymbol{K}_{AC}$ is the left hand side stiffness matrix for the Allen-Cahn equation consisting of inertia, convection, diffusion, reaction and stabilization terms.

As derived in \cite{NIFC_1}, the idea of partitioning is to eliminate the off-diagonal term 
$\boldsymbol{A}^\mathrm{Is}$ to facilitate 
the staggered sequential updates for strongly coupled fluid-structure system. Through static condensation, Eq.~(\ref{CLMF}) can be written as
\begin{align} \label{CLMFv2}
	\begin{bmatrix} \boldsymbol{A}^\mathrm{ss} & \boldsymbol{0} & \boldsymbol{0} & \boldsymbol{0} \\ \noalign{\vspace{4pt}}
					\boldsymbol{A}^\mathrm{sI} & \mathbf{I} & \boldsymbol{0} & \boldsymbol{0} \\ \noalign{\vspace{4pt}}
					\boldsymbol{0} & \boldsymbol{A}^\mathrm{If} & \boldsymbol{A}^\mathrm{ff} & \boldsymbol{0} \\ \noalign{\vspace{4pt}}
					\boldsymbol{0} & \boldsymbol{0} & \boldsymbol{0} & \boldsymbol{A}^\mathrm{II} \end{bmatrix}  
					\begin{Bmatrix} \Delta \boldsymbol{\eta}^\mathrm{s} \\ \noalign{\vspace{4pt}}
									\Delta \boldsymbol{\eta}^\mathrm{I} \\ \noalign{\vspace{4pt}}
									\Delta \boldsymbol{q}^\mathrm{f} \\ \noalign{\vspace{4pt}}
									\Delta \boldsymbol{f}^\mathrm{I} \end{Bmatrix}
				= \begin{Bmatrix} {\boldsymbol{\mathcal{R}}}^\mathrm{s} \\ \noalign{\vspace{4pt}}
								  {\boldsymbol{\mathcal{R}}}^\mathrm{I}_\mathrm{D} \\ \noalign{\vspace{4pt}}
								  {\boldsymbol{\mathcal{R}}}^\mathrm{f} \\ \noalign{\vspace{4pt}}
								  \widetilde{\boldsymbol{\mathcal{R}}}^\mathrm{I}_\mathrm{N} \end{Bmatrix}
\end{align}
In the nonlinear interface force correction, we form the iterative scheme of 
the following matrix-vector product form
\begin{equation} 
\Delta \vec{f}^\mathrm{I} =
{\big({\vec{A}^\mathrm{II}}\big)^{-1}} \widetilde{\boldsymbol{\mathcal{R}}}^\mathrm{I}_\mathrm{N},
\end{equation}
where ${\big({\vec{A}^\mathrm{II}}}\big)^{-1}$ is not constructed explicitly. Instead,
the force correction vector $\Delta\boldsymbol{f}^\mathrm{I}$ at the nonlinear iteration (subiteration) $\mathrm{k}$ can be constructed by 
successive matrix-vector products. This process essentially provides the control  for the interface fluid force 
$\vec{f}^I=\int_{\Gamma^\mathrm{fs}} \boldsymbol{\sigma}^\mathrm{f}\cdot \mathbf{n}^\mathrm{f} \mathrm{d\Gamma}$ 
to stabilize strong fluid-structure interaction at low structure-to-fluid mass ratio. 
The scheme proceeds in a similar fashion as the 
predictor-corrector schemes by constructing the iterative interface force correction at each iteration. 
Consider the system of equations for the interface force residual:
\begin{align}
\vec{E}^\mathrm{I} = \Delta \vec{f}^\mathrm{I} - {\big({\vec{A}^\mathrm{II}}\big)^{-1}} \widetilde{\boldsymbol{\mathcal{R}}}^\mathrm{I}_\mathrm{N} =\vec{0},
\end{align}
which can be recast as the following quasi-Newton update
\begin{align}
\Delta\vec{f}^\mathrm{I}_\mathrm{(k+1)}  = \Delta\vec{f}^\mathrm{I}_\mathrm{(k)} + \Lambda_\mathrm{(k)} \Delta \vec{E}^\mathrm{I}_\mathrm{(k)},  \qquad \mathrm{k}=0,1,2, ...
\end{align}
where ${\Lambda}_\mathrm{(k)}$ is an $n \times n$ matrix and $\vec{E}^\mathrm{I}_\mathrm{(k)}$ is the residual vector. There are 
three possible alternatives for the matrix $\Lambda_\mathrm{(k)}$, namely, scalar, diagonal and full matrix. 
We consider ${\Lambda}_\mathrm{(k)} =  \alpha_\mathrm{k} \mathbf{I}$ for the iterative quasi-Newton update, which can 
be considered as a minimal residual iteration method as follows:
Let the vector $\Delta \vec{y} = \Delta \vec{f}^\mathrm{I} + \Lambda_\mathrm{(k)} \Delta \vec{E}^\mathrm{I}_\mathrm{(k)}$ 
and the scalar $\alpha_\mathrm{k}$ can be selected such that 
\begin{align}
\alpha_\mathrm{k} = -\frac{( \vec{y}, \Delta \vec{f}^\mathrm{I}_\mathrm{(k-1)} )  } {( \vec{y}, \Delta \vec{E}^\mathrm{I}_\mathrm{(k-1)}  )},
\end{align}
where $(\cdot ,\cdot)$ denotes the standard inner product. The choice of $\vec{y}= \Delta \vec{f}^\mathrm{I}_\mathrm{(k)}$ 
minimizes $\|\Delta \vec{y}_\mathrm{(k)} \| =  \|\Delta \vec{f}^\mathrm{I}_\mathrm{(k-1)} + \alpha_\mathrm{k} \Delta \vec{E}^\mathrm{I}_\mathrm{(k-1)}  \|$. 
The above quasi-Newton update is a nonlinear generalization of the steepest descent method \cite{Brezinski2003,Brezinski}.

\subsection{Implementation details}
The two-phase flow system in Eq.~(\ref{TPFS_LM}) is decoupled into two subsystems: Navier-Stokes and Allen-Cahn solves, for which the linear system of equations can be summarized as
\begin{align} \label{LS_NS}
\begin{bmatrix}
\boldsymbol{K}_{\Omega^\mathrm{f}}& & \boldsymbol{G}_{\Omega^\mathrm{f}}\\ \noalign{\vspace{4pt}}
-\boldsymbol{G}^T_{\Omega^\mathrm{f}}& &\boldsymbol{C}_{\Omega^\mathrm{f}}
\end{bmatrix}
\begin{Bmatrix}
\Delta\boldsymbol{u}^\mathrm{f} \\ \noalign{\vspace{4pt}}
\Delta p
\end{Bmatrix}
&=
\begin{Bmatrix}
\widetilde{\boldsymbol{\mathcal{R}}}_\mathrm{m} \\ \noalign{\vspace{4pt}}
\widetilde{\boldsymbol{\mathcal{R}}}_\mathrm{c}
\end{Bmatrix}\\
\begin{bmatrix}
\boldsymbol{K}_{AC}
\end{bmatrix}
\begin{Bmatrix}
\Delta\phi
\end{Bmatrix}
&=
\begin{Bmatrix}
\widetilde{\mathcal{R}}(\phi)
\end{Bmatrix} \label{LS_AC}
\end{align}
where $\widetilde{\boldsymbol{\mathcal{R}}}_\mathrm{m}$, $\widetilde{\boldsymbol{\mathcal{R}}}_\mathrm{c}$ and $\widetilde{\mathcal{R}}(\phi)$ represent the weighted residuals of the stabilized momentum, continuity and the Allen-Cahn equations respectively. Notice that the terms forming the matrices $\boldsymbol{D}_{\Omega^\mathrm{f}}$ and $\boldsymbol{G}_{AC}$ do not exist after the decoupling since we decouple the equations in a partitioned iterative manner which is described below.  
Using a Newton-Raphson technique, the resulting two-phase flow variables and the ALE mesh displacement coming from the finite element discretization 
are evaluated by solving the linear system of equations via the Generalized Minimal RESidual (GMRES) algorithm proposed in \cite{saad1986}. 
To  form the linear matrix system, we only construct the required matrix-vector products of each block matrix  for the GMRES algorithm, 
instead of  constructing the left-hand side matrix explicitly. 

The algorithm for the partitioned iterative coupling of the implicit two-phase fluid structure solver is presented in Algorithm \ref{algorithm_1}. It consists of seven steps in a nonlinear iteration for the exchange of data between the different blocks of the solver. In a typical nonlinear iteration $\mathrm{k}$, the first step involves the solution of the structure equation to get the updated structural displacements $\boldsymbol{\eta}^\mathrm{s,n+1}_\mathrm{(k+1)}$. These displacements are transferred to the Navier-Stokes solve by satisfying the ALE compatibility condition at the fluid-structure interface $\Gamma^\mathrm{fs}$ in the second step. This is accomplished as follows: let the updated mesh displacement be denoted by $\boldsymbol{\eta}^\mathrm{m,n+1}_\mathrm{(k+1)}$. This mesh displacement is equated to the structural displacement at the interface $\Gamma^\mathrm{fs}$ to prevent any overlaps between the fluid and the structural domains,
\begin{align}
	\boldsymbol{\eta}^\mathrm{m,n+1}_\mathrm{(k+1)} = \boldsymbol{\eta}^\mathrm{s,n+1}_\mathrm{(k+1)},\ \mathrm{on}\ \Gamma^\mathrm{fs}.
\end{align}	 
Moreover, the conservation property between the moving elements in the fluid domain is satisfied by equating the fluid velocity to the mesh velocity at the interface, i.e., 
\begin{align}
	\boldsymbol{u}^\mathrm{f,n+\alpha^f}_\mathrm{(k+1)} = \boldsymbol{u}^\mathrm{m,n+\alpha^f}_\mathrm{(k+1)},\  \mathrm{on}\ \Gamma^\mathrm{fs},
\end{align}
where the mesh velocity is written as,
\begin{align} \label{mesh_vel_eqn}
	\boldsymbol{u}^\mathrm{m,n+\alpha^f}_\mathrm{(k+1)} = \frac{\boldsymbol{\eta}^\mathrm{m,n+1}_\mathrm{(k+1)}-\boldsymbol{\eta}^\mathrm{m,n}_\mathrm{(k+1)}}{\Delta t} = \boldsymbol{u}^\mathrm{s,n+\alpha^s}_\mathrm{(k+1)}\ \mathrm{on}\ \Gamma^\mathrm{fs}.
\end{align}
This ensures that the no-slip condition is satisfied at the fluid-structure interface (Eq.~(\ref{kin_cond})). The mesh displacement for each spatial point $\boldsymbol{x}^\mathrm{f} \in \Omega^\mathrm{f}(t)$ is obtained by solving a Poisson equation assuming the mesh to act as a hyper-elastic material \cite{Ogden, Bonet}. The mesh velocity for the spatial points $\boldsymbol{x}^\mathrm{f}\in \Omega^\mathrm{f}(t)$ is then evaluated using the first equality in Eq.~(\ref{mesh_vel_eqn}). The convection velocity is adjusted by subtracting the mesh velocity $\boldsymbol{u}^\mathrm{m,n+\alpha^f}_\mathrm{(k+1)}$ from the fluid velocity $\boldsymbol{u}^\mathrm{f,n+\alpha^f}_\mathrm{(k+1)}$ and transferred to the Navier-Stokes solve.  In the third step, the Navier-Stokes equations are solved in the ALE reference coordinate system (Eq.~(\ref{LS_NS})), thus solving for updated velocity $\boldsymbol{u}^\mathrm{f,n+1}_\mathrm{(k+1)}$ and pressure $p^\mathrm{n+1}_\mathrm{(k+1)}$. This updated fluid velocity with the mesh velocity (to obtain the adjusted convection velocity) is then transferred to the Allen-Cahn solve in the fourth step. The Allen-Cahn equation (Eq.~(\ref{LS_AC})) is solved to evolve the fluid-fluid interface $\Gamma^\mathrm{ff}$ in the updated mesh configuration in the fifth step. The physical properties of the fluid such as its density, viscosity and surface tension are then updated with the help of the updated order parameter values $\phi^\mathrm{n+1}_\mathrm{(k+1)}$ in the sixth step. With the help of all the updated fluid variables, the hydrodynamic forces on the fluid-structure interface $\Gamma^\mathrm{fs}$ is evaluated by integrating the stress tensor over the structural surface. The force corrected by the nonlinear iterative force correction (NIFC) \cite{NIFC_1, NIFC_2} procedure, $\boldsymbol{f}^\mathrm{I}_\mathrm{(k+1)}$ at the fluid-structure interface is equated with the structural force in the final step, thus satisfying the dynamic equilibrium (Eq.~(\ref{dyn_cond})) at the fluid-structure interface,
\begin{align}
	\boldsymbol{f}^\mathrm{s,n+\alpha^s}_\mathrm{(k+1)} = \boldsymbol{f}^\mathrm{I}_\mathrm{(k+1)}\ \mathrm{on}\ \Gamma^\mathrm{fs}.
\end{align}
\begin{algorithm}
\caption{Partitioned coupling of implicit two-phase fluid-structure interaction solver}
\label{algorithm_1}
\begin{tabbing}
\     Given $\boldsymbol{u}^\mathrm{f,0}$, $p^0$, $\phi^0$, $\boldsymbol{\eta}^\mathrm{s,0}$ \\
\qquad    Loop over time steps, $\mathrm{n}=0,1,\cdots$ \\
\qquad\	  Start from known variables $\boldsymbol{u}^\mathrm{f,n}$, $p^\mathrm{n}$, $\phi^\mathrm{n}$, $\boldsymbol{\eta}^\mathrm{s,n}$\\
\qquad\   Predict the solution: \\
\qquad\qquad $\boldsymbol{u}^\mathrm{f,n+1}_{(0)}=\boldsymbol{u}^\mathrm{f,n}$;\quad $p^\mathrm{n+1}_{(0)} = p^\mathrm{n}$;\quad $\phi^\mathrm{n+1}_{(0)} = \phi^\mathrm{n}$;\quad $\boldsymbol{\eta}^\mathrm{s,n+1}_{(0)} = \boldsymbol{\eta}^\mathrm{s,n}$\\ 
\qquad\   Loop over the nonlinear iterations, $\mathrm{k}=0,1,\cdots$ until convergence \\
\qquad \begin{tikzpicture}
	\draw (0.75,0.5) node(E){[5]};
	\draw (0.75,0) node(E1){\textbf{Allen-Cahn}};
	\draw (0.75,-0.5) node(E2){ \textbf{Implicit Solve}};
	\draw (0.75,-1) node(E2){ Solve Eq.~(\ref{LS_AC}) on $\Omega^\mathrm{f}(t)$};
	\node[fit = (E) (E1) (E2),style={block3}](AC){};
\end{tikzpicture}
\hspace{0cm}
\begin{tikzpicture}
	\node (0,0){};
	\draw [->,>=stealth'] (1,0) to [out=210,in=-30] (-1,0);
	\draw [->,>=stealth'] (-1,1.5) to [out=30,in=150] (1,1.5);
	\draw (0,-0.5) node[anchor=north]{[4] $\boldsymbol{u}^\mathrm{f,n+\alpha^f}_\mathrm{(k+1)}$};
	\draw (0,2) node[anchor=south]{[6] $\phi^\mathrm{n+\alpha^f}_\mathrm{(k+1)}$};
	\draw (0,0.5) node[anchor=south]{$\Gamma^\mathrm{ff}$};
\end{tikzpicture}
\begin{tikzpicture}
	\centering
	\draw (0.75,0.5) node(E){[3]};
	\draw (0.75,0) node(E1){\textbf{Navier-Stokes}};
	\draw (0.75,-0.5) node(E2){\textbf{Implicit Solve}};
	\draw (0.75,-1) node(E2){ Solve Eq.~(\ref{LS_NS}) on $\Omega^\mathrm{f}(t)$};
	\node[fit = (E) (E1) (E2),style={block3}](NS){};
\end{tikzpicture}
\hspace{0cm}
\begin{tikzpicture}
	\node (0,0){};
	\draw [->,>=stealth'] (1,0) to [out=210,in=-30] (-1,0);
	\draw [->,>=stealth'] (-1,1.5) to [out=30,in=150] (1,1.5);
	\draw (0,-0.5) node[anchor=north]{[2] $\boldsymbol{u}^\mathrm{f,n+\alpha^f}_\mathrm{(k+1)} = \boldsymbol{u}^\mathrm{s,n+\alpha^s}_\mathrm{(k+1)}$};
	\draw (0,2) node[anchor=south]{[7] $\boldsymbol{f}^\mathrm{s,n+\alpha^s}_\mathrm{(k+1)} = \boldsymbol{f}^\mathrm{I}_\mathrm{(k+1)}$};
	\draw (0,0.5) node[anchor=south]{$\Gamma^\mathrm{fs}$};
\end{tikzpicture}
\begin{tikzpicture}
	\centering
	\draw (0.75,0.5) node(E){[1]};
	\draw (0.75,0) node(E1){\textbf{Structure}};
	\draw (0.75,-0.5) node(E2){ \textbf{Implicit Solve}};
	\draw (0.75,-1) node(E2){on $\Omega^\mathrm{s}$};
	\node[fit = (E) (E1) (E2),style={block3}](SS){};
\end{tikzpicture}
\end{tabbing}
\end{algorithm}

\subsection{General remarks}
The exact surface tracking of the fluid-structure interface via the ALE technique along with the interface capturing phase-field technique for the fluid-fluid interface renders the formulation hybrid. 
While the phase-field model approximates the interface by a smeared surface using the internal length scale parameter, the present Allen-Cahn based phase-field formulation is derived from the thermodynamic arguments and has a theoretical basis in the minimization of the Ginzburg-Landau energy functional. Unlike the level-set and volume-of-fluid techniques, the interface evolution by the phase-field description simplifies the formulation by avoiding any re-initialization or geometric reconstruction of the interface. Furthermore, the PPV formulation to solve the nonlinear Allen-Cahn equation helps to establish the positivity condition nonlinearly at the local element matrix level resulting in the positivity preserving and monotone scheme, which has been shown in \cite{AC_adaptive_JCP}. 

The ability of the solver to handle low structure-to-fluid mass ratio can be attributed to the NIFC procedure based on quasi-Newton updates. The idea behind the procedure is to construct the cross-coupling effect of strong fluid-structure interaction along the interface without forming the off-diagonal Jacobian term ($\boldsymbol{A}^\mathrm{Is}$ in Eq.~(\ref{CLMF})) via nonlinear iterations. The correction relies on an input-output relationship between the structural displacement and the force transfer at each nonlinear iteration. The input-output feedback process can be also considered as a nonlinear generalization of the steepest descent method to transform a divergent fixed-point iteration to a stable and convergent update of the approximate forces associated with the interface degrees of freedom \cite{NIFC_1}. Unlike the brute-force iterations in the strongly coupled FSI which lead to severe numerical instabilities for low structure-to-fluid mass ratios, the NIFC procedure provides a desired stability to the partitioned fluid-structure coupling, without the explicit evaluation of the off-diagonal Jacobian term. 
Further details about the NIFC formulation can be found in \cite{NIFC_1, NIFC_2}.
The above mentioned characteristics of the proposed partitioned coupling between the two-phase fluid and the structure lead to a robust and stable formulation. Moreover, the partitioned-block type feature of the solver leads to flexibility and ease in its implementation to the existing variational solvers. These desirable features of the proposed formulation are analyzed and assessed through various numerical tests in the next section.

\section{Numerical tests}
\label{tests}
In this section, we present some numerical tests to assess the coupling between the two-phase Allen-Cahn based solver and the structural solver. To accomplish this, we perform the decay tests by examining the interaction of the free-surface with a rigid circular cylinder under translation and a rectangular barge under pure rotation.

\subsection{Heave decay test under translation}
\label{cyl_test}
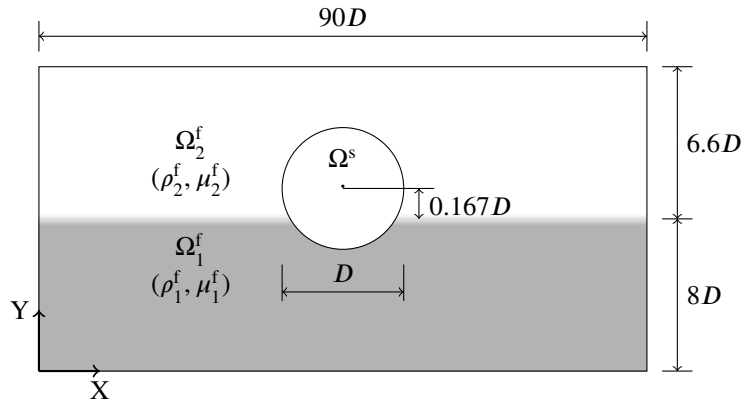
\begin{figure}[h]
\centering
\begin{tikzpicture}[decoration={markings,mark=at position 1.0 with {\arrow{>}}},scale=4]
	\draw[fill={lightBlack},draw=none] (0,0) -- (2,0)-- (2,0.48) -- (0,0.48) -- cycle;
	\draw[fill=white,draw=none] (0,0.52) -- (0,1) --(2,1) --(2,0.52) --(0,0.52);
	\foreach \i[evaluate={\col=(\i+1)/10*100}] in {0,...,10}
      \fill[color=white!\col!lightBlack]
      (0,0.48+\i*0.004) -- (2,0.48+\i*0.004) -- (2,0.48+\i*0.004+0.004) -- (0,0.48+\i*0.004+0.004) -- cycle  ;
    \draw (0,0) -- (2,0) --(2,1) --(0,1) --(0,0);
	\draw[fill=white] (1,0.6) circle (0.2cm);
	\draw (0.5,0.4) node(A){$\Omega^\mathrm{f}_1$};
	\node [below = -0.2cm of A]{$(\rho_1^\mathrm{f}, \mu_1^\mathrm{f})$};
	\draw (0.5,0.75) node(B){$\Omega^\mathrm{f}_2$};
	\node [below = -0.2cm of B]{$(\rho_2^\mathrm{f}, \mu_2^\mathrm{f})$};
	\draw (1,0.7) node(C){$\Omega^\mathrm{s}$};
	\draw[thick,postaction={decorate}] (0,0) to (0.2,0);
	\draw[thick,postaction={decorate}] (0,0) to (0,0.2);
	\draw (0.2,0) node[anchor=north]{X};
	\draw (0,0.2) node[anchor=east]{Y};
	\draw[postaction={decorate}] (2.1,0.75) to (2.1,1);
	\draw[postaction={decorate}] (2.1,0.75) to (2.1,0.5);
	\draw[postaction={decorate}] (2.1,0.25) to (2.1,0.5);
	\draw[postaction={decorate}] (2.1,0.25) to (2.1,0);
	\draw (2.05,0) -- (2.15,0);
	\draw (2.05,1) -- (2.15,1);
	\draw (2.05,0.5) -- (2.15,0.5);
	\draw (2.1,0.75) node[anchor=west]{$6.6D$};
	\draw (2.1,0.25) node[anchor=west]{$8D$};
	\draw[postaction={decorate}] (0.5,1.1) to (2,1.1);
	\draw[postaction={decorate}] (0.5,1.1) to (0,1.1);
	\draw (0,1.05) -- (0,1.15);
	\draw (2,1.05) -- (2,1.15);
	\draw (1,1.1) node[anchor=south]{$90D$};
	\draw (0.8,0.35) -- (0.8,0.23);
	\draw (1.2,0.35) -- (1.2,0.23);
	\draw[<->] (0.8,0.26) -- (1.2,0.26);
	\draw (1,0.26) node[anchor=south] {$D$};
	\draw[<->] (1.25,0.6) -- (1.25,0.5);
	\draw(1,0.6) node(D){$\boldsymbol{\cdot}$};
	\draw (1,0.6) -- (1.3,0.6) ;
	\draw (1.42, 0.48) node[anchor=south]{$0.167D$};
\end{tikzpicture}
\caption{Schematic of the decay response of a cylinder of diameter $D$ under gravity in the $X$-$Y$ cross-section. The computational domain extends a distance of $17D$ in the $Z$-direction. } 
\label{cyl}
\end{figure}

We herein consider the free heave motion of a circular cylinder at the free-surface of water. The schematic of the computational domain, $\Omega \in [0,90D]\times [0,14.6D] \times [0,17D]$ considered in this study is shown in Fig. \ref{cyl},  where a circular cylinder of diameter $D=0.1524$ is placed initially at an offset of $0.167D$ m from the free-surface of water. The density of the cylinder is half that of the denser fluid, i.e., $\rho^\mathrm{s}=500$, $\rho^\mathrm{f}_1=1000$ and $\rho^\mathrm{f}_2=1.2$. The dynamic viscosities of the two phases are $\mu^\mathrm{f}_1=10^{-3}$ and $\mu^\mathrm{f}_2=1.8\times 10^{-5}$. The acceleration due to gravity is $\boldsymbol{g} = (0,-9.81,0)$. Apart from the high density ratio between the two phases, $\rho^*=\rho^\mathrm{f}_{1}/\rho^\mathrm{f}_{2}=833.3$, a low structure-to-fluid density ratio $(\rho^\mathrm{s}/\rho^\mathrm{f}_{1}=0.5)$ based on the denser fluid has been chosen. The initial condition for the order parameter is given as
\begin{align}
	\phi(x,y,0) =  -\mathrm{tanh}\bigg( \frac{y}{\sqrt{2}\varepsilon} \bigg).
\end{align}
We have employed the hybrid RANS/LES model discussed in \cite{DDES_CAF} for modeling the turbulent effects. The Reynolds number is defined based on the maximum velocity achieved by the cylinder and its diameter with respect to the denser fluid, i.e., $Re=\rho^\mathrm{f}_{1}U_{cyl}D/\mu^\mathrm{f}_1 \approx 30,000$. 

\begin{figure}[h]
\centering
	\begin{subfigure}[b]{0.45\textwidth}
		\includegraphics[trim={1cm 1cm 1cm 1cm},clip,width=8cm]{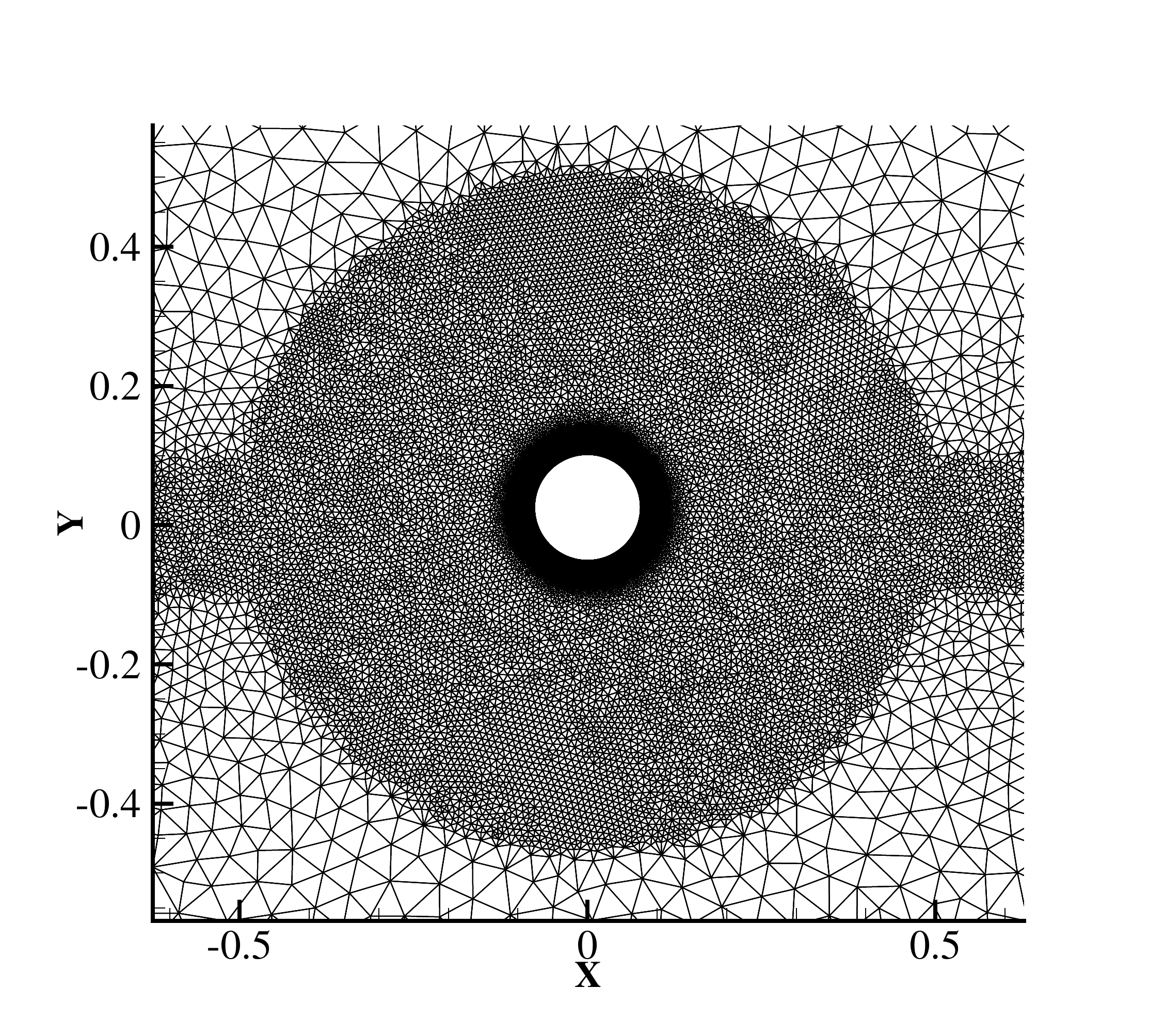}
	\caption{}
	\end{subfigure}%
	\begin{subfigure}[b]{0.45\textwidth}
		\includegraphics[trim={1cm 1cm 1cm 1cm},clip,width=8cm]{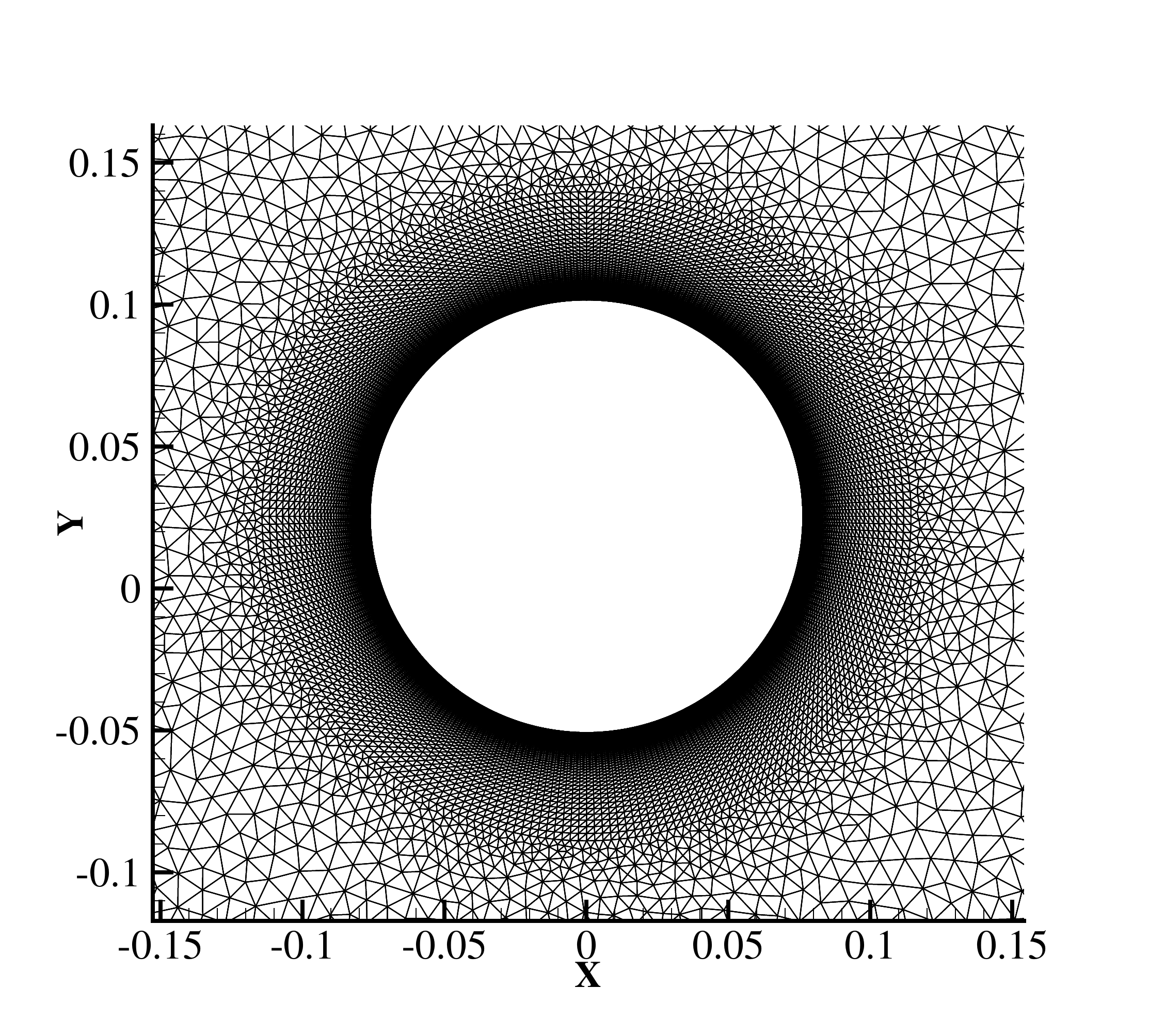}
	\caption{}
	\end{subfigure}		
\caption{Mesh employed for the decay response of a cylinder interacting with the free-surface: (a) the refined mesh around the cylinder to capture the flow vortices near the structure, and (b) the boundary layer mesh around the cylinder with $y^+\sim 1$ for hybrid RANS/LES modeling.} 
\label{cyl_mesh}
\end{figure}

A typical computational mesh prepared for the decay test is shown in Fig.~\ref{cyl_mesh}. A boundary layer covers the structural cylinder with the first layer at a distance from the cylindrical surface such as to maintain $y^+\sim 1$ (Fig. \ref{cyl_mesh}(b)). Moreover, a refined mesh consisting of a cylinder with radius $3.3D$ is constructed around the boundary layer region to capture the vortices produced due to the heave motion at the free-surface (Fig. \ref{cyl_mesh}(a)). To capture the air-water interface accurately, the interfacial region is refined in accordance with the suggestions in \cite{AC_JCP} such that at least $4$ elements lie in the equilibrium interface region. The mesh is then extruded in the $Z$-direction consisting of $7$ layers. The no-slip boundary condition is satisfied at the cylindrical surface while the slip boundary condition is set on all other boundaries.

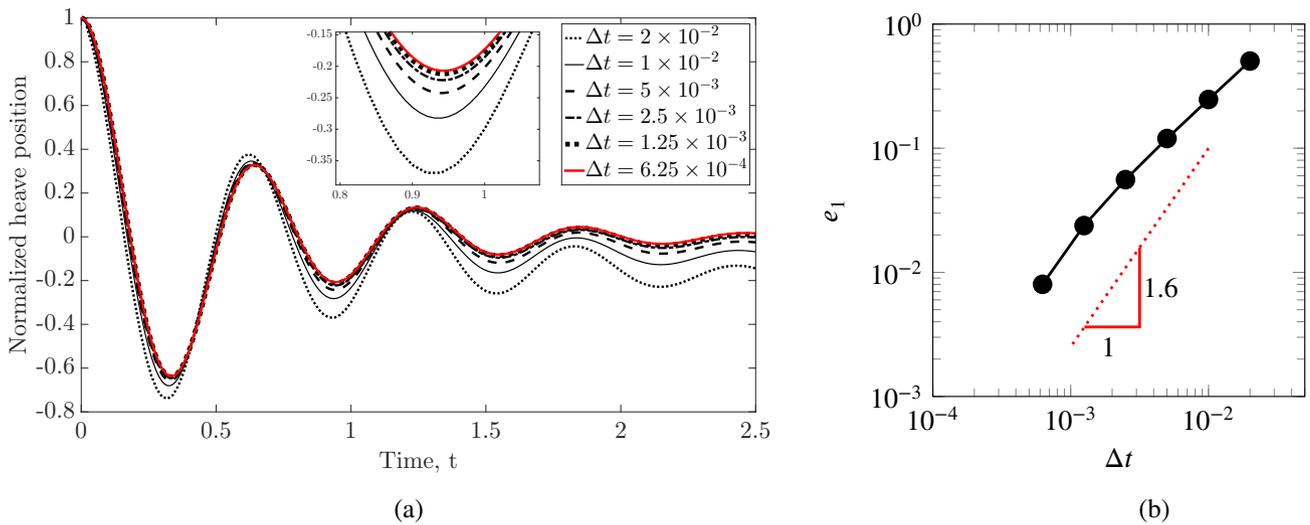
\begin{figure}[h]
\centering
\begin{subfigure}[b]{0.6\textwidth}
		\includegraphics[trim={2cm 0 0cm 0cm},clip,width=11cm]{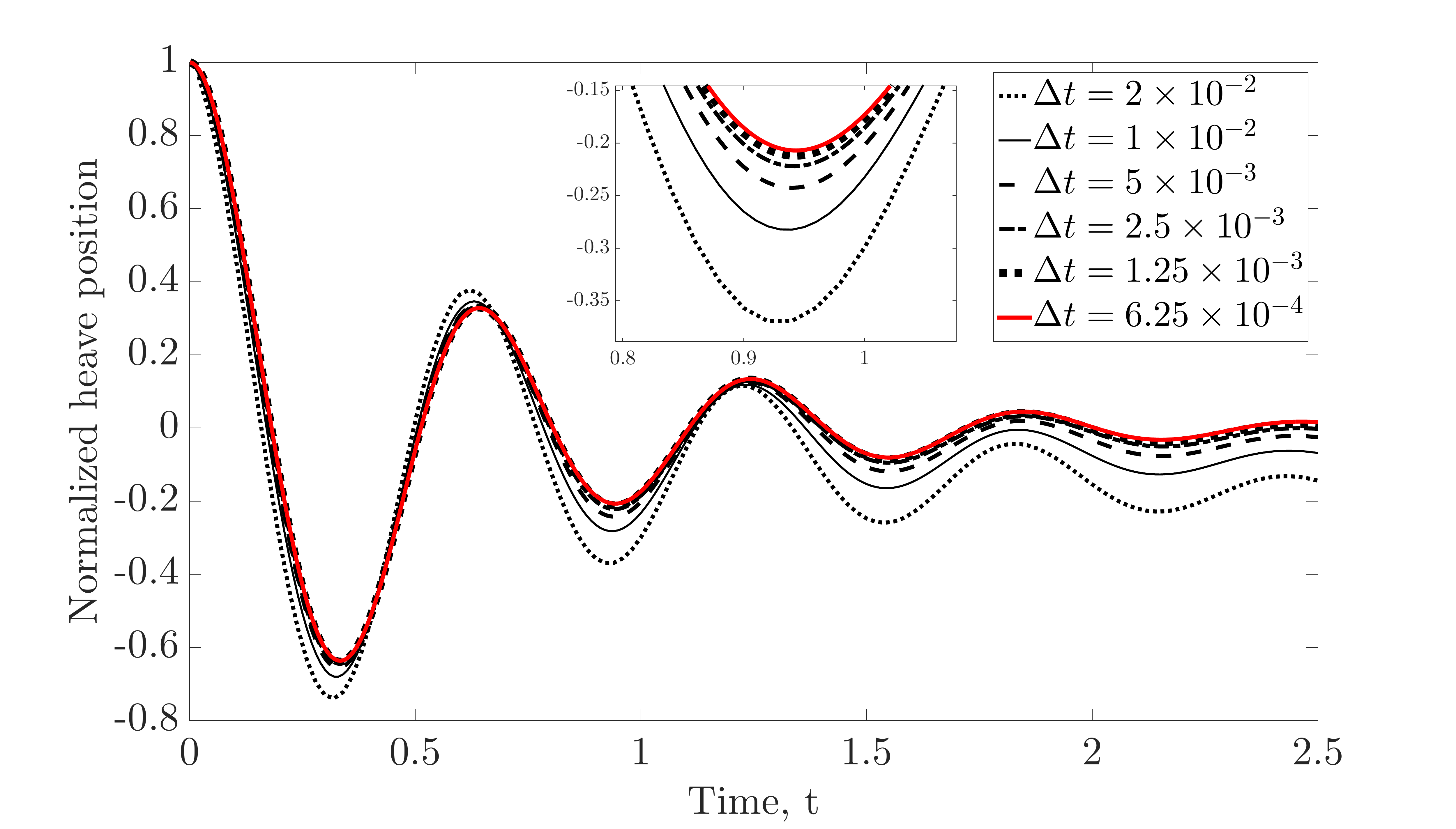}
	\caption{}
\end{subfigure}%
\begin{subfigure}[b]{0.5\textwidth}
\begin{tikzpicture}[scale=1.05]
\begin{loglogaxis}[
    width=0.7\textwidth,
    height=0.7\textwidth,
    xlabel={$\Delta t$},
    ylabel={$e_1$},
    xmin=0.0001, xmax=0.05,
    ymin=0.001, ymax=1,
    legend pos=north west,
]
\addplot[mark=*,mark options={solid}, black,mark size={3pt},line width={1pt}]
    coordinates {
    (0.02,0.5036)(0.01,0.24702)(0.005,0.11972)(0.0025,0.05582)(0.00125,0.02381)(0.000625,0.00801)
    };
\addplot[dotted,red,mark size={3pt},line width={1pt}]
    coordinates {
    (0.01,0.1)(0.001,2.5119e-3)
    }
    coordinate [pos=0.5] (A)
    coordinate [pos=0.9] (B)
    coordinate [pos=0.7] (C)
    coordinate [pos=1.0] (D);
\draw[red,line width={1pt}] (A) |- (B) ;
\draw (C) node[anchor=west,black]{$\ \ \ 1.6$};
\draw (D) node[anchor=west,black]{$\ \ \ 1$};
\end{loglogaxis}
\end{tikzpicture}
	\caption{}
\end{subfigure}
\caption{Temporal convergence study for the decay test of a circular cylinder: (a) heave motion employing temporal refinement, and (b) the dependence of non-dimensionalized $L^2$ error $(e_1)$ as a function of uniform temporal refinement $\Delta t$.}
\label{time_conv}
\end{figure}

We first present the convergence studies for which we considered the two-dimensional domain for the current case.  For the temporal convergence, a time step of $\Delta t=2\times 10^{-2}$ is decreased by a factor of 2 till $\Delta t=6.25\times 10^{-4}$. The decaying heave motion of the cylinder under different time steps is plotted in Fig. \ref{time_conv}(a). A non-dimensional error for quantifying the convergence is established which is defined as 
\begin{align}
	e_1 = \frac{|| \boldsymbol{\eta}-\boldsymbol{\eta}_\mathrm{ref}||_2}{||\boldsymbol{\eta}_\mathrm{ref}||_2},
\end{align}
where $\boldsymbol{\eta}$ represents the temporal evolution of the heave motion of the cylinder for corresponding time step, $\boldsymbol{\eta}_\mathrm{ref}$ is the heave motion evolution with time for the finest time step ($\Delta t=6.25\times 10^{-4}$) and $||\cdot||_2$ is the standard Euclidean $L^2$ norm. Figure \ref{time_conv}(b) shows the plot for the error $e_1$ with the time step size $\Delta t$, which gives a temporal convergence of $1.6$. 

\begin{figure}[h]
\centering
		\includegraphics[trim={0cm 0 0cm 0cm},clip,width=14cm]{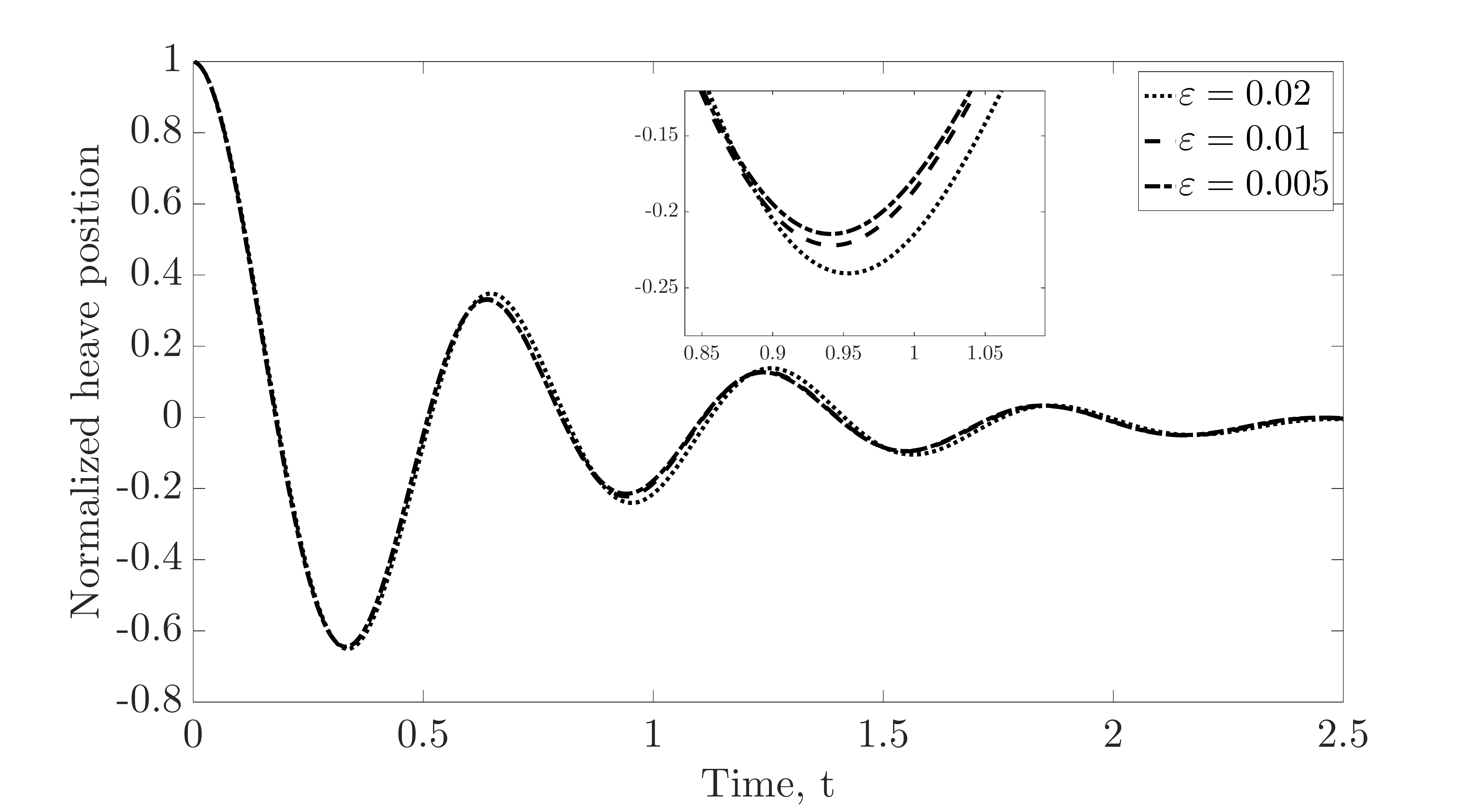}
\caption{Dependence of spatial grid convergence on  the interfacial thickness parameter $\varepsilon$ for the decay test of a circular cylinder at free-surface.} 
\label{spatial_conv}
\end{figure}

The spatial convergence is studied based on the interface thickness parameter $\varepsilon$. Three values of the parameter were selected, viz., $\varepsilon\in [0.02,0.01,0.005]$. The heave motion is shown in Fig. \ref{spatial_conv} for different values of $\varepsilon$. We observe a minor difference in the response of the cylinder interacting with the free-surface. Therefore, based on the convergence studies, we select the interface thickness parameter of $\varepsilon=0.01$ and time step of $\Delta t=0.0025$ for further assessment.

\begin{figure}[h]
\centering
\begin{subfigure}[b]{0.6\textwidth}
		\includegraphics[trim={0cm 0cm 0cm 0cm},clip,width=10cm]{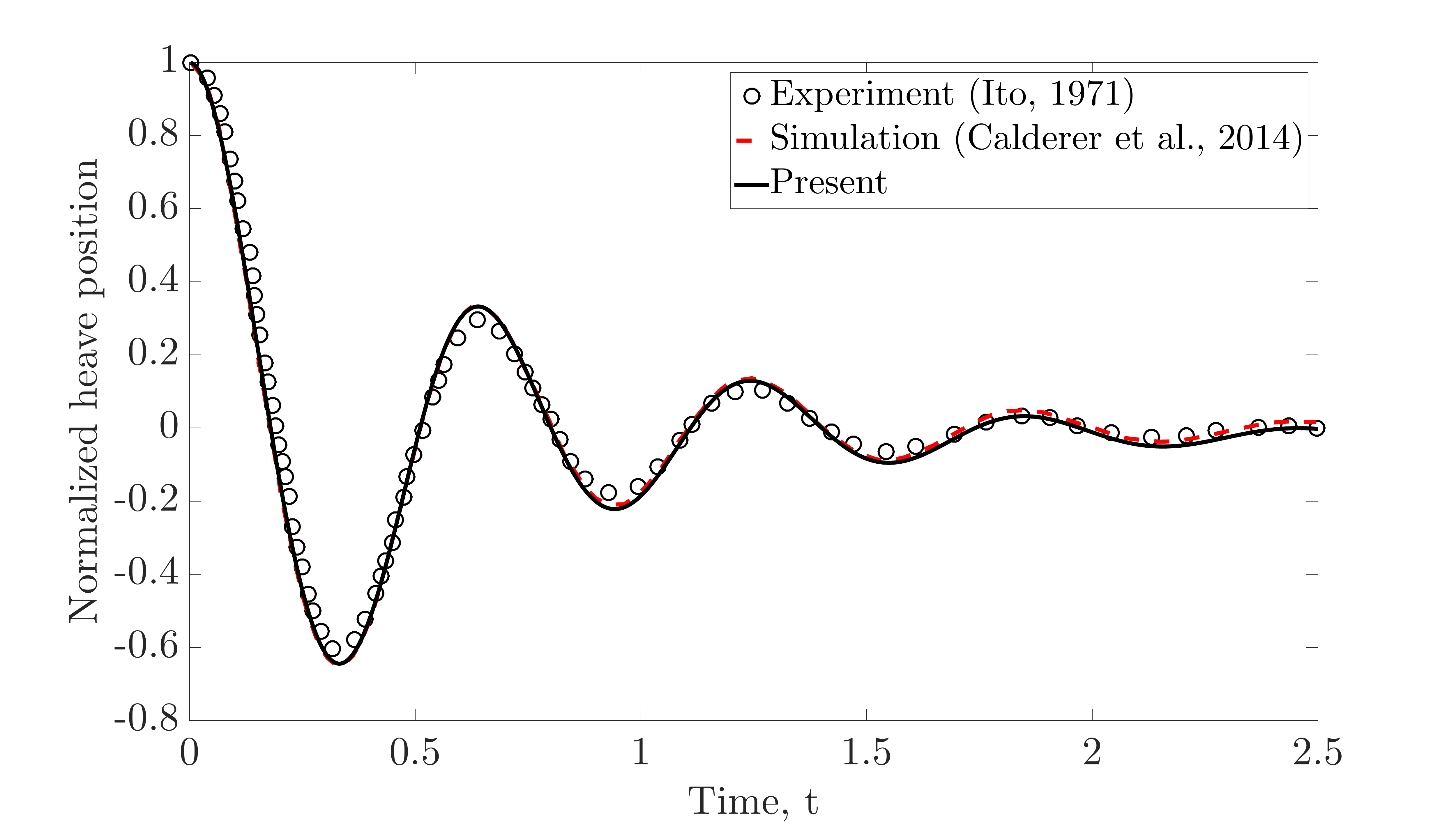}
\caption{} 
\end{subfigure}%
\begin{subfigure}[b]{0.35\textwidth}
		\includegraphics[trim={0cm 2cm 0cm 2cm},clip,width=6cm]{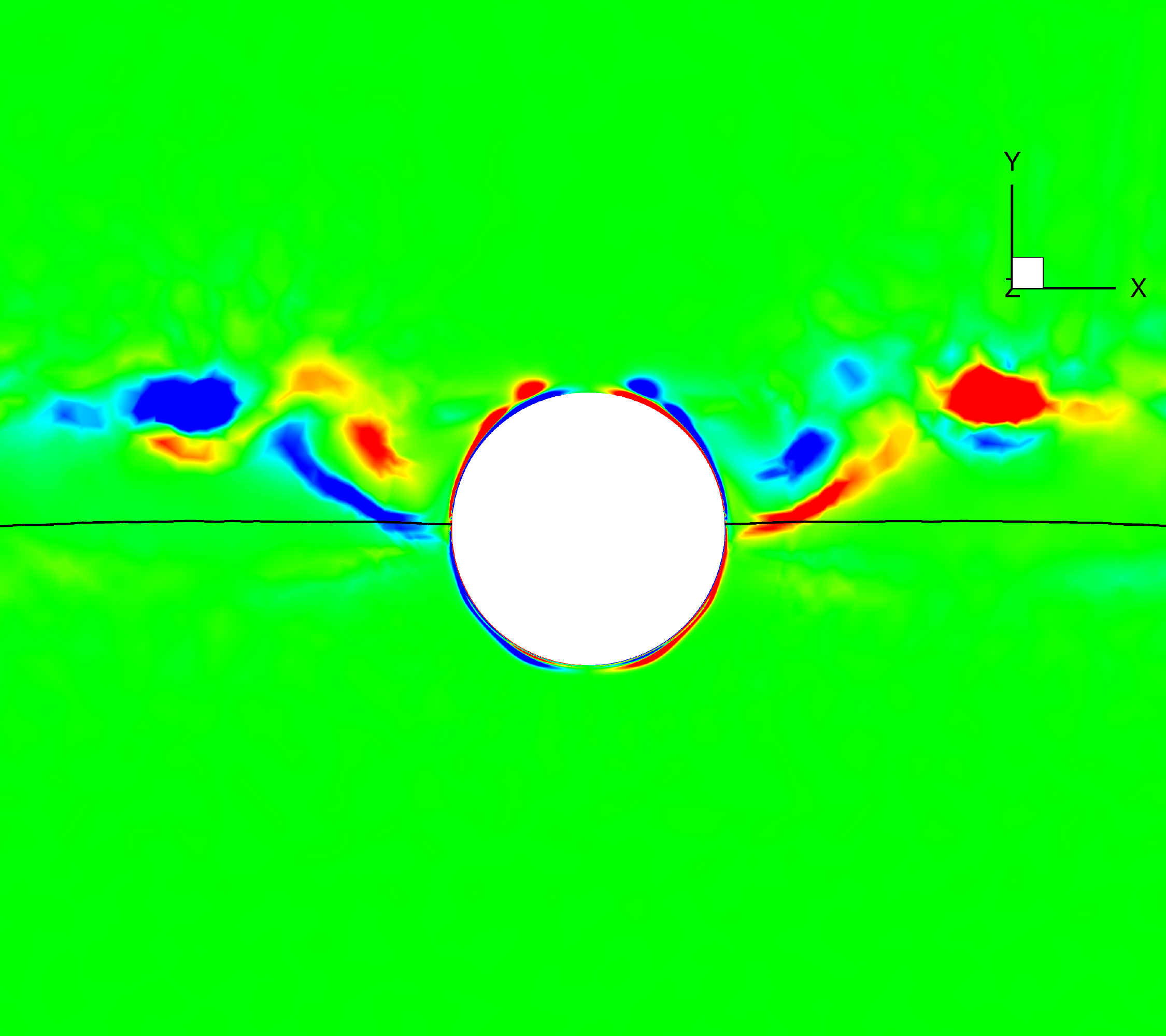}
		\includegraphics[trim={4cm 8cm 5cm 10cm},clip,width=6cm]{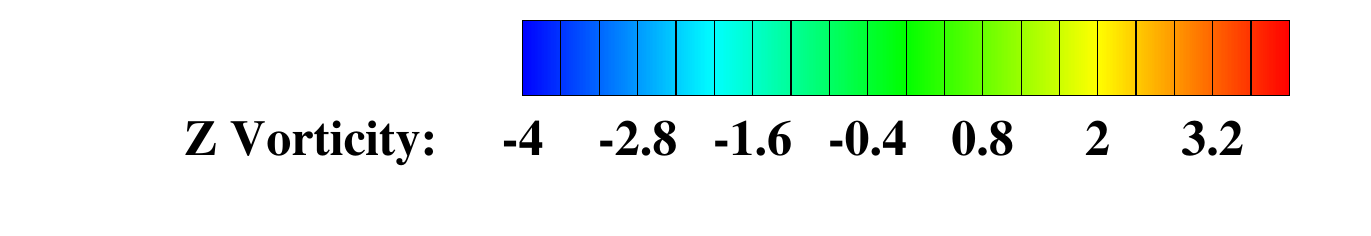}
		\caption{} 
\end{subfigure}
\caption{Decay test of a circular cylinder under translation along the free-surface: (a) validation of the heave motion of the cylinder at the free-surface with the experimental \cite{Ito_thesis} and simulation \cite{Calderer} studies, and (b) $Z$-vorticity contours around the cylinder at $t=1.5$ s with the free-surface indicated at $\phi=0$.} 
\label{cyl_val}
\end{figure}

In what follows, we perform the three-dimensional computation of the fluid-structure interaction problem with the selected spatial and temporal convergence parameters for validation with the experiment \cite{Ito_thesis} and the simulation \cite{Calderer}. The three-dimensional mesh consists of $580,000$ nodes with $1.01$ million six-node wedge elements. The simulation was carried out with $48$ processors which took a total computational time of $11.63$ hours. The solver performed $4$ nonlinear iterations to achieve a nonlinear convergence tolerance of $5\times 10^{-4}$. The results of the evolution of the heave of the cylinder are shown in Fig. \ref{cyl_val}(a) where we find a very good agreement with the literature. The $Z$-vorticity contours with the interface shown at $\phi=0$ is depicted in Fig. \ref{cyl_val}(b). A finer grid surrounding the cylinder can increase the resolution of the vortices around the cylinder.  

\subsection{Decay test under rotation along free-surface}
\begin{figure}[h]
\centering
	\begin{subfigure}[b]{0.5\textwidth}
\begin{tikzpicture}[decoration={markings,mark=at position 1.0 with {\arrow{>}}},scale=4]
	\draw[fill={lightBlack},draw=none] (0,0) -- (2,0)-- (2,0.48) -- (0,0.48) -- cycle;
	\draw[fill=white,draw=none] (0,0.52) -- (0,1) --(2,1) --(2,0.52) --(0,0.52);
	\foreach \i[evaluate={\col=(\i+1)/10*100}] in {0,...,10}
      \fill[color=white!\col!lightBlack]
      (0,0.48+\i*0.004) -- (2,0.48+\i*0.004) -- (2,0.48+\i*0.004+0.004) -- (0,0.48+\i*0.004+0.004) -- cycle  ;
    \draw (0,0) -- (2,0) --(2,1) --(0,1) --(0,0);
	\draw[fill=white,rotate=-15,shift={(-0.15cm,0.27cm)}] (0.8,0.7)--(1.2,0.7)--(1.2,0.4)--(0.8,0.4)--(0.8,0.7);
	\draw (1.5,0.4) node(A){$\Omega^\mathrm{f}_1$};
	\node [below = -0.2cm of A]{$(\rho_1^\mathrm{f}, \mu_1^\mathrm{f})$};
	\draw (1.5,0.75) node(B){$\Omega^\mathrm{f}_2$};
	\node [below = -0.2cm of B]{$(\rho_2^\mathrm{f}, \mu_2^\mathrm{f})$};
	\draw (1,0.65) node(C){$\Omega^\mathrm{s}$};
	\draw[thick,postaction={decorate}] (0,0) to (0.2,0);
	\draw[thick,postaction={decorate}] (0,0) to (0,0.2);
	\draw (0.2,0) node[anchor=north]{X};
	\draw (0,0.2) node[anchor=east]{Y};
	\draw[postaction={decorate}] (2.1,0.75) to (2.1,1);
	\draw[postaction={decorate}] (2.1,0.75) to (2.1,0.5);
	\draw[postaction={decorate}] (2.1,0.25) to (2.1,0.5);
	\draw[postaction={decorate}] (2.1,0.25) to (2.1,0);
	\draw (2.05,0) -- (2.15,0);
	\draw (2.05,1) -- (2.15,1);
	\draw (2.05,0.5) -- (2.15,0.5);
	\draw (2.1,0.75) node[anchor=west]{$3.33L$};
	\draw (2.1,0.25) node[anchor=west]{$3L$};
	\draw[postaction={decorate}] (0.5,1.1) to (2,1.1);
	\draw[postaction={decorate}] (0.5,1.1) to (0,1.1);
	\draw (0,1.05) -- (0,1.15);
	\draw (2,1.05) -- (2,1.15);
	\draw (1,1.1) node[anchor=south]{$58.3L$};
	\draw(1,0.5) node(D){$\boldsymbol{\cdot}$};
	\draw[<->,rotate=-15,shift={(-0.15cm,0.27cm)}] (0.8,0.75) -- (1.2,0.75) node [midway, above, sloped]{$L$};
	\draw[<->,rotate=-15,shift={(-0.15cm,0.27cm)}] (0.75,0.47) -- (0.75,0.7) node [midway, left]{$0.5L$};
	\draw[thick,dotted,rotate=-15,shift={(-0.2cm,-0.005cm)}] (0.6,0.75) -- (1.055,0.75) node [midway, above, sloped]{};
	\draw[] (0.75,0.57) arc (130:220:0.05);
	\draw[] (0.65,0.55) node {$\theta$};
\end{tikzpicture}
	\caption{}
	\end{subfigure}%
	\begin{subfigure}[b]{0.5\textwidth}
	\hspace{1cm}
		\includegraphics[trim={1cm 1cm 1cm 1cm},clip,width=7cm]{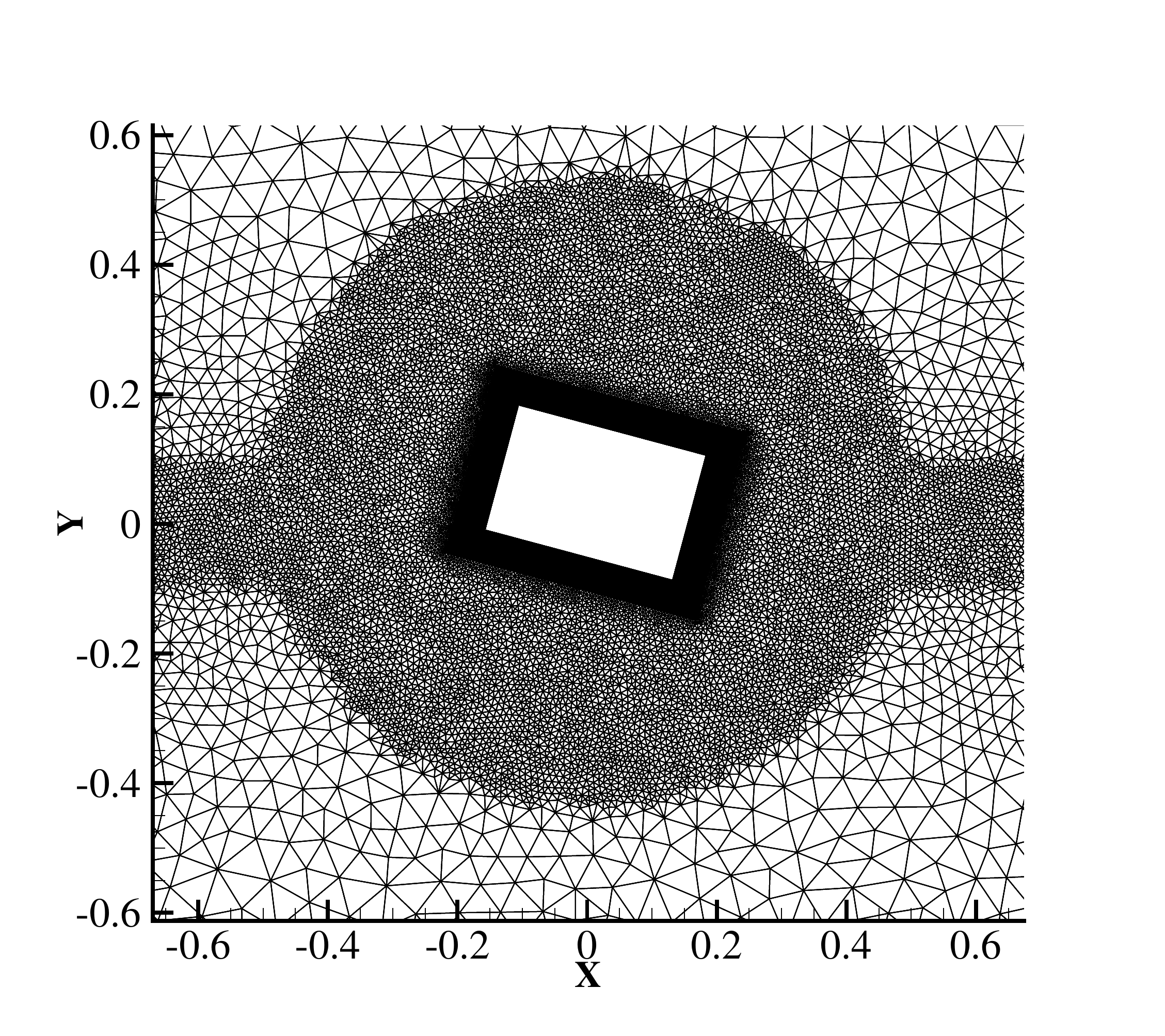}
	\caption{}
	\end{subfigure}	
\caption{Rotation of a rectangular barge under gravity: (a) schematic of the computational setup of a barge of length $L=0.3$, height $H=0.2$ and width $W=3L$ in the $X$-$Y$ cross-section, and (b) zoom-in view of the computational mesh near the barge with boundary layer. The computational domain extends a distance of $3L$ in the $Z$ direction.} 
\label{domain_barge}
\end{figure}
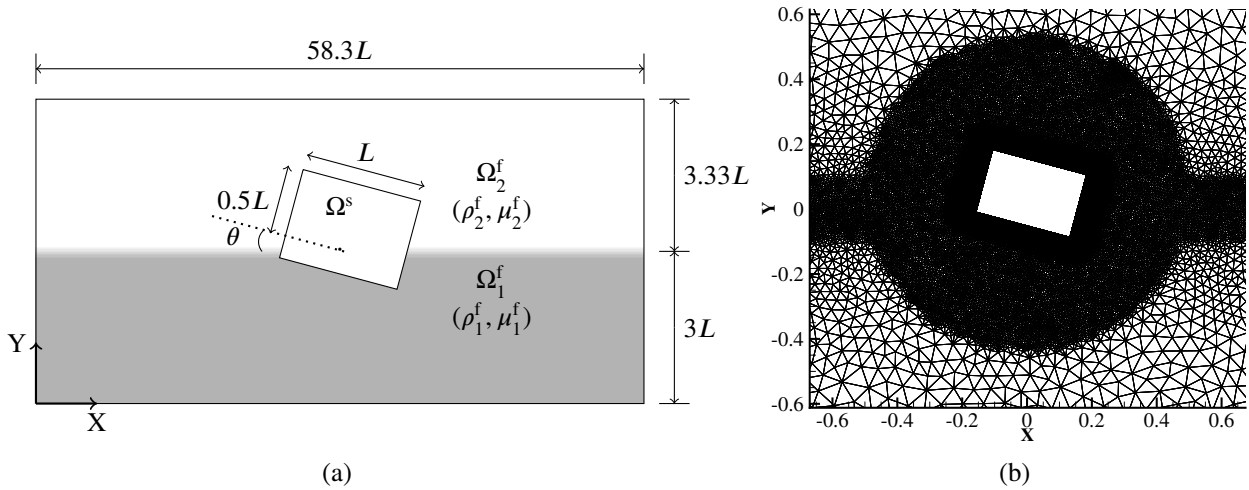

For further validation and robustness assessment, we consider the pure rotation of a rectangular barge of length $L=0.3$, height $H=0.2$ and width $W=3L$ at the free-surface of water. The computational domain $\Omega\in [0,58.3L]\times [0,6.3L]\times [0,3L]$ with the barge inclined at an angle of $\theta=15^{\circ}$ from the free-surface level is shown in Fig. \ref{domain_barge}(a). The centre of gravity of the barge is at the free-surface level with its mass moment of inertia and the rotational damping matrices respectively as
\begin{align}
 	\boldsymbol{I}^\mathrm{s}= \begin{bmatrix} 0 & 0 & 0 \\
 												 0 & 0 & 0 \\
 												 0 & 0 & 0.236 \end{bmatrix} \mathrm{kg\cdot m^2},\qquad \boldsymbol{C}_\mathrm{\theta}= \begin{bmatrix} 0 & 0 & 0 \\
 												 0 & 0 & 0 \\
 												 0 & 0 & 0.275 \end{bmatrix} \mathrm{kg\cdot \frac{m^2}{s}}.
 \end{align}
The physical properties of the fluid domain are $\rho^\mathrm{f}_1=1000$, $\rho^\mathrm{f}_2=1.2$, $\mu^\mathrm{f}_1=10^{-3}$, $\mu^\mathrm{f}_2=1.8\times 10^{-5}$ and $\boldsymbol{g}=(0,-9.81,0)$. The initial condition for the order parameter is given as
\begin{align}
	\phi(x,y,0) =  -\mathrm{tanh}\bigg( \frac{y}{\sqrt{2}\varepsilon} \bigg).
\end{align}

The Reynolds number is defined based on the maximum velocity achieved by the upper corner of the barge and its length with respect to the denser fluid, i.e., $Re=\rho^\mathrm{f}_1 U_{corner}L/\mu^\mathrm{f}_1\approx 99,500$. The computational mesh employed for the simulation shown in Fig. \ref{domain_barge}(b) is constructed with similar characteristics as the mesh in section \ref{cyl_test}. A boundary layer envelops the rectangular barge with a refined mesh around the boundary layer to capture the vortices produced at the free-surface. Moreover, the evolution in the free-surface is captured by a refined region around the interfacial region. The mesh extrudes in the $Z$-direction and consists of $10$ layers.
\begin{figure}[h]
\centering
\begin{subfigure}[b]{0.6\textwidth}
		\includegraphics[trim={0cm 0cm 0cm 0cm},clip,width=10cm]{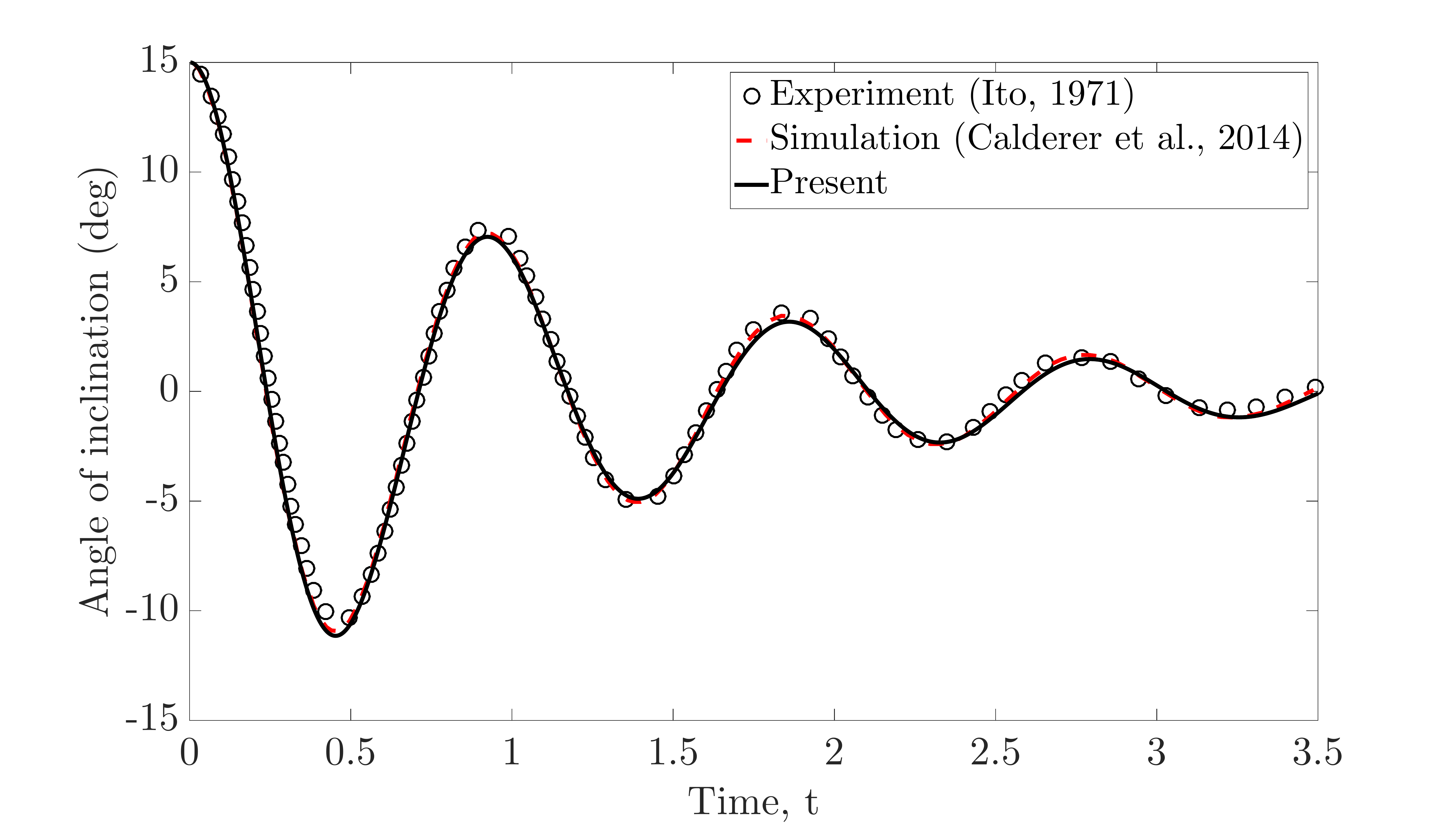}
\caption{} 
\end{subfigure}%
\begin{subfigure}[b]{0.35\textwidth}
		\includegraphics[trim={0cm 2cm 0cm 2cm},clip,width=6cm]{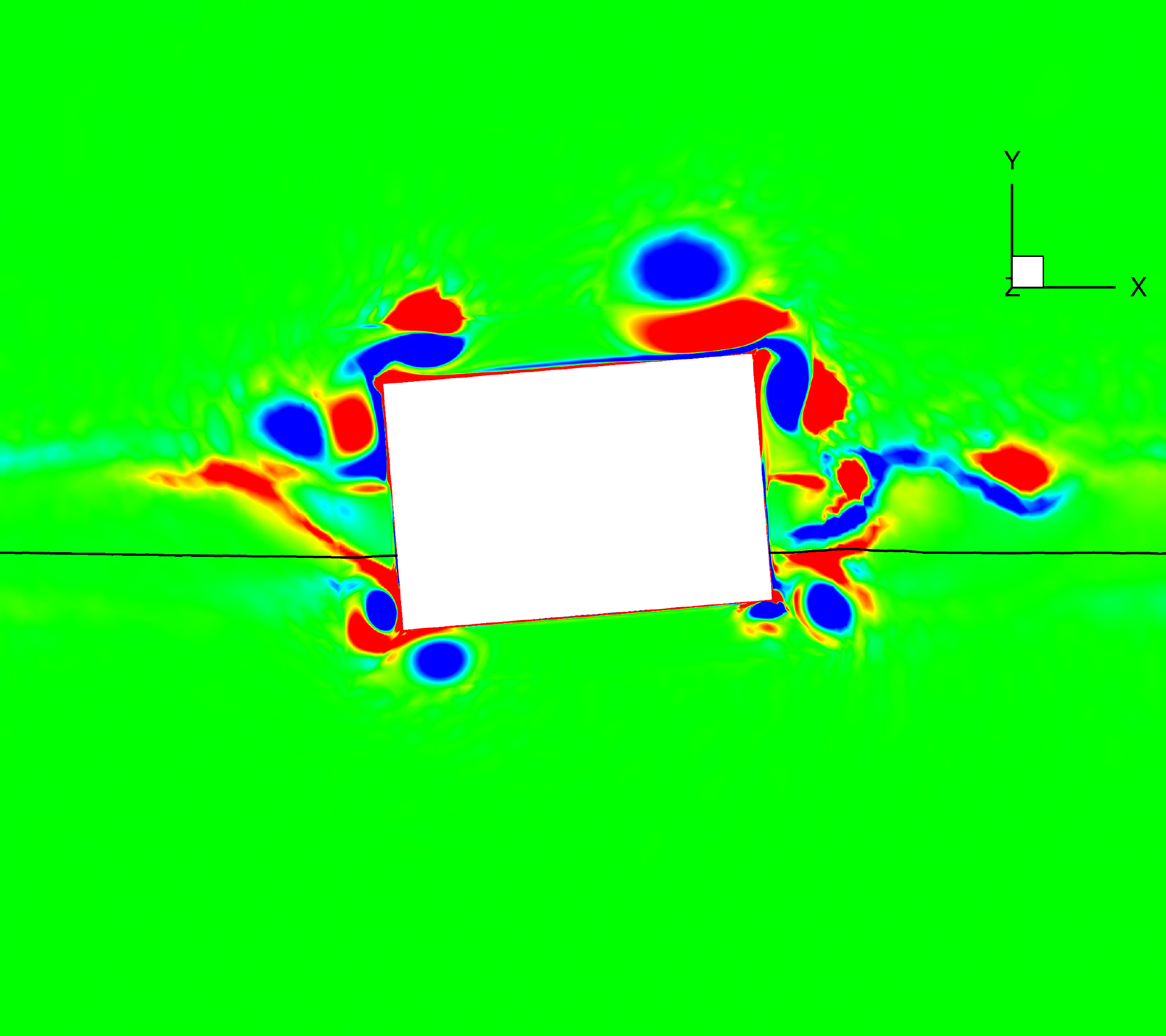}
		\includegraphics[trim={4cm 8cm 5cm 10cm},clip,width=6cm]{VorZ_legend.eps}
		\caption{} 
\end{subfigure}
\caption{Decay test under rotation: (a) validation of the rotational motion of the rectangular barge at the free-surface with the experimental \cite{Kwang} and simulation \cite{Calderer} studies, and (b) $Z$-vorticity contours around the barge at $t=1.4$ s with the free-surface indicated at $\phi=0$.} 
\label{barge_val}
\end{figure}

With the converged spatial and temporal parameters, we validate the rotational response of the rectangular barge with that of the experiment \cite{Kwang} and the computational data \cite{Calderer}. The mesh consists of 1,025,684 nodes with 1,856,930 six-node wedge elements. The simulation was carried out by 72 processors with a computational time of 22.2 hours. On an average, the solver performed $4$ nonlinear iterations to achieve a nonlinear convergence tolerance of $5\times 10^{-4}$. The rotational motion is plotted in Fig. \ref{barge_val}(a) where we observe good agreement with the available results in the literature. The $Z$-vorticity contours are shown in Fig. \ref{barge_val}(b) at $t=1.4$ s. This concludes the validation and convergence studies for the coupled FSI two-phase solver. We next demonstrate a practical problem of a uniform flow across a flexible pipeline with internal two-phase flow flowing inside it.

\section{Application to flexible riser FSI with internal two-phase flow}
\label{demonstration}
We next demonstrate the capability of the developed phase-field FSI solver in handling a practical problem of a riser with an internal two-phase flow and exposed to external uniform current flow. A typical schematic for the problem is shown in Fig. \ref{riser_IF_schematic}. The riser has an outer diameter of $D$ and span of $L=20D$. The inflow and outflow boundaries are at a distance of $10D$ and $30D$ from the center of the riser respectively. The side walls are equidistant from the riser center at $15D$ on either side. The outer surface of the riser is exposed to a uniform inflow current of $\boldsymbol{u}^\mathrm{f}=(U_{\infty},0,0)$. The no-slip boundary condition is satisfied at the outer surface of the riser. All other boundaries are slip boundaries except the outflow where the stress-free condition is satisfied. The fluid domain exposed to the external part of the riser is denoted by $\Omega^\mathrm{f}_1$. The interior of the riser has internal diameter $2r_2$ with an initial concentric profile for the two phases with the interface at a radius of $r_1$ from the riser axis separating the two phases $\Omega^\mathrm{f}_2$ and $\Omega^\mathrm{f}_1$. A prescribed profile for the $Z$-velocity is imposed at the inlet and outlet of the riser for the internal flow. The velocity is such that no-slip condition is satisfied at the internal surface of the riser. We consider the profile of the velocity for a co-annular, laminar and fully developed flow regime consisting of immiscible Newtonian fluids, given by \cite{Nogueira1990}
\begin{align}
	\boldsymbol{u}^\mathrm{f} &= (0,0,\mathrm{w}),\\
	\mathrm{w}(R) &= \begin{dcases} \frac{C[1-(r^*)^2 + \mu^*((r^*)^2-R^2)]}{(r^*)^{n+3}(\mu^*-1)+1},\ &0\leq R\leq r^*, \\ \frac{C[1-R^2]}{(r^*)^{n+3}(\mu^*-1)+1},\ &r^*\leq R\leq 1, \end{dcases}
\end{align}
where $R= \sqrt{x^2 + y^2}/r_2$, $r^* = r_1/r_2$, $\mu^* = \mu^\mathrm{f}_1/\mu^\mathrm{f}_2$ and $C=(n+3)/2$, where $n=1$ for a circular tube. The physical parameters employed for the demonstration are $r_1=0.2$, $r_2=0.4$ and $\boldsymbol{g}=(0,0,0)$. The initial condition for the order parameter is given as
\begin{align}
	\phi(x,y,0) = \begin{dcases} \mathrm{tanh}\bigg( \frac{\sqrt{x^2 + y^2} - r_1}{\sqrt{2}\varepsilon} \bigg),\ &\sqrt{x^2 + y^2} \leq r_2,\\ 1,\ &\mathrm{elsewhere}    \end{dcases}
\end{align}
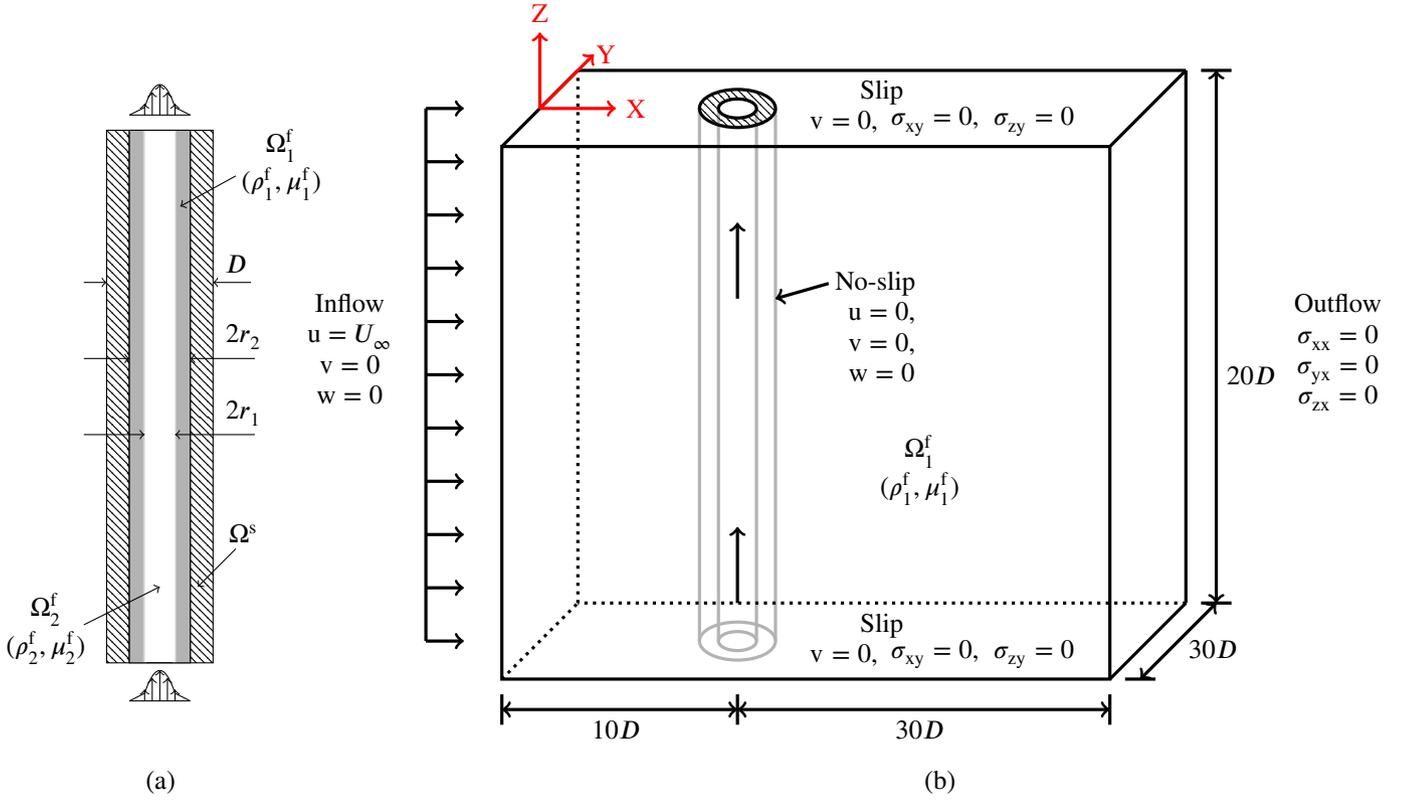
\begin{figure}[h]
\centering
	\begin{subfigure}[b]{0.25\textwidth}
	\hspace{0cm}
		\begin{tikzpicture}[decoration={markings,mark=at position 0.5 with {\arrow{>}}},scale=1]
	\draw [] (-0.4,0) to (-0.4,7);
	\draw [] (0.4,0) to (0.4,7);
	\draw [] (-0.4,0) to (0.4,0);
	\draw [] (-0.4,7) to (0.4,7);
	
	\draw [pattern=north west lines] (0.4,0) rectangle (0.7,7);
	\draw [pattern=north west lines] (-0.4,0) rectangle (-0.7,7);
	
	\draw [fill={lightBlack},draw=none] (0.4,0) rectangle (0.22,7);
	\draw [fill={lightBlack},draw=none] (-0.4,0) rectangle (-0.22,7);
	
	\foreach \i[evaluate={\col=(\i+1)/10*100}] in {0,...,10}
      \fill[color=white!\col!lightBlack]
      (0.22-\i*0.004,0) -- (0.22-\i*0.004,7) -- (0.22-\i*0.004-0.004,7) -- (0.22-\i*0.004-0.004,0) -- cycle  ;
    \foreach \i[evaluate={\col=(\i+1)/10*100}] in {0,...,10}
      \fill[color=white!\col!lightBlack]
      (-0.22+\i*0.004,0) -- (-0.22+\i*0.004,7) -- (-0.22+\i*0.004+0.004,7) -- (-0.22+\i*0.004+0.004,0) -- cycle  ;
    \draw (0.4,0) -- (0.4,7) --(-0.4,7) --(-0.4,0) --(0.4,0);	

	\draw [->] (-1,5) to (-0.7,5);
	\draw [<-] (0.7,5) to (1.2,5);
	\draw (1,5) node[above]{$D$};
	
	\draw [->] (-1,4) to (-0.4,4);
	\draw [<-] (0.4,4) to (1.25,4);
	\draw (1.1,4) node[above]{$2r_2$};
	
	\draw [->] (-1,3) to (-0.2,3);
	\draw [<-] (0.2,3) to (1.25,3);
	\draw (1.1,3) node[above]{$2r_1$};

	\draw (-0.4,-0.5) to (0.4,-0.5);
	\draw (0,-0.1) parabola (-0.2,-0.4);
	\draw (-0.2,-0.4) to (-0.4,-0.5);
	\draw (0,-0.1) parabola (0.2,-0.4);
	\draw (0.2,-0.4) to (0.4,-0.5);
	\draw [->] (0,-0.5) to (0,-0.1);
	\draw [->] (0.1,-0.5) to (0.1,-0.2);
	\draw [->] (-0.1,-0.5) to (-0.1,-0.2);
	\draw [->] (0.2,-0.5) to (0.2,-0.35);
	\draw [->] (-0.2,-0.5) to (-0.2,-0.35);
	
	\draw (-0.4,7.2) to (0.4,7.2);
	\draw (0,7.6) parabola (-0.2,7.3);
	\draw (-0.2,7.3) to (-0.4,7.2);
	\draw (0,7.6) parabola (0.2,7.3);
	\draw (0.2,7.3) to (0.4,7.2);
	\draw [->] (0,7.2) to (0,7.6);
	\draw [->] (0.1,7.2) to (0.1,7.5);
	\draw [->] (-0.1,7.2) to (-0.1,7.5);
	\draw [->] (0.2,7.2) to (0.2,7.35);
	\draw [->] (-0.2,7.2) to (-0.2,7.35);
	
	\draw [<-] (0,1) to (-1,0.5);
	\draw (-1.5,0.7) node[]{$\Omega^\mathrm{f}_2$};
	\draw (-1.5,0.2) node[]{$(\rho^\mathrm{f}_2,\mu^\mathrm{f}_2)$};
	\draw [<-] (0.27,6) to (1,6.4);
	\draw (1.6,6.8) node[]{$\Omega^\mathrm{f}_1$};
	\draw (1.6,6.3) node[]{$(\rho^\mathrm{f}_1,\mu^\mathrm{f}_1)$};
	\draw [<-] (0.5,1) to (1,1.5);
	\draw (1.1,1.7) node[]{$\Omega^\mathrm{s}$};
 
 	\draw (0,-1) node{};
\end{tikzpicture}
\caption{}
	\end{subfigure}%
\begin{subfigure}[b]{0.9\textwidth}
\hspace{-0.5cm}
\begin{tikzpicture}[very thick,decoration={markings,mark=at position 0.5 with {\arrow{>}}},scale=1]
	\draw[{lightBlack}] (3.1,0.5) ellipse (0.5cm and 0.25cm);
	\draw[{lightBlack}] (3.1,0.5) ellipse (0.25cm and 0.125cm);
	\draw[{lightBlack}] (3.35,7.5) to (3.35,0.5);
	\draw[{lightBlack}] (3.6,7.5) to (3.6,0.5);
	\draw[{lightBlack}] (2.85,7.5) to (2.85,0.5);
	\draw[{lightBlack}] (2.6,7.5) to (2.6,0.5);
	\draw[pattern=north west lines] (3.1,7.5) ellipse (0.5cm and 0.25cm);
	\draw[fill=white] (3.1,7.5) ellipse (0.25cm and 0.125cm);

	\draw (0,0) node[left]{} -- (0,7) node[right]{} -- (8,7) node[above]{} -- (8,0) node[above]{} -- cycle;
	\draw[black,dotted] (1,1) to (1,8);
	\draw[black] (1,8) to (9,8);
	\draw[black] (9,8) to (9,1);
	\draw[black,dotted] (9,1) to (1,1);
	\draw[black] (0,7) to (1,8);
	\draw[black] (8,7) to (9,8);
	\draw[black] (8,0) to (9,1);
	\draw[black,dotted] (0,0) to (1,1);

	\draw[->](-1.0,7.5) to (-0.5,7.5);
	\draw[->](-1.0,0.5) to (-0.5,0.5);
	\draw[->](-1.0,1.2) to (-0.5,1.2);
	\draw[->](-1.0,1.9) to (-0.5,1.9);
	\draw[->](-1.0,2.6) to (-0.5,2.6);
	\draw[->](-1.0,3.3) to (-0.5,3.3);
	\draw[->](-1.0,4) to (-0.5,4);
	\draw[->](-1.0,4.7) to (-0.5,4.7);
	\draw[->](-1.0,5.4) to (-0.5,5.4);
	\draw[->](-1.0,6.1) to (-0.5,6.1);
	\draw[->](-1.0,6.8) to (-0.5,6.8);
	\draw (-1,0.5) to (-1,7.5);
	\draw (-2,5.2) node[anchor=north]{Inflow};
	\draw (-2,4.8) node[anchor=north]{$\mathrm{u}=U_{\infty}$};
	\draw (-2,4.4) node[anchor=north]{$\mathrm{v}=0$};
	\draw (-2,4.0) node[anchor=north]{$\mathrm{w}=0$};

	\draw (9.2,1) to (9.6,1);
	\draw (9.2,8) to (9.6,8);
	\draw[<->] (9.4,1) to (9.4,8);
	\draw (11,5.2) node[anchor=north]{Outflow};
	\draw (11,4.8) node[anchor=north]{$\sigma_{\mathrm{xx}}=0$};
	\draw (11,4.4) node[anchor=north]{$\sigma_{\mathrm{yx}}=0$};
	\draw (11,4.0) node[anchor=north]{$\sigma_{\mathrm{zx}}=0$};
	\draw (9.4,4) node[anchor=west]{$20D$};

	\draw (5,8) node[anchor=north]{Slip};
	\draw (4.5,7) node[anchor=south]{$\mathrm{v}_{}=0$,};
	\draw (5.7,7) node[anchor=south]{$\sigma_\mathrm{xy}=0,$};
	\draw (7.0,7) node[anchor=south]{$\sigma_\mathrm{zy}=0$};

	\draw (5,1) node[anchor=north]{Slip};
	\draw (4.5,0) node[anchor=south]{$\mathrm{v}_{}=0$,};
	\draw (5.7,0) node[anchor=south]{$\sigma_\mathrm{xy}=0,$};
	\draw (7.0,0) node[anchor=south]{$\sigma_\mathrm{zy}=0$};

	\draw[->] (4.3,5.2) to (3.6,5.0);
	\draw (4.25,5.2) node[anchor=west]{No-slip};
	\draw (5,5.1) node[anchor=north]{$\mathrm{u}=0$,};
	\draw (5,4.7) node[anchor=north]{$\mathrm{v}=0$,};
	\draw (5,4.3) node[anchor=north]{$\mathrm{w}=0$};

	\draw (0,-0.2) to (0,-0.6);
	\draw (3.1,-0.2) to (3.1,-0.6);
	\draw (8,-0.2) to (8,-0.6);
	\draw[<->] (0,-0.4) to (3.1,-0.4);
	\draw (1.5,-0.4) node[anchor=north]{$10D$};
	\draw[<->] (3.1,-0.4) to (8,-0.4);
	\draw (5.5,-0.4) node[anchor=north]{$30D$};
	\draw (8.2,0) to (8.6,0);
	\draw[<->] (8.4,0) to (9.4,1);
	\draw (8.9,0.4) node[anchor=west]{$30D$};

	\draw[->,red] (0.5,7.5) to (1.5,7.5);
	\draw (1.5,7.5) node[anchor=west,red]{X};
	\draw[->,red] (0.5,7.5) to (0.5,8.5);
	\draw (0.5,8.5) node[anchor=south,red]{Z};
	\draw[->,red] (0.5,7.5) to (1.21,8.21);
	\draw (1.1,8.21) node[anchor=west,red]{Y};
	
	\draw (5.5,3) node {$\Omega^\mathrm{f}_1$};
	\draw (5.5,2.5) node {$(\rho^\mathrm{f}_1,\mu^\mathrm{f}_1)$};
	\draw[->] (3.1,1) to (3.1,2);
	\draw[->] (3.1,5) to (3.1,6);
\end{tikzpicture}
	\caption{}
	\end{subfigure}%
\caption{Schematic of the uniform flow past a flexible riser with internal two-phase flow: (a) the $X-Z$ cross-section of the flexible riser with the internal flow velocity profile at the inlet and outlet of the pipe, and (b) the computational setup and boundary conditions employed for the demonstration. Here, ${\boldsymbol{u}}^\mathrm{f} = (\mathrm{u},\mathrm{v},\mathrm{w})$ denotes the components of the fluid velocity and the hatched area shows the flexible structure (riser).} 
\label{riser_IF_schematic}
\end{figure}

\begin{figure}[h]
\centering
\begin{subfigure}[b]{0.5\textwidth}
		\includegraphics[trim={0.01cm 1cm 0.01cm 0.5cm},clip,width=8cm]{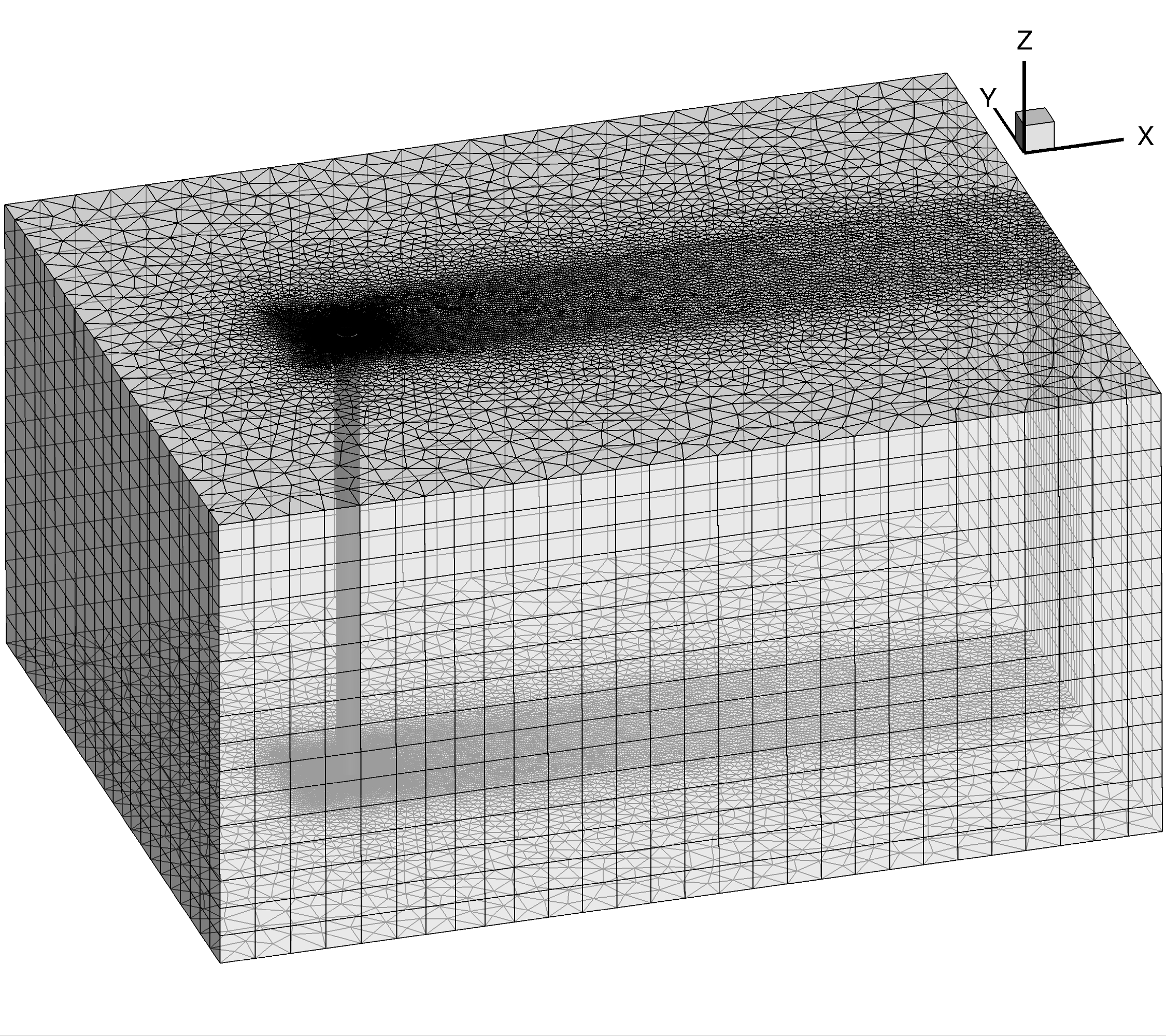}
\caption{} 
\end{subfigure}%
\begin{subfigure}[b]{0.5\textwidth}
		\includegraphics[trim={1cm 1.5cm 1cm 2cm},clip,width=9cm]{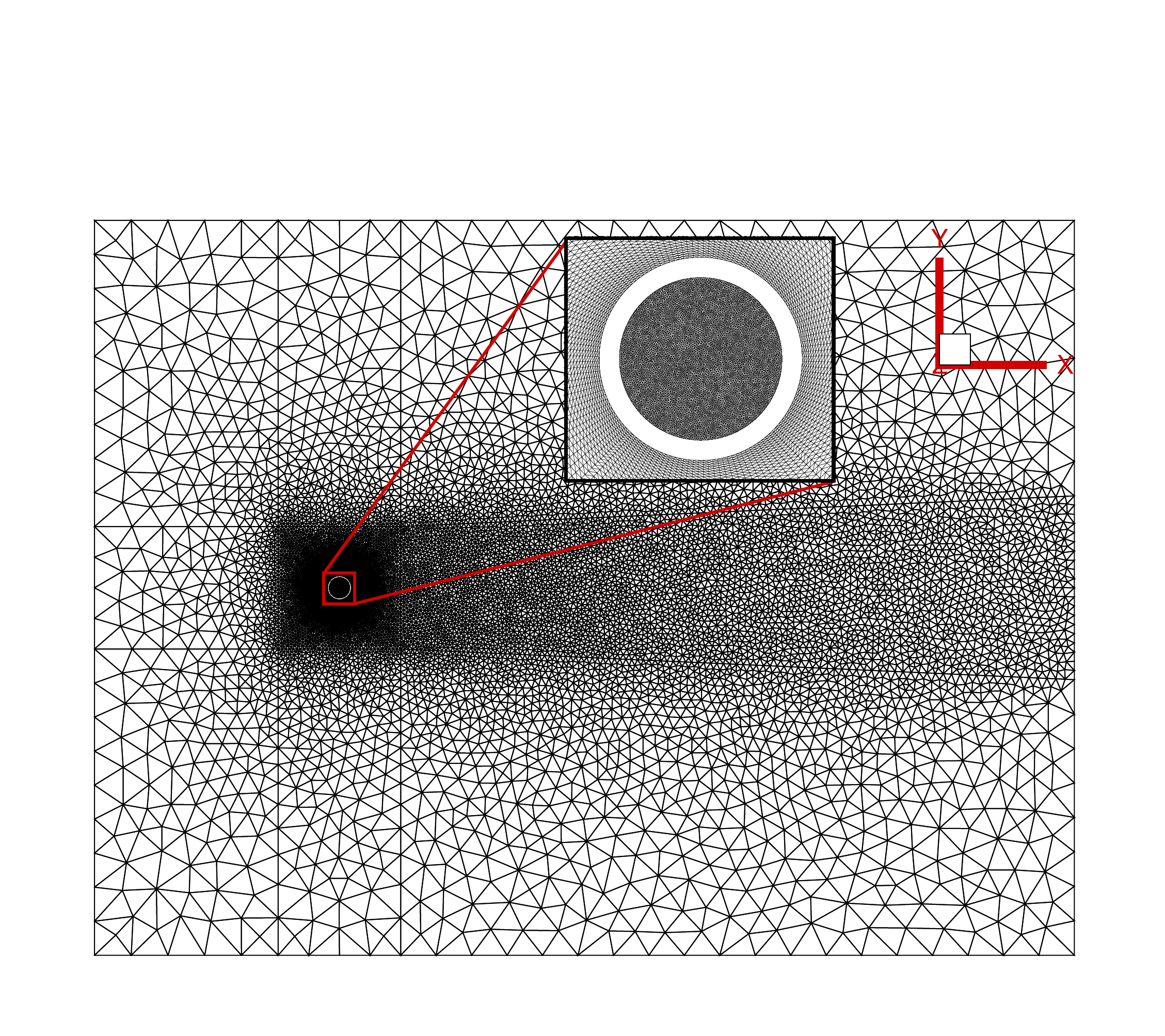}
		\caption{} 
\end{subfigure}
\caption{Computational mesh for the VIV of riser with internal two-phase flow: (a) three-dimensional view of the mesh, and (b) two-dimensional cross-section of the mesh with refined wake region behind the riser and refined internal region of the riser to capture the interface between the two phases in the internal flow accurately.} 
\label{riser_IF_mesh}
\end{figure}
In the present demonstration, we employ a linear flexible body solver for solving the structural equation. The Euler-Bernoulli beam equation is solved in the eigenspace with the structural displacement represented as a linear combination of the eigenmodes. This analysis has been explained in Appendix \ref{FMBD_eqn}. Following the notations from the analysis in Appendix \ref{FMBD_eqn}, the non-dimensional parameters for the VIV of riser with the internal flow are defined as follows:
\begin{align}
	Re = \frac{\rho^\mathrm{f}_1 U_{\infty}D}{\mu^\mathrm{f}_1} &= \frac{\rho^\mathrm{f}_2 U_{\infty}D}{\mu^\mathrm{f}_2},\qquad m^* = \frac{m^\mathrm{s}}{\pi D^2L\rho^\mathrm{f}_1 /4},\qquad \rho^* = \frac{\rho^\mathrm{f}_1}{\rho^\mathrm{f}_2},\qquad \mu^* = \frac{\mu^\mathrm{f}_1}{\mu^\mathrm{f}_2}, \nonumber \\
	&U_r = \frac{U_{\infty}}{f_1 D},\qquad P^* = \frac{P}{\rho^\mathrm{f}_1 U_{\infty}^2 D^2},\qquad EI^* = \frac{EI}{\rho^\mathrm{f}_1 U_{\infty}^2 D^4},
\end{align} 
where $Re$, $m^*$, $\rho^*$, $\mu^*$, $U_r$, $P^*$ and $EI^*$ denote the non-dimensional quantities, viz., Reynolds number, mass ratio, density ratio, viscosity ratio, reduced velocity, axial tension and flexural rigidity of the riser modeled as a beam. Here, $f_1$ denotes the frequency of the first eigenmode calculated by Eq.~(\ref{FMBD_freq}). We consider two cases with different Reynolds number in the present demonstration. The non-dimensional numbers related to the cases are shown in Table \ref{parameters_internalflow}.
\renewcommand{\arraystretch}{0.5}
\begin{table}[h]
\caption{Non-dimensional parameters for the two cases considered for the present study}
\centering
\begin{tabular}{  M{1.5cm}  M{1cm}  M{1cm}  M{1cm}  M{1cm}  M{1cm} M{1cm} M{1cm} M{1cm} N }
	\hline
\centering
	\textbf{Cases} & $Re$ & $m^*$ & $\rho^*$ & $\mu^*$ & $U_r$ & $P^*$ & $EI^*$ & $\rho^\mathrm{s}/\rho^\mathrm{f}_{1}$ &\\[10pt]
	\hline
\centering
	 Case 1 & 100  &  2.89 & 100 & 100  & 5 & 0.34 & 5872.8 & 6.68  &\\[10pt]

\centering
	 Case 2 & 1000  &  2.89 & 100 & 100  & 5 & 0.34 & 5872.8 & 6.68 &\\[10pt]
	\hline
\end{tabular}
\label{parameters_internalflow}
\end{table} 

The computational mesh for the demonstration is depicted in Fig. \ref{riser_IF_mesh}. The mesh is divided into two parts: external and internal. The external mesh consists of the fluid domain external to the riser. A two-dimensional cross-section is shown in Fig. \ref{riser_IF_mesh}(b) where we find a refined wake region to capture the flow structures. This external mesh is extruded in the third dimension consisting of $16$ layers which is evident from Fig. \ref{riser_IF_mesh}(a). The internal mesh comprises of the internal volume of the riser. It is much more refined than the external region to capture the two-phase interface and is extruded $500$ layers in the third dimension. The complete three-dimensional mesh contains $3.62$ million grid points with $7.07$ million six-node wedge elements. The solver performed $4-5$ nonlinear iterations to achieve a nonlinear convergence tolerance of $5\times 10^{-4}$. 

\begin{figure}[h]
\centering
\begin{subfigure}[b]{0.54\textwidth}
		\includegraphics[trim={1cm 0cm 3cm 0.0cm},clip,width=8.5cm]{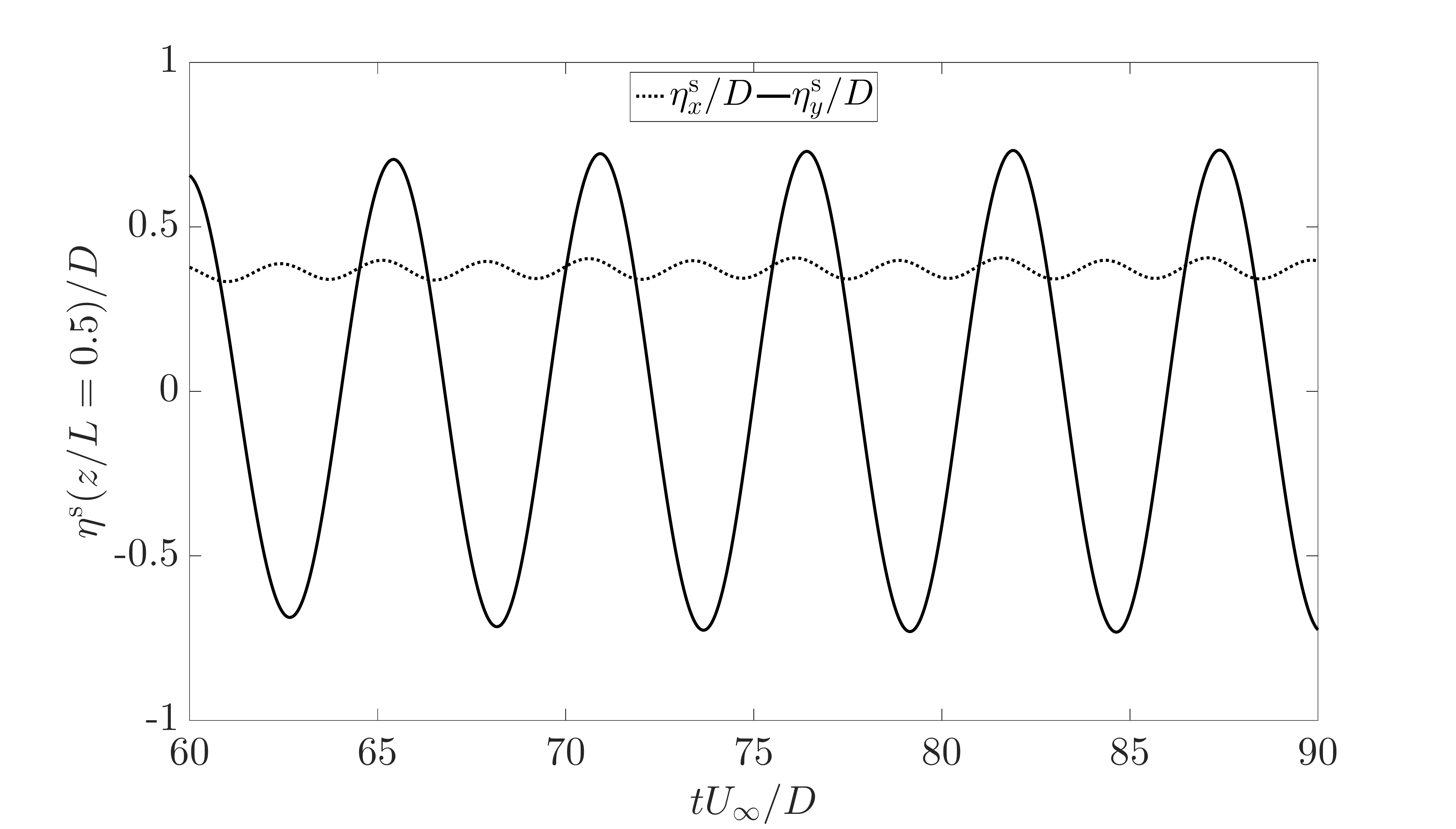}
\caption{} 
\end{subfigure}%
\begin{subfigure}[b]{0.54\textwidth}
		\includegraphics[trim={1cm 0cm 3cm 0cm},clip,width=8.5cm]{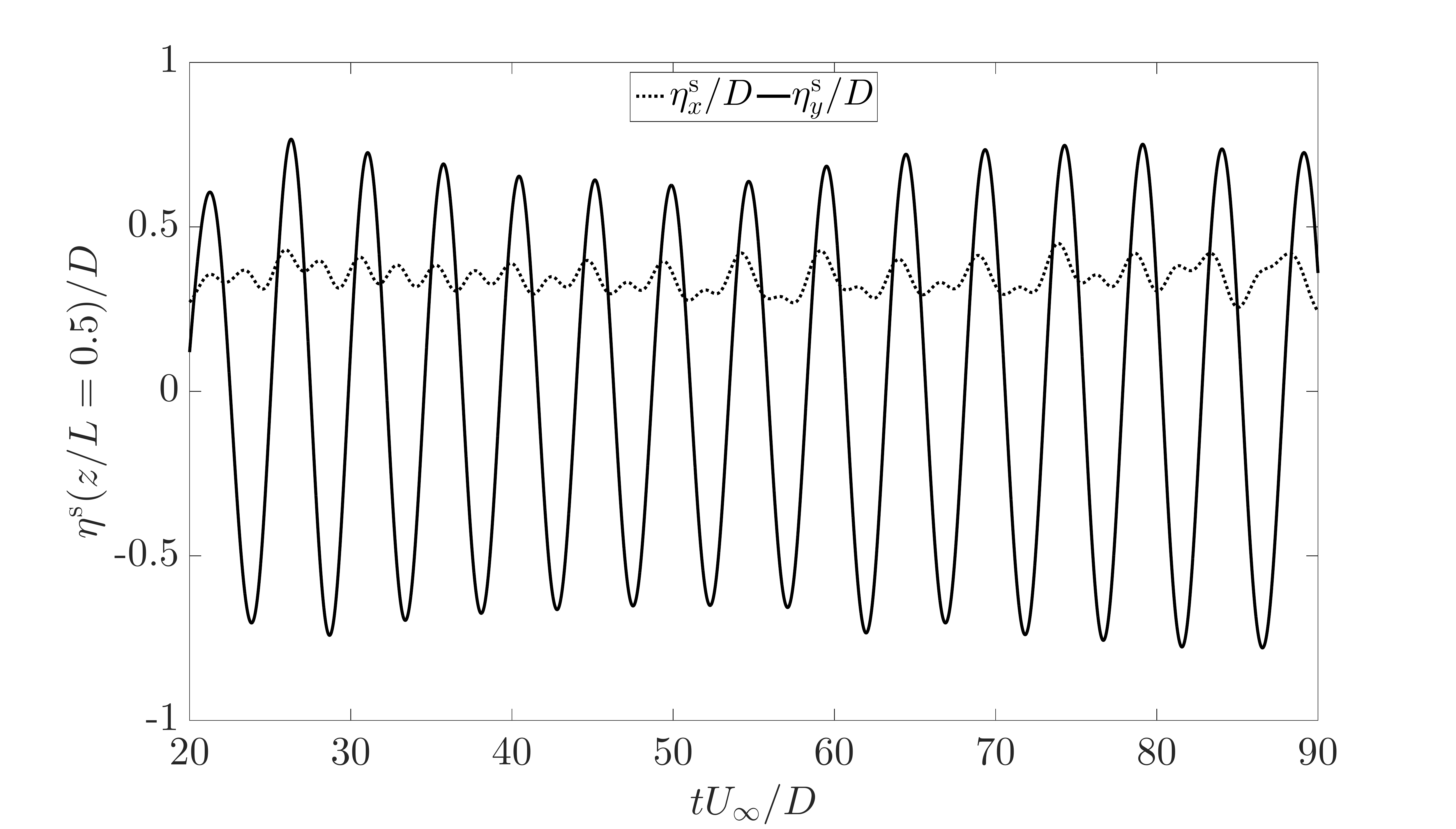}
		\caption{} 
\end{subfigure}
\caption{The response amplitude at the mid-point of the riser ($z/L=0.5$) exposed to external uniform flow with internal two-phase flow: 
(a) Case 1 ($Re=100$), and (b) Case 2 ($Re=1000$).} 
\label{riser_IF_disp}
\end{figure}

\begin{figure}[h]
\centering
\begin{subfigure}[b]{0.5\textwidth}
		\includegraphics[trim={10cm 0cm 10cm 0.0cm},clip,width=8cm]{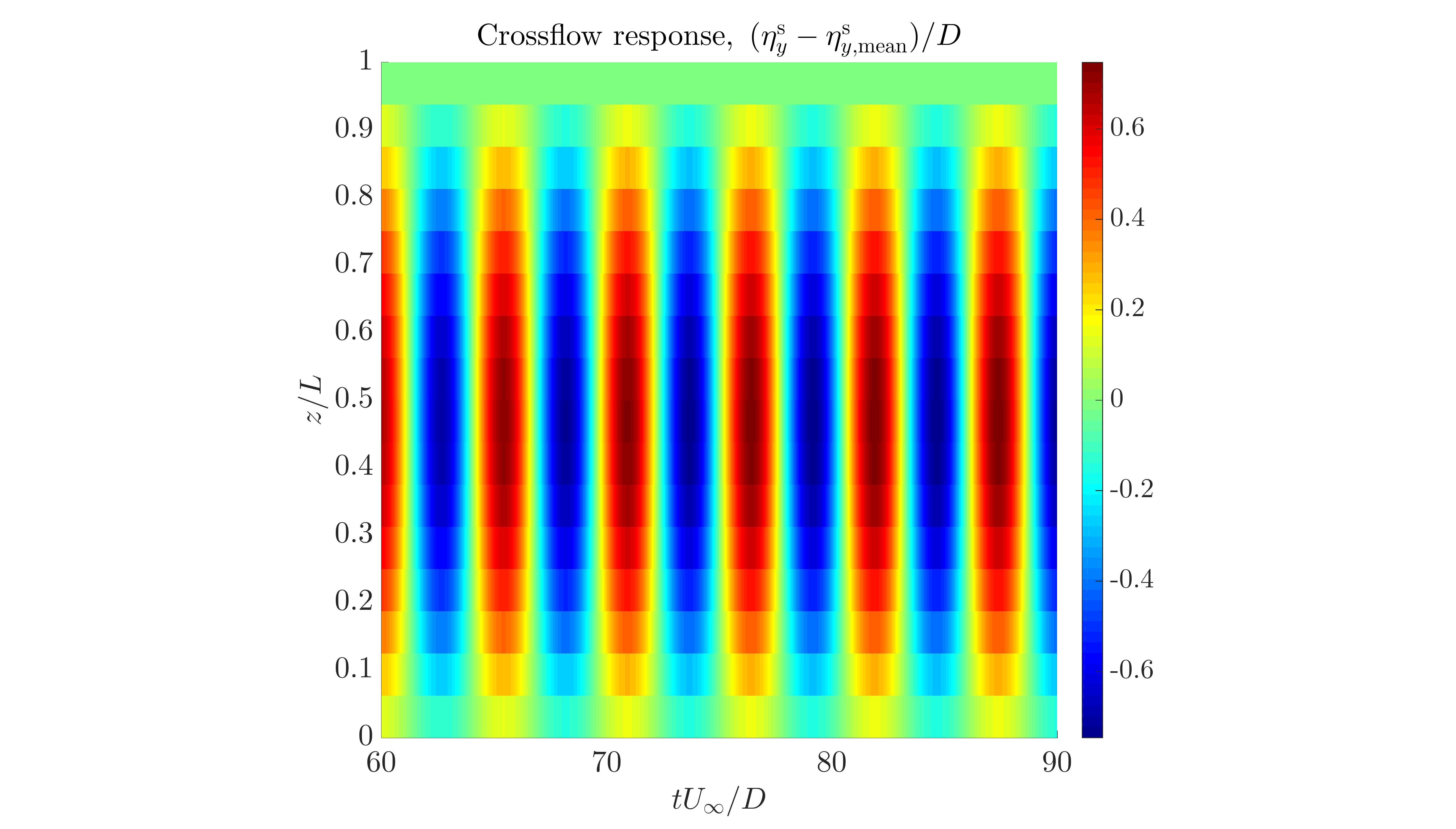}
\caption{} 
\end{subfigure}%
\begin{subfigure}[b]{0.5\textwidth}
		\includegraphics[trim={10cm 0cm 10cm 0.0cm},clip,width=8cm]{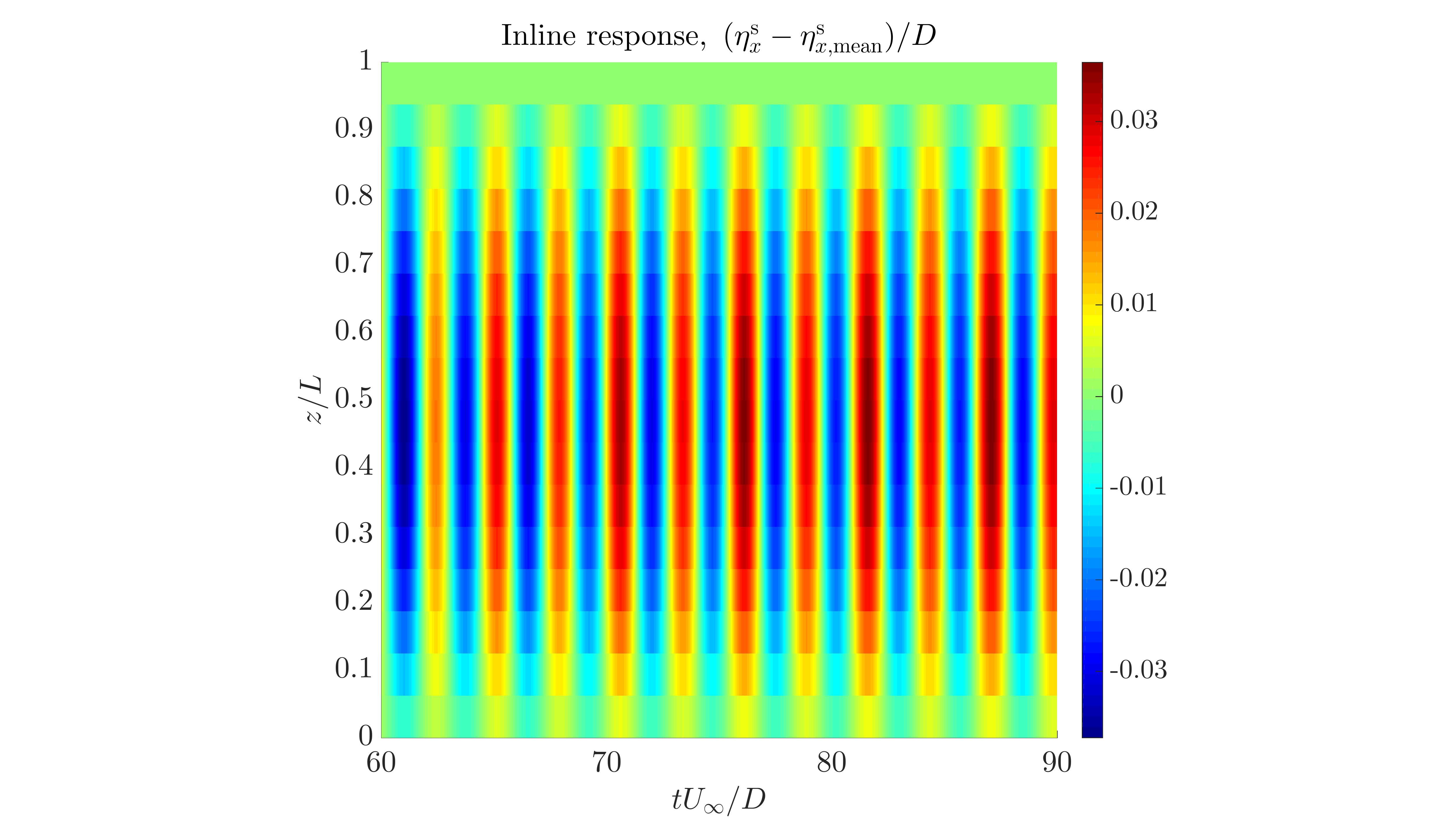}
		\caption{} 
\end{subfigure}
\caption{Response amplitudes of the riser along the span with non-dimensional time $tU_{\infty}/D$ for Case 1: (a) cross-flow, and (b) in-line.} 
\label{riser_IF_waves}
\end{figure}

\begin{figure}[h]
\centering
\begin{subfigure}[b]{0.4\textwidth}
		\includegraphics[trim={4cm 0.1cm 0.1cm 0.1cm},clip,width=6cm]{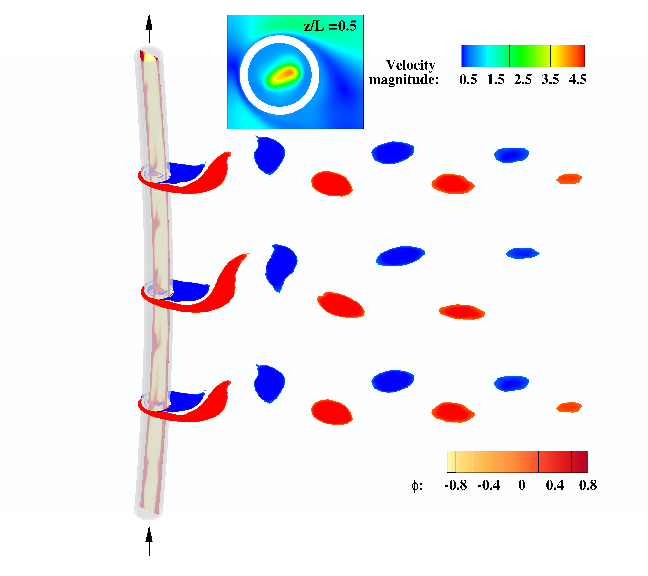}
\caption{} 
\end{subfigure}%
\begin{subfigure}[b]{0.4\textwidth}
		\includegraphics[trim={4cm 0.1cm 0.1cm 0.1cm},clip,width=6cm]{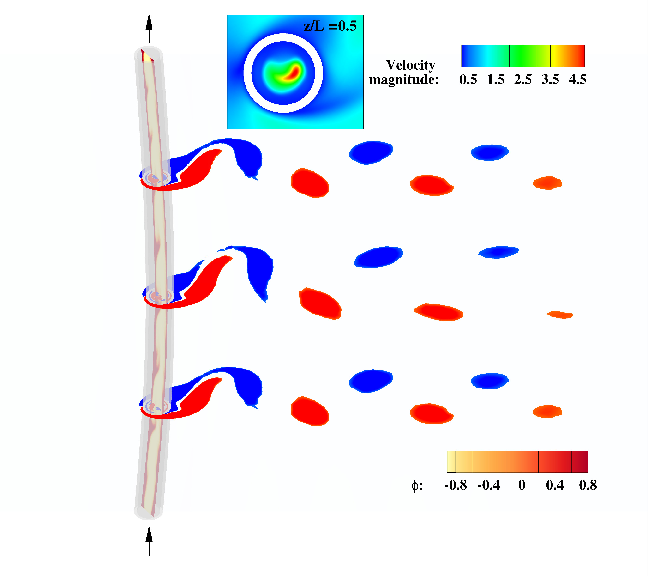}
		\caption{} 
\end{subfigure}%
\begin{subfigure}[b]{0.2\textwidth}
		\includegraphics[trim={6cm 0.5cm 6cm 0.5cm},clip,width=3.5cm]{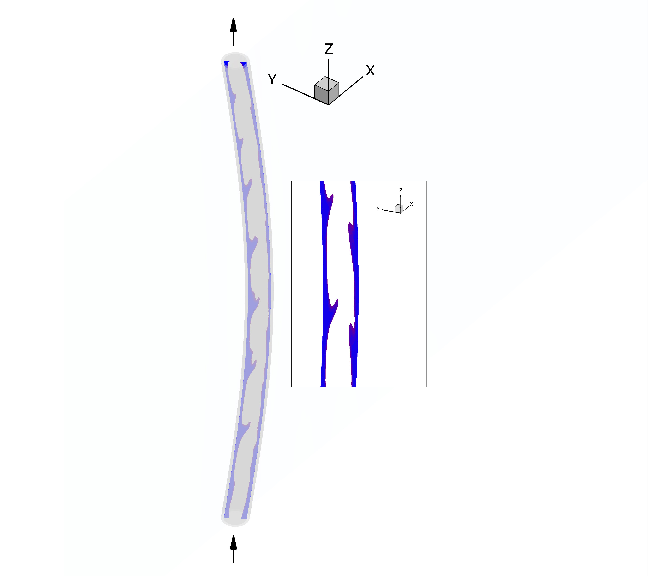}
		\caption{} 
\end{subfigure}
\caption{Contour plots for the VIV of a riser at $Re=100$ with internal two-phase flow  at $tU_{\infty}/D$: (a) $80$, (b) $90$ and (c) the internal flow along the riser. The $Z$-vorticity contours are shown at three cross-sections along the riser, viz., $z/L\in [0.25,0.5,0.75]$ and are colored with red for positive vorticity and blue for negative vorticity. The inset figure provides the velocity magnitude at the mid-section of the riser. The interior two-phase flow of the riser is visualized by the contours of order parameter $\phi>0$ at an arbitrary plane passing through the axis of the deformed riser.} 
\label{riser_IF_contourRe100}
\end{figure}

\begin{figure}[h]
\centering
\begin{subfigure}[b]{0.4\textwidth}
		\includegraphics[trim={4cm 0.1cm 0.1cm 0.1cm},clip,width=6cm]{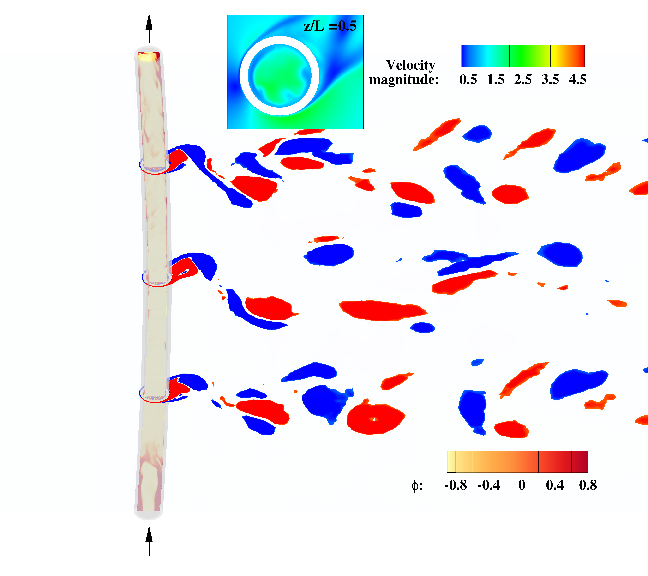}
\caption{} 
\end{subfigure}%
\begin{subfigure}[b]{0.4\textwidth}
		\includegraphics[trim={4cm 0.1cm 0.1cm 0.1cm},clip,width=6cm]{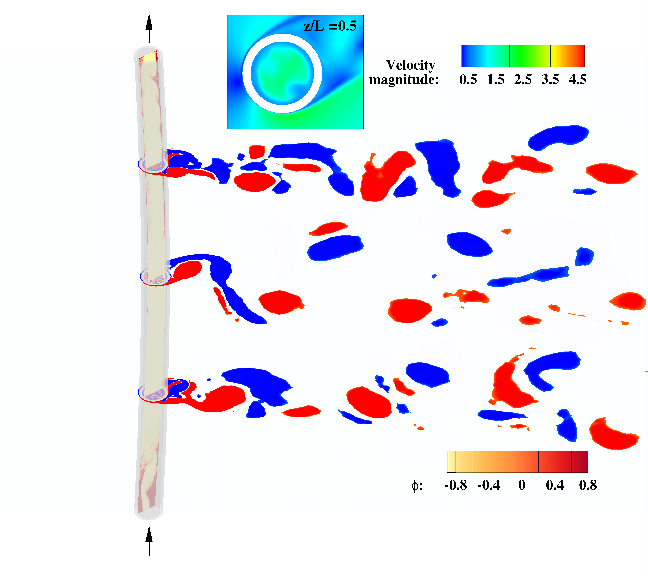}
		\caption{} 
\end{subfigure}%
\begin{subfigure}[b]{0.2\textwidth}
		\includegraphics[trim={6cm 0.2cm 6cm 0.5cm},clip,width=3.5cm]{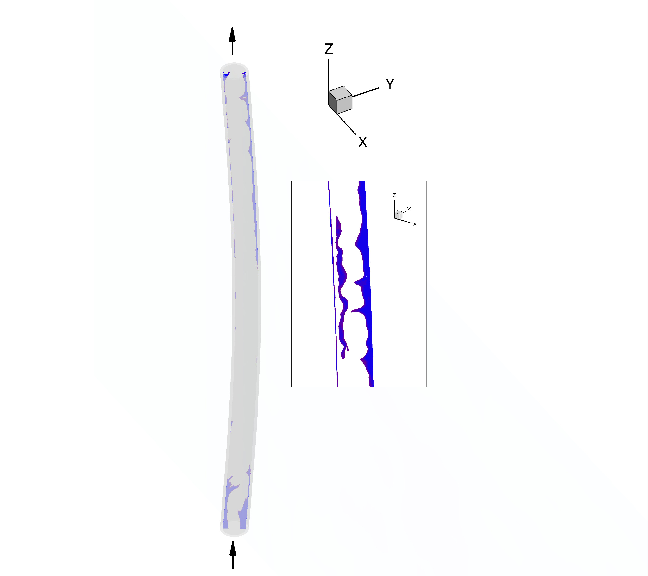}
		\caption{} 
\end{subfigure}
\caption{Contour plots for the VIV of a riser at $Re=1000$ with internal two-phase flow  at $tU_{\infty}/D$: (a) $80$, (b) $90$ and (c) the internal flow along the riser. The $Z$-vorticity contours are shown at three cross-sections along the riser, viz., $z/L\in [0.25,0.5,0.75]$ and are colored with red for positive vorticity and blue for negative vorticity. The inset figure provides the velocity magnitude at the mid-section of the riser. The interior two-phase flow of the riser is visualized by the contours of order parameter $\phi>0$ at an arbitrary plane passing through the axis of the deformed riser.} 
\label{riser_IF_contourRe1k}
\end{figure}

The amplitude of the riser is found maximum at the mid-point along its span and has been plotted in Fig. \ref{riser_IF_disp}. We also observe a standing wave pattern along the riser in Fig. \ref{riser_IF_waves} where the variation of the response with time along the riser is shown for the Case 1 at $Re=100$. The flow contours of the $Z$-vorticity along the riser span with the visualization of the internal flow via the order parameter $\phi$ is shown in Fig. \ref{riser_IF_contourRe100} for Case 1 and Fig. \ref{riser_IF_contourRe1k} for Case 2. The irregularity in the vortex patterns suggests the onset of turbulent wake for Case 2. It is found that the topological changes in the fluid-fluid interface are captured qualitatively in the current simulation. It is also observed that the co-annular initial two-phase flow pattern is transitioning into elongated bubble/slug flow pattern. This type of flow pattern prediction through a fully-coupled two-phase FSI can be advantageous to improve multiphase flow assurance. Further analysis is required to quantify the effect of the internal flow on the VIV or vice-versa which forms a topic for future study.

\section{Conclusions}
\label{conclusion}
In this study, a novel variational phase-field FSI formulation has been developed for the coupled analysis of fluid-structure interaction in two-phase flow. The one-fluid formulation for the two-phase flows with a diffuse interface description of the fluid-fluid interface offers an advantage to capture the interface with topological changes without involving 
any remeshing or complex geometric manipulations of unstructured mesh. 
The conservative Allen-Cahn equation has been utilized to evolve the fluid-fluid interface. 
The fluid-structure interface is considered in the Lagrangian manner whereas the two-phase flow equations are formulated in the ALE framework. This produces a hybrid Allen-Cahn/ALE scheme which can accurately capture the fluid-structure interface with phase-field based interface capturing of the fluid-fluid interface. 
The governing equations are solved via the nonlinear partitioned iterative technique which can be easily implemented in variational solvers with little effort. This type of coupling considering the Allen-Cahn equation to model the two-phases has been carried out for the first time. 
Such FSI simulations via Allen-Cahn phase-field model can be very powerful for highly complex three-dimensional evolving fluid-fluid and fluid-structure interfaces.

The desirable features of the proposed phase-field/ALE formulation are examined for two-phase flow interacting with rigid and flexible structures.
The robustness and generality of the proposed formulation have been demonstrated for increasing complexity of problems involving high density $\sim\mathcal{O}(10^3)$ and viscosity ratios $\sim\mathcal{O}(10^2)$ of the two fluid phases and low structure-to-fluid mass ratios. For the decay test, the structure-to-fluid density ratio is $\rho^\mathrm{s}/\rho^\mathrm{f}_{1}=0.5$ while the mass ratio for the three-dimensional flexible riser is $m^*=2.89$. These tests establish the high density/viscosity ratio and low mass ratio handling capability of the coupling. 
The convergence tests reveal almost second-order of temporal accuracy for the phase-field FSI formulation. It is also noticed that the coupled two-phase FSI solver performed around $4-5$ nonlinear iterations to achieve a nonlinear convergence tolerance of $5\times 10^{-4}$. With regard to spatial accuracy, the interfacial thickness parameter $\varepsilon=0.01$ is observed to be sufficient to capture the topological changes in the fluid-fluid interface accurately.   
The numerical results are validated with the available experimental as well as computational results from the literature for the two decay test problems 
namely free translational heave decay of a circular cylinder  and 
free rotation  of a rectangular barge.  
Finally, a practical problem of an internal flow through a flexible riser exposed to external flow vortex-induced vibration has been demonstrated. A detailed analysis of the effect of the internal flow on the VIV of the riser or vice-versa is a subject for future study.   

{\bf{Acknowledgements}}
\ack{The first author would like to thank for the financial support from  National Research Foundation through Keppel-NUS Corporate Laboratory. The conclusions put forward reflect the views of the authors alone, and not necessarily those of the institutions. }

\appendix
\setcounter{equation}{0} 
\setcounter{figure}{0}
\section{Structural equations}
\label{app_structure}
\subsection{Rigid body}
\label{MBD_eqn}
We first consider the six degress of freedom motion of a rigid body. Let the center of mass of the body in the reference configuration $\boldsymbol{x}^\mathrm{s}$ and the current configuration $\boldsymbol{\varphi}^\mathrm{s}$ be $\boldsymbol{x}^\mathrm{s}_{0}$ and $\boldsymbol{\varphi}^\mathrm{s}_{0}$ respectively and $\boldsymbol{\eta}^\mathrm{s}_{0}$ denote the displacement of the center of mass due to the translation of the body. Therefore, the rigid body kinematics is given by
\begin{align} \label{RBD_eqn}
	\boldsymbol{\varphi}^\mathrm{s} = \boldsymbol{Q}(\boldsymbol{x}^\mathrm{s}-\boldsymbol{x}^\mathrm{s}_{0}) + \boldsymbol{\varphi}^\mathrm{s}_{0} = \boldsymbol{Q}(\boldsymbol{x}^\mathrm{s}-\boldsymbol{x}^\mathrm{s}_{0}) + \boldsymbol{x}^\mathrm{s}_{0} + \boldsymbol{\eta}^\mathrm{s}_{0},
\end{align}
where $\boldsymbol{Q}$ is a rotation matrix. Using Eqs.~(\ref{varphi_eqn}) and (\ref{RBD_eqn}), 
\begin{align}
	\boldsymbol{\eta}^\mathrm{s} = (\boldsymbol{Q}-\boldsymbol{I})(\boldsymbol{x}^\mathrm{s}-\boldsymbol{x}^\mathrm{s}_{0}) + \boldsymbol{\eta}^\mathrm{s}_{0}, \label{etaeqn}\\
	\frac{\partial\boldsymbol{\eta}^\mathrm{s}}{\partial t} = \frac{\partial\boldsymbol{Q}}{\partial t}(\boldsymbol{x}^\mathrm{s}-\boldsymbol{x}^\mathrm{s}_{0}) + \frac{\partial\boldsymbol{\eta}^\mathrm{s}_{0}}{\partial t}, \label{detadteqn}
\end{align}
where $\boldsymbol{I}$ is the identity matrix and Eq.~(\ref{detadteqn}) is obtained by differentiating Eq.~(\ref{etaeqn}) with respect to time. Suppose the rotational degrees of freedom for the body are given by $\boldsymbol{\theta}^\mathrm{s}$. Equation (\ref{detadteqn}) can be restructured in terms of the angular velocity of the body denoted by $\boldsymbol{\omega}^\mathrm{s} = \partial\boldsymbol{\theta}^\mathrm{s}/dt$ as
\begin{align}
	\frac{\partial\boldsymbol{\eta}^\mathrm{s}}{\partial t} = \boldsymbol{\omega}^\mathrm{s}\times(\boldsymbol{\varphi}^\mathrm{s}-\boldsymbol{\varphi}^\mathrm{s}_{0}) + \frac{\partial\boldsymbol{\eta}^\mathrm{s}_{0}}{\partial t}
\end{align}
The rigid body equations are thus given by
\begin{align} \label{LS_MBD_1}
	\boldsymbol{M}^\mathrm{s}\frac{\partial^2\boldsymbol{\eta}^\mathrm{s}_{0}}{\partial t^2} + \boldsymbol{C}_{\eta}\frac{\partial\boldsymbol{\eta}^\mathrm{s}_{0}}{\partial t} + \boldsymbol{K}_{\eta}\boldsymbol{\eta}^\mathrm{s}_{0} &= \boldsymbol{f}^\mathrm{s},\ \mathrm{on}\ \Omega^\mathrm{s},\\
	\boldsymbol{I}^\mathrm{s}\frac{\partial^2\boldsymbol{\theta}^\mathrm{s}}{\partial t^2} + \boldsymbol{C}_{\theta}\frac{\partial\boldsymbol{\theta}^\mathrm{s}}{\partial t} + \boldsymbol{K}_{\theta}\boldsymbol{\theta}^\mathrm{s} &= \boldsymbol{\tau}^\mathrm{s},\ \mathrm{on}\ \Omega^\mathrm{s},	\label{LS_MBD_2}
\end{align}
where $\boldsymbol{M}^\mathrm{s}$, $\boldsymbol{C}_\eta$ and $\boldsymbol{K}_\eta$ denote the mass, damping and stiffness matrices for the translational degrees of freedom respectively, $\boldsymbol{I}^\mathrm{s}$, $\boldsymbol{C}_\theta$ and $\boldsymbol{K}_\theta$ represent the moment of inertia, damping and stiffness matrices for the rotational degrees of freedom respectively, and $\boldsymbol{f}^\mathrm{s}$ and $\boldsymbol{\tau}^\mathrm{s}$ denote the forces and the moments applied on the body respectively.  

\subsection{Linear flexible body}
\label{FMBD_eqn}
For modeling the flexible body dynamics, we consider a linear modal analysis by solving the Euler-Bernoulli beam equation. Suppose the axis of the beam is parallel to the $Z$-direction along which the coordinates are given by $z$. We solve for the lateral displacements (denoted by $\boldsymbol{\eta}^\mathrm{s}(z,t)$) along the beam as
\begin{align} \label{beam_eqn}
	m^\mathrm{s}\frac{\partial^2 \boldsymbol{\eta}^\mathrm{s}}{\partial t^2} + \frac{\partial^2}{\partial z^2}\bigg( EIL\frac{\partial^2\boldsymbol{\eta}^\mathrm{s}}{\partial z^2} \bigg) - PL\frac{\partial^2\boldsymbol{\eta}^\mathrm{s}}{\partial z^2} = \boldsymbol{f}^\mathrm{s},
\end{align}
where $m^\mathrm{s}$, $E$, $I$, $P$, and $\boldsymbol{f}^\mathrm{s}$ denote the mass, the Young's modulus, the second moment of the cross-sectional area of the beam, the applied axial tension and the external applied force on the beam of span $L$ respectively.

A beam under pinned-pinned condition has to satisfy the following boundary conditions at its ends
\begin{align}
	\boldsymbol{\eta}^\mathrm{s}|_{z=0} = \boldsymbol{0},\qquad & \frac{\partial^2\boldsymbol{\eta}^\mathrm{s}}{\partial z^2}\bigg|_{z=0} = \boldsymbol{0},\\
	\boldsymbol{\eta}^\mathrm{s}|_{z=L} = \boldsymbol{0},\qquad & \frac{\partial^2\boldsymbol{\eta}^\mathrm{s}}{\partial z^2}\bigg|_{z=L} = \boldsymbol{0}.
\end{align} 
To solve Eq.~(\ref{beam_eqn}), we employ a mode superposition procedure where the frequency of $n$th mode is given by
\begin{align} \label{FMBD_freq}
	f_n = \frac{1}{2\pi}\sqrt{\frac{\frac{\pi^4}{L^3}\bigg( n^4EI + \frac{n^2PL^2}{\pi^2}\bigg)}{m^\mathrm{s}}}
\end{align}
The structural displacements are represented by the superposition of linear eigenmodes which are obtained using the eigenvalue analysis. For the current configuration, the eigenmodes are assumed to be sinusoidal so that the eigenmode shape for mode $n$ is given by
\begin{align} \label{eigenmodes}
	S^n(z) = \mathrm{sin}\bigg( \frac{n\pi z}{L} \bigg).
\end{align}

Equation (\ref{beam_eqn}) can be recast into a matrix form as
\begin{align} \label{LBE_matrix}
	\boldsymbol{M}^\mathrm{s}\frac{\partial^2 \boldsymbol{\eta}^\mathrm{s}}{\partial t^2} + \boldsymbol{K}^\mathrm{s}\boldsymbol{\eta}^\mathrm{s} = \boldsymbol{f}^\mathrm{s},
\end{align}
where $\boldsymbol{M}^\mathrm{s}$ and $\boldsymbol{K}^\mathrm{s}$ are the mass and stiffness matrices respectively, $\boldsymbol{\eta}^\mathrm{s}$ is the vector of unknown displacements along the beam and $\boldsymbol{f}^\mathrm{s}$ is the vector of the force applied on the beam. Now, we project the above equation in the eigenspace with eigenmodes defined by Eq.~(\ref{eigenmodes}), which transforms Eq.~(\ref{LBE_matrix}) to a system of linear equations with $n$ degrees of freedom (modes) as
\begin{align} \label{LS_FMBD}
	\widetilde{\boldsymbol{M}^\mathrm{s}}\frac{\partial^2 \boldsymbol{\xi}^\mathrm{s}}{\partial t^2} + \widetilde{\boldsymbol{K}^\mathrm{s}}\boldsymbol{\xi}^\mathrm{s} = \widetilde{\boldsymbol{f}^\mathrm{s}},
\end{align}
where $\widetilde{\boldsymbol{M}^\mathrm{s}}=\boldsymbol{S}^T\boldsymbol{M}^\mathrm{s}\boldsymbol{S}$ and $\widetilde{\boldsymbol{K}}=\boldsymbol{S}^T\boldsymbol{K}^\mathrm{s}\boldsymbol{S}$ are the projected matrices on the eigenspace where $\boldsymbol{S}$ is the matrix containing the eigenvectors and $\boldsymbol{\xi}^\mathrm{s}$ is the vector of the modal responses along the beam. The projected force vector $\widetilde{\boldsymbol{f}^\mathrm{s}}=\boldsymbol{S}^T\boldsymbol{f}^\mathrm{s}$ and $\boldsymbol{\eta}^\mathrm{s}=\boldsymbol{S}\boldsymbol{\xi}^\mathrm{s}$.

\bibliographystyle{unsrt}
\bibliography{citations}

\end{document}